\documentclass[useAMS,usenatbib]{mn2e}
\usepackage{times}
\usepackage{booktabs}
\usepackage{cellspace}
\usepackage{dcolumn}

\usepackage{amsfonts}
\usepackage{amssymb}
\usepackage{amsmath}
\usepackage{graphicx}
\usepackage{multirow}
\usepackage{subfigure}
\usepackage{times}
\usepackage{sidecap}
\usepackage{url}

\usepackage[hyperfigures=false, colorlinks=true, pagecolor=black,
linkcolor=black, citecolor=black, urlcolor=black, pagebackref=false,
hyperfigures=true, bookmarks=true, bookmarksopen=true, pdfauthor={A.
Danehkar, Q. A. Parker and
B. Ercolano}, pdftitle={Observations and three-dimensional ionization structure of the planetary nebula SuWt 2},
pdfsubject={ISM: abundances, planetary nebulae: individual: PN SuWt 2}, pdfkeywords={planetary nebulae, ISM,
SuWt 2}]{hyperref}%

\def\aj{AJ}
\def\araa{ARA\&A}
\def\apj{ApJ}

\def\apjs{ApJS}

\def\apss{Ap\&SS}
\def\aap{A\&A}

\def\aaps{A\&AS}

\def\mnras{MNRAS}

\def\pasa{PASA}
\def\pasp{PASP}

\def\rmxaa{RMxAA}

\def\arcmin{\hbox{$^\prime$}}
\def\arcsec{\hbox{$^{\prime\prime}$}}

\def\ha{H$\alpha$}

\def\p0{\phantom{0}}

\def\lessim{\raise-.5ex\hbox{$\buildrel<\over{\scriptstyle\mathtt{\sim}}$}}
\def\grtsim{\raise-.5ex\hbox{$\buildrel>\over{\scriptstyle\mathtt{\sim}}$}}

\title[3-D structure of SuWt~2]{Observations and three-dimensional ionization structure \\of the planetary nebula SuWt~2\thanks{Based on observations made with the ANU 2.3-m Telescope at the Siding Spring Observatory
under programs 
2090064 and
3120158.
}}
\author[A.~Danehkar, Q.\,A.~Parker and B.~Ercolano]{A.~Danehkar,$^{\,1}$\thanks{E-mail: ashkbiz.danehkar@mq.edu.au}
Q.\,A.~Parker$^{\,1,2}$ and 
B.~Ercolano$^{\,3,4}$ 
\\
$^{1}$Research Centre in Astronomy, Astrophysics \& Astrophotonics, Macquarie University,
Sydney, NSW 2109, Australia\\
$^{2}$Australian Astronomical Observatory, PO Box 296, Epping, NSW
1710, Australia\\
$^{3}$Universit\"{a}ts-Sternwarte M\"{u}nchen, Ludwig-Maxmilians Universit\"{a}t M\"{u}nchen, Scheinerstr. 1, D-81679 M\"{u}nchen, Germany\\
$^{4}$Excellence Cluster Universe, Boltzmannstr. 2, D-85748 Garching, Germany
}

\begin{document}

\date{Accepted 2013 June 17. Received 2013 June 16; in original form 2013 March 20}

\pagerange{\pageref{firstpage}--\pageref{lastpage}} \pubyear{2013}

\maketitle

\begin{abstract}

The planetary nebula SuWt~2 (PN~G311.0+02.4), is an unusual object with a prominent, inclined central emission ellipse and faint bipolar extensions. It has two A-type stars in a proven binary system at the centre. However, the radiation from these two central stars is too soft to ionize the surrounding material leading to a so far fruitless search for the responsible ionizing source. Such a source is clearly required and has already been inferred to exist via an observed temporal variation of the centre-of-mass velocity of the A-type stars. Moreover, the ejected nebula is nitrogen-rich which raises question about the mass-loss process from a likely intermediate-mass progenitor. 
We use optical integral-field spectroscopy to study the emission lines of the inner nebula ring. This has enabled us to perform an empirical analysis of the optical collisionally excited lines, together with a fully three-dimensional photoionization modelling. Our empirical results are used to constrain the photoionization models, which determine the evolutionary stage of the responsible ionizing source and its likely progenitor. 
The time-scale for the evolutionary track of a hydrogen-rich model atmosphere is inconsistent with the dynamical age obtained for the ring. This suggests that the central star has undergone a very late thermal pulse. We conclude that the ionizing star could be hydrogen-deficient and compatible with what is known as a PG 1159-type star. The evolutionary tracks for the very late thermal pulse models imply a central star mass of $\sim 0.64 {\rm M}_{\bigodot}$, which originated from a $\sim 3 {\rm M}_{\bigodot}$  progenitor. The evolutionary time-scales suggest that the central star left the asymptotic giant branch about 25,000 years ago, which is consistent with the nebula's age.

\end{abstract}

\label{firstpage}


\section{Introduction}
\label{suwt2:sec:introduction}

The southern planetary nebula (PN) SuWt~2 (PN~G311.0+02.4) is a particularly exotic object. 
It appears as an elliptical ring-like nebula with much fainter bipolar lobes extending perpendicularly to the ring, and with what appears to be an obvious, bright central star. The inside of the ring is apparently empty, but brighter than the nebula's immediate surroundings. An overall view of this ring-shaped structure and its surrounding environment can be seen in the H$\alpha$ image available from the SuperCOSMOS H$\alpha$ Sky Survey \citep[SHS;][]{Parker2005}. 
\citet{West1976} classified SuWt~2 as of intermediate excitation class \citep[EC; $p=6$--$7$;][]{Aller1968} based on the strength of the He\,\textsc{ii} $\lambda$4686 and [O\,\textsc{ii}] $\lambda$3728 doublet lines. The line ratio of [N \textsc{ii}] $\lambda$6584 and H$\alpha$ illustrated by \cite{Smith2007} showed a nitrogen-rich nebula that most likely originated from post-main-sequence mass-loss of an intermediate-mass progenitor star.

Over a decade ago, \citet{Bond2000} discovered that the apparent central star of SuWt~2 (NSV~19992) is a detached double-lined eclipsing binary consisting of two early A-type stars of nearly identical type. 
Furthermore, \citet{Bond2002} suggested that this is potentially a triple system consisting of the two A-type stars
and a hot, unseen PN central star. However, to date, optical and UV studies have failed to find any signature of the nebula's true ionizing source
\citep[e.g.][]{Bond2002,Bond2003,Exter2003,Exter2010}. 
Hence the putative hot (pre-)white dwarf 
would have to be in a wider orbit
around the close eclipsing pair. 
\citet{Exter2010} recently derived a period of 4.91 d from time series photometry and spectroscopy of the eclipsing pair, and concluded that the centre-of-mass velocity of the central binary varies with time, based on different systemic velocities measured over the period from 1995 to 2001. This suggests the presence of an unseen third orbiting body, which they concluded is a white dwarf of $\sim 0.7 {\rm M}_{\bigodot}$, and is the source of ionizing radiation for the PN shell. 

There is also a very bright B1Ib star, SAO~241302 (HD 121228), located 73 arcsec northeast of the nebula. \citet{Smith2007} speculated that this star is the ionizing source for SuWt~2. However, the relative strength of He\,\textsc{ii}\,$\lambda$4686 in our spectra (see later) shows that the ionizing star must be very hot, $T$ $>$ 100,000\,K, so the B1 star is definitively ruled out as the ionizing source. 

Narrow-band H$\alpha$+[N\,\textsc{ii}] and [O\,\textsc{iii}]\,5007 images of SuWt 2 obtained by 
\citet{Schwarz1992} show that the angular dimensions of the bright elliptical ring are about $86.5$\,arcsec\,$\times$\,$43.4$\,arcsec at the 10\% of maximum surface brightness isophote \citep{Tylenda2003}, and are used throughout this paper. 
\citet{Smith2007} used the MOSAIC2 camera on the Cerro Tololo Inter-American Observatory
(CTIO) 4-m telescope to obtain a more detailed H$\alpha$+[N\,\textsc{ii}] image,
which hints that the ring is possibly the inner edge of a swept-up disc. 
The [N\,\textsc{ii}] image also shows 
the bright ring structure and much fainter bipolar lobes extending perpendicular
to the ring plane. We can see similar structure in the images taken by Bond and Exter in 1995 with the CTIO 1.5 m telescope using an H$\alpha$+[N\,\textsc{ii}] filter. Fig.\,\ref{suwt2:fig1} shows both narrow-band 
[N\,\textsc{ii}] 6584\,{\AA} and H$\alpha$ images taken in 1995 with the ESO 3.6 m
New Technology Telescope at the La Silla Paranal Observatory using the ESO Multi-Mode Instrument (EMMI).  
The long-slit emission-line spectra
also obtained with the EMMI (programme ID 074.D-0373) 
in 2005 revealed much more detail of the nebular morphology. The first
spatio-kinematical model using the EMMI long-slit data by
\citet{Jones2010} suggested the existence of 
a bright torus with a systemic heliocentric radial velocity of $-25\pm5$ km\,s${}^{-1}$ encircling the waist of an extended bipolar nebular shell.

In this paper, we aim to uncover the properties of the hidden hot ionizing source in SuWt~2.  We aim to do this by applying a self-consistent three-dimensional photoionization model using the \textsc{mocassin 3d} code by \citet{Ercolano2003a,Ercolano2005,Ercolano2008}.
In Section~\ref{suwt2:sec:observations}, we describe our optical integral field observations 
as well as the data reduction process and the corrections for interstellar extinction.
In  Section~\ref{suwt2:sec:kinematic}, we describe the kinematics. 
In Section~\ref{suwt2:sec:tempdens}, we present our derived electron temperature and density, together with
our empirical ionic abundances in Section~\ref{suwt2:sec:abundances}. 
In Section~\ref{suwt2:sec:photoionization}, we present derived
ionizing source properties and distance from our self-consistent photoionization models, followed by a conclusion 
in Section~\ref{suwt2:sec:conclusions}. 

\section{Observations and data reduction}
\label{suwt2:sec:observations}

Integral-field spectra of SuWt~2 were obtained during two observing runs in 2009 May and 2012 August with 
the Wide Field Spectrograph \citep[WiFeS;][]{Dopita2007}.
WiFeS is an image-slicing Integral Field Unit (IFU) developed and built for the ANU 2.3-m telescope at the Siding Spring Observatory, feeding a double-beam spectrograph. WiFeS samples 0.5 arcsec along each of twenty five $38$\,arcsec \,$\times$ \,$1$\,arcsec slits, which
provides a field-of-view of $25$\,arcsec\,$\times$\,$38$\,arcsec  and a spatial resolution element of  $1.0$\,arcsec\,$\times$\,$0.5$\,arcsec (or $1\arcsec \times 1\arcsec$ for {y-binning=2}). 
The spectrograph uses volume phase holographic gratings to provide a spectral resolution of $R=3000$ (100 km\,s${}^{-1}$ full width at half-maximum, FWHM),  
and $R=7000$ (45 km\,s${}^{-1}$ FWHM) for the red and blue arms, respectively.  
Each grating has a different wavelength coverage. 
It can operate two data accumulation modes: classical and nod-and-shuffle (N\&S). 
The N\&S accumulates both object and nearby sky-background data in either equal exposures or unequal exposures. 
The complete performance of the WiFeS has been fully described by
\citet{Dopita2007,Dopita2010}.

\renewcommand{\baselinestretch}{0.9}
\begin{figure}
\begin{center}
\includegraphics[width=3.3in]{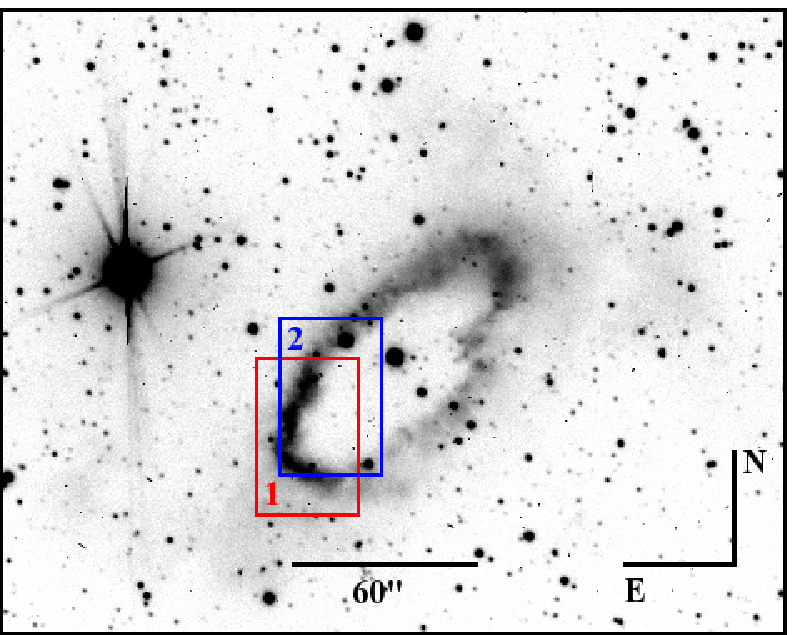}\\
\includegraphics[width=1.7in]{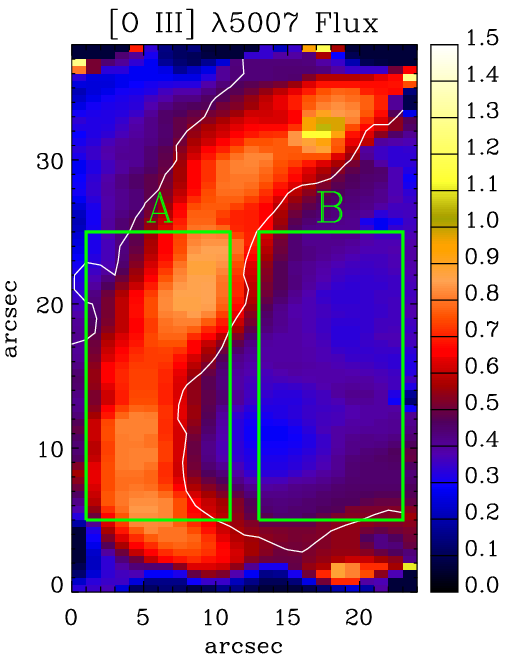}%
\includegraphics[width=1.7in]{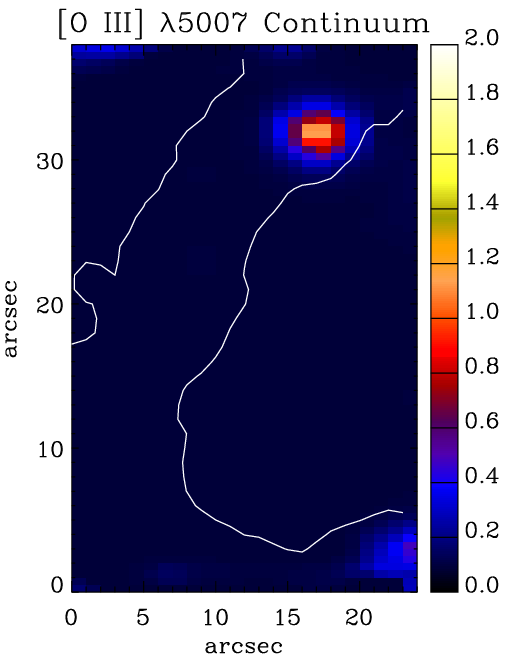}%
\caption{
Top panel: narrow-band filter image of the PN~SuWt~2 on a logarithmic scale in 
H$\alpha$ and [N\,\textsc{ii}]\,6584\,{\AA} taken 
with the European Southern Observatory (ESO) 3.6-m telescope
(programme ID 055.D-0550). The rectangles 
correspond to the WiFeS fields of view used for our study: 1 (red) and 2 (blue); see Table~\ref{suwt2:tab:observations} for more details. 
Bottom panels: Spatial distribution maps of flux intensity and continuum of $[$O~{\sc iii}$]$ $\lambda$5007 for field 2 and locations of apertures ($10$\,arcsec\,$\times$\,$20$\,arcsec) used to 
integrate fluxes, namely `A' the ring and `B' the interior structure.
The white contour lines show the distribution of the above narrow-band H$\alpha$ emission in arbitrary unit.
North is up and east is towards the left-hand side. 
}%
\label{suwt2:fig1}%
\end{center}
\end{figure}
\renewcommand{\baselinestretch}{1.5}

Our observations were carried out with the B3000/R3000  grating combination and the RT\,560 dichroic using N\&S mode in 2012 August; and 
the B7000/R7000 grating combination and the RT\,560 dichroic using the classical mode in 2009 May. This covers  $\lambda\lambda$3300--5900 {\AA} 
in the blue channel and $\lambda\lambda$5500--9300 {\AA}
in the red channel.
As summarized in Table~\ref{suwt2:tab:observations}, we took two different WiFeS exposures from different positions of SuWt~2; see Fig.~\ref{suwt2:fig1} (top).
The sky field was collected about 1 arcmin away from the object.   
To reduce and calibrate the data, it is necessary to take the usual bias frames, dome flat-field frames, twilight sky flats, `wire' frames and arc calibration lamp frames. 
Although wire, arc, bias and dome flat-field frames were collected during the afternoon prior to observing, arc and bias frames were also taken through the night. Twilight sky flats were taken in the evening.
For flux calibration, we also observed some spectrophotometric standard stars.

\renewcommand{\baselinestretch}{1.2}
\begin{table}
\begin{center}
\caption{Journal of SuWt 2 Observations at the ANU 2.3-m Telescope.}
\begin{tabular}{lccc}
\hline
\hline
Field  & 1 & 2 \\
\hline
Instrument  & WiFeS & WiFeS \\
Wavelength Resolution   & $\sim 7000$ & $\sim 3000$ \\
\multirow{2}{*} {Wavelength Range ({\AA})}   & 4415--5589, & 3292--5906,\\
   & 5222--7070 & 5468--9329\\
Mode & Classical & N\&S \\
Y-Binning & 1 & 2 \\
Object Exposure (s) & $900$ & $1200$ \\
Sky Exposure (s) & -- & $600$ \\
Standard Star & LTT\,3218  & LTT\,9491, \\ 
              &            & HD\,26169 \\ 
$v_{\rm LSR}$ correction & $-5.51$ & $-25.77$ \\
Airmass  & $1.16$ & $1.45$ \\
\multirow{2}{*} {Position (see Fig.\,\ref{suwt2:fig1})} 
  & 13:55:46.2 & 13:55:45.5 \\
  & $-59$:22:57.9 & $-59$:22:50.3 \\
Date (UTC) & 16/05/09 & 20/08/12 \\
\hline \label{suwt2:tab:observations}
\end{tabular}
\end{center}
\end{table}
\renewcommand{\baselinestretch}{1.5}

\subsection{WiFeS data reduction}
 
The WiFeS data were reduced using the \textsc{wifes} pipeline (updated on 2011 November 21), which is based 
on the {Gemini} \textsc{iraf}\footnote{The Image Reduction and Analysis Facility (\textsc{iraf}) software is distributed by the National Optical Astronomy Observatory. 
} package (version 1.10; \textsc{iraf} version 2.14.1)
developed by the Gemini Observatory for the integral-field spectroscopy.  

Each CCD pixel in the WiFeS camera has a slightly different sensitivity, giving pixel-to-pixel variations in the spectral direction. This effect is corrected using the dome flat-field frames taken with a quartz iodine (QI) lamp. Each slitlet is corrected for slit transmission variations using the twilight sky frame taken at the beginning of the night.
The wavelength calibration was performed using Ne--Ar arc exposures taken at the beginning of the night and throughout the night. 
For each slitlet the corresponding arc spectrum is extracted, and then wavelength solutions for each slitlet are obtained from the extracted arc lamp spectra using low-order polynomials. 
The spatial calibration was accomplished by using so called `wire' frames obtained by diffuse illumination of the coronagraphic aperture with a QI lamp. This procedure locates only the centre of each slitlet, since small spatial distortions by the spectrograph are corrected by the WiFeS cameras.
Each wavelength slice was also corrected for the differential atmospheric refraction by relocating each slice in $x$ and $y$ to its correct spatial position.

In the N\&S mode, the sky spectra are accumulated in the unused 80 pixel spaces between the adjacent object slices. The sky subtraction is conducted by subtracting the image shifted by 80 pixels from the original image. 
The cosmic rays and bad pixels were removed from the raw data set prior to sky subtraction using the \textsc{iraf} task LACOS\_IM of the cosmic ray identification procedure of  \citet{Dokkum2001}, which is based on a Laplacian edge detection algorithm. 
However, a few bad pixels and cosmic rays still remained in raw data, and these were manually removed by 
the \textsc{iraf/stsdas} task IMEDIT. 

We calibrated the science data to absolute flux units using observations of spectrophotometric standard stars 
observed in classical mode (no N\&S), so sky regions within the object data cube were used for sky subtraction. 
An integrated flux standard spectrum
is created by summing all spectra in 
a given aperture. 
After manually removing absorption features, an absolute calibration curve is fitted to 
the integrated spectrum using third-order polynomials. 
The flux calibration curve was then applied to the object data 
to convert to an absolute flux scale. The $[$O~{\sc i}$]\,\lambda$5577{\AA} night sky line was compared in the sky spectra of 
the red and blue arms to determine a difference in the flux levels, which was used to scale the blue spectrum of the science data. 
Our analysis using different spectrophotometric standard stars (LTT\,9491 and HD\,26169) revealed that the spectra at the extreme blue
have an uncertainty of about 30\% and are particularly unreliable for faint objects due to the CCD's poor sensitivity in this area.

\subsection{Nebular spectrum and reddening}

\begin{figure*}
\begin{center}
\includegraphics[width=6.9in]{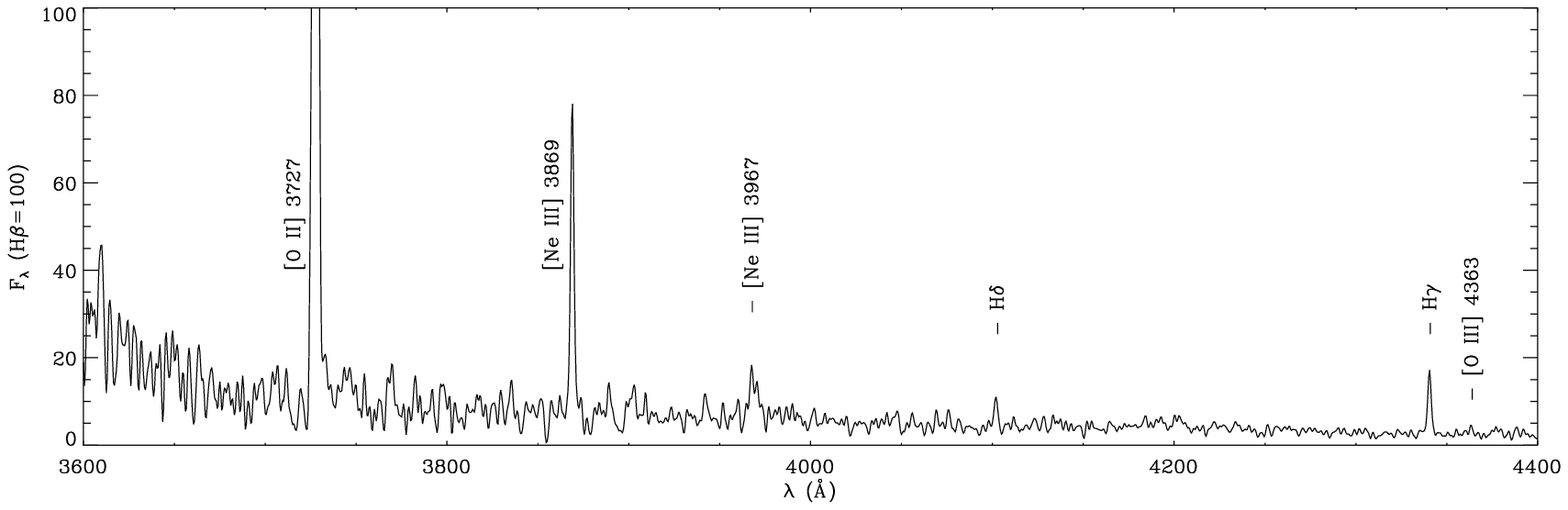}\\
\includegraphics[width=6.9in]{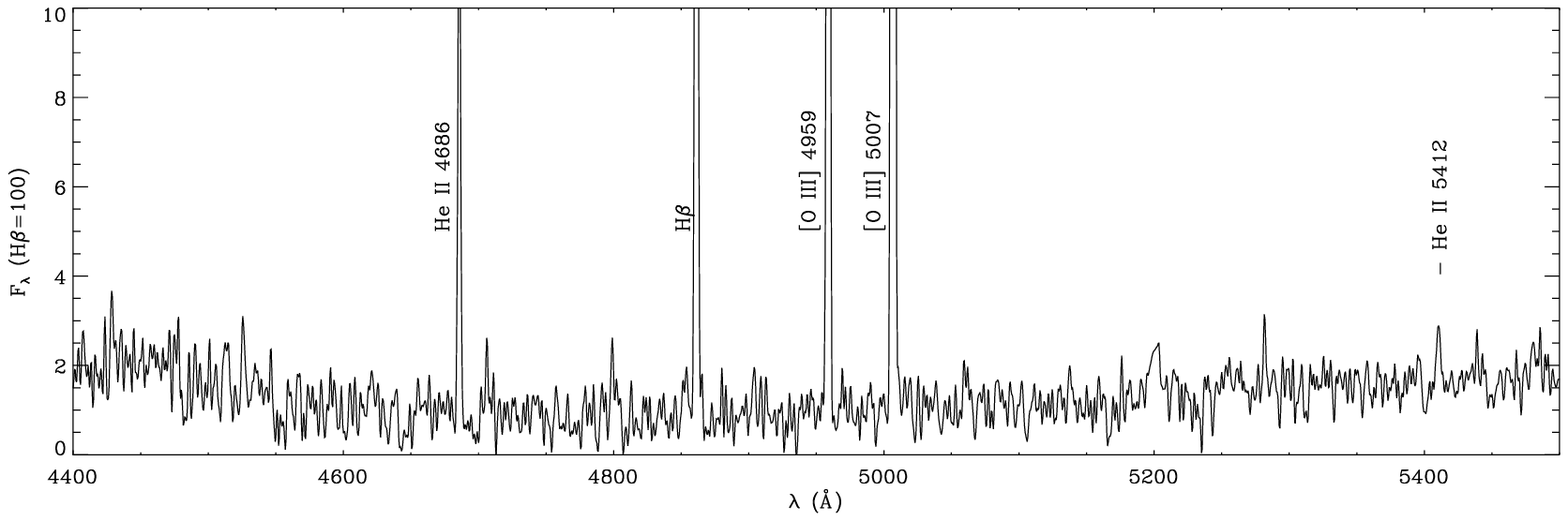}\\
\includegraphics[width=6.9in]{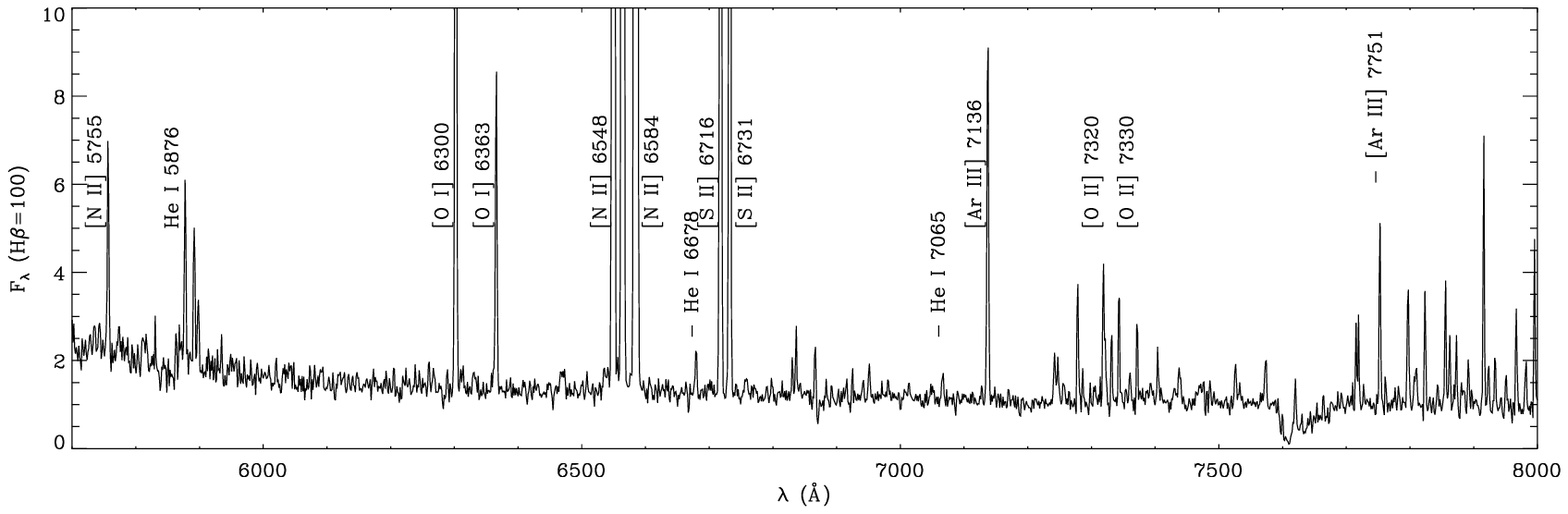}%
\caption{
The observed optical spectrum from an aperture $10$\,arcsec\,$\times$\,$20$\,arcsec taken from field 2 located on the east ring of the PN~SuWt~2 and normalized such that $F({\rm H}\beta)=100$.
}
\label{suwt2:fig2}%
\end{center}
\end{figure*}

Table~\ref{suwt2:tab:obslines} represents a full list of observed lines and their measured fluxes from different apertures ($10$\,arcsec\,$\times$\,$20$\,arcsec) taken from field 2: (A) 
the ring and (B) the inside of the shell. Fig.~\ref{suwt2:fig1} (bottom panel) shows the location and area of each aperture in the nebula. The top and bottom panels of Fig.~\ref{suwt2:fig2} show the extracted blue and red spectra after integration over the aperture located on the ring with the strongest lines truncated so the weaker features can be seen. The emission line identification, laboratory wavelength, multiplet number,  the transition with the lower- and upper-spectral terms, are given in columns 1--4 of Table~\ref{suwt2:tab:obslines}, respectively.
The observed fluxes of the interior and ring, and the fluxes after correction for interstellar extinction are given in columns 5--8.
Columns 9 and 10 present the integrated and dereddened fluxes after integration over two apertures (A and B). All fluxes are given relative to H$\beta$, on a scale where ${\rm H}\beta=100$. 

\renewcommand{\baselinestretch}{1.2}
\begin{table*}
\begin{center}
\caption{Observed and dereddened relative line fluxes, on a scale where ${\rm H}\beta = 100$. 
The integrated observed H($\beta$) flux was dereddened using $c({\rm H}\beta)$ to give an integrated dereddened flux.
Uncertain (errors of 20\%) and very uncertain (errors of 30\%) values are followed by ``:'' and ``::'', respectively.   
The symbol `*' denotes doublet emission lines.  
}
\begin{tabular}{lccccccccccccc}
\hline
\hline
\multicolumn{4}{l}{ Region}  & \multicolumn{2}{c}{Interior} &  \multicolumn{2}{c}{Ring} & \multicolumn{2}{c}{Total}  \\
\hline
Line  & $\lambda_{\rm lab}$({\AA}) & {Mult} & {Transition}&  $F(\lambda)$  & $I(\lambda)$ & $F(\lambda)$  & $I(\lambda)$   & $F(\lambda)$  & $I(\lambda)$  \\
{(1)} & {(2)} & {(3)} & {(4)} & {(5)} & {(6)} & 
{(7)} & {(8)} & {(9)} & {(10)} \\
\hline
3726 $[$O~{\sc ii}$]$ & 3726.03 & {F1}  & {$2{\rm p}^{3}\,{}^{4}{\rm S}_{3/2}-2{\rm p}^{3}\,{}^{2}{\rm D}_{3/2}$}  & $183 \pm 54$ & $307 \pm 91$ &  $576 \pm 172$ & $815 \pm 244$  &  $479 \pm 143$ & $702\pm 209$ \\
3729 $[$O~{\sc ii}$]$ &  3728.82 & {F1} & {$2{\rm p}^{3}\,{}^{4}{\rm S}_{3/2}-2{\rm p}^{3}\,{}^{2}{\rm D}_{5/2}$} & *  & *  & *  & *  &  * &  * \\
3869 $[$Ne~{\sc iii}$]$ & 3868.75 & {F1}  & {$2{\rm p}^{4}\,{}^{3}{\rm P}_{2}-2{\rm p}^{4}\,{}^{1}{\rm D}_{2}$} & 128.93:: & 199.42:: &  144.31:: & 195.22::  & 145.82:: & 204.57:: \\
3967 $[$Ne~{\sc iii}$]$  & 3967.46 & {F1} & { $2{\rm p}^{4}\,{}^{3}{\rm P}_{1}-2{\rm p}^{4}\,{}^{1}{\rm D}_{2}$}  & -- & -- & 15.37:: & 20.26:: & --  & --  \\
4102 H$\delta$ & 4101.74 &  {H6} & {$2{\rm p}\,{}^{2}{\rm P}-6{\rm d}\,{}^{2}{\rm D}$}  & -- &  --  & 16.19: & 20.55: &  16.97: & 22.15:  \\
4340 H$\gamma$ & 4340.47 &  {H5} & {$2{\rm p}\,{}^{2}{\rm P}-5{\rm d}\,{}^{2}{\rm D}$}  &  24.47:: & 31.10:: &  30.52: & 36.04:  & 31.69: & 38.18: \\
4363 $[$O~{\sc iii}$]$ & 4363.21 &  {F2} & { $2{\rm p}^{2}\,{}^{1}{\rm D}_{2}-2{\rm p}^{2}\,{}^{1}{\rm S}_{0}$} & 37.02:: & 46.58:: &  5.60 &  6.57  & 5.15 & 6.15 \\
4686 He~{\sc ii} & 4685.68 &  {3-4}  & { $3{\rm d}\,{}^{2}{\rm D}-4{\rm f}\,{}^{2}{\rm F}$} & 80.97 & 87.87 & 29.98 & 31.72  & 41.07 & 43.76\\
4861 H$\beta$ & 4861.33 &  {H4} & { $2{\rm p}\,{}^{2}{\rm P}-4{\rm d}\,{}^{2}{\rm D}$} & 100.00 & 100.00 &  100.00   &  100.00  &  100.00 & 100.00  \\
4959 $[$O~{\sc iii}$]$ & 4958.91 & {F1} & { $2{\rm p}^{2}\,{}^{3}{\rm P}_{1}-2{\rm p}^{2}\,{}^{1}{\rm D}_{2}$} & 390.90 & 373.57 &  173.63 & 168.27  &  224.48 & 216.72 \\
5007 $[$O~{\sc iii}$]$ & 5006.84 & {F1} & { $2{\rm p}^{2}\,{}^{3}{\rm P}_{2}-2{\rm p}^{2}\,{}^{1}{\rm D}_{2}$} &  1347.80 & 1259.76 &  587.22 & 560.37  & 763.00 & 724.02 \\
5412 He~{\sc ii} &  5411.52  & {4-7} & { $4{\rm f}\,{}^{2}{\rm F}-7{\rm g}\,{}^{2}{\rm G}$} &  19.33 & 15.01 &  5.12 & 4.30 & 6.90 & 5.68  \\
5755 $[$N~{\sc ii}$]$  & 5754.60  & {F3} & { $2{\rm p}^{2}\,{}^{1}{\rm D}_{2}-2{\rm p}^{2}\,{}^{1}{\rm S}_{0}$} & 7.08:  & 4.90: &  13.69 & 10.61 &  10.17 & 7.64 \\
5876 He~{\sc i} &  5875.66 & {V11} & { $2{\rm p}\,{}^{3}{\rm P}-3{\rm d}\,{}^{3}{\rm D}$} &  --  & -- & 11.51 &  8.69 & 8.96 & 6.54 \\
6548 $[$N~{\sc ii}$]$ & 6548.10 & {F1} & { $2{\rm p}^{2}\,{}^{3}{\rm P}_{1}-2{\rm p}^{2}\,{}^{1}{\rm D}_{2}$} & 115.24 & 63.13 &  629.36 &  414.79 &  513.64 & 321.94  \\
6563 H$\alpha$ &  6562.77 & {H3} & { $2{\rm p}\,{}^{2}{\rm P}-3{\rm d}\,{}^{2}{\rm D}$} & 524.16 & 286.00 &  435.14 & 286.00  &  457.70 & 286.00 \\
6584 $[$N~{\sc ii}$]$ & 6583.50 & {F1} & { $2{\rm p}^{2}\,{}^{3}{\rm P}_{2}-2{\rm p}^{2}\,{}^{1}{\rm D}_{2}$} & 458.99 & 249.05 & 1980.47 & 1296.67  & 1642.12 & 1021.68 \\
6678 He~{\sc i} & 6678.16 & {V46} & { $2{\rm p}\,{}^{1}{\rm P}_{1}-3{\rm d}\,{}^{1}{\rm D}_{2}$} & -- & -- & 3.30 & 2.12 & 2.68  &  1.63 \\
6716 $[$S~{\sc ii}$]$ & 6716.44 & {F2} & { $3{\rm p}^{3}\,{}^{4}{\rm S}_{3/2}-3{\rm p}^{3}\,{}^{2}{\rm D}_{5/2}$} & 60.63 & 31.77 & 131.84 & 84.25  & 116.21 &  70.36 \\
6731 $[$S~{\sc ii}$]$ & 6730.82 & {F2} & { $3{\rm p}^{3}\,{}^{4}{\rm S}_{3/2}-3{\rm p}^{3}\,{}^{2}{\rm D}_{3/2}$} & 30.08 & 15.70  & 90.39 & 57.61 & 76.98 & 46.47  \\
7005 [Ar V]  &  7005.40& {F1}   &   {$3{\rm p}^{2}\,{}^{3}{\rm P}-3{\rm p}^{2}\,{}^{1}{\rm D}$} & 5.46: &  2.66: & -- & -- & -- &  -- \\
7136 $[$Ar~{\sc iii}$]$ &  7135.80 & {F1} & { $3{\rm p}^{4}\,{}^{3}{\rm P}_{2}-3{\rm p}^{4}\,{}^{1}{\rm D}_{2}$} & 31.81 & 15.03 &  26.22 & 15.59  &  27.75 & 15.51 \\
7320 $[$O~{\sc ii}$]$ &  7319.40 & {F2} & { $2{\rm p}^{3}\,{}^{2}{\rm D}_{5/2}-2{\rm p}^{3}\,{}^{2}{\rm P}$} & 18.84 &  8.54 & 9.00 & 5.20  &  10.96 & 5.93 \\
7330 $[$O~{\sc ii}$]$ &  7329.90 & {F2} & { $2{\rm p}^{3}\,{}^{2}{\rm D}_{3/2}-2{\rm p}^{3}\,{}^{2}{\rm P}$} & 12.24 & 5.53  &  4.50 & 2.60  &  6.25 & 3.37 \\
7751 $[$Ar~{\sc iii}$]$ & 7751.43 & {F1} & { $3{\rm p}^{4}\,{}^{3}{\rm P}_{1}-3{\rm p}^{4}\,{}^{1}{\rm D}_{2}$} & 46.88 & 19.38  & 10.97  & 5.95 & 19.05 & 9.60  \\
9069 $[$S~{\sc iii}$]$ &  9068.60 & {F1} &  { $3{\rm p}^{2}\,{}^{3}{\rm P}_{1}-3{\rm p}^{2}\,{}^{1}{\rm D}_{2}$} & 12.32 & 4.07 & 13.27 & 6.16  & 13.34 & 5.65 \\
\hline
$c({\rm H}\beta)$ &  & & & -- & 0.822 &  --  & 0.569  &  -- & 0.638 \\
\hline \label{suwt2:tab:obslines}
\end{tabular}
\end{center}
\end{table*}
\renewcommand{\baselinestretch}{1.5}

\begin{figure*}
\begin{center}
\includegraphics[width=1.75in]{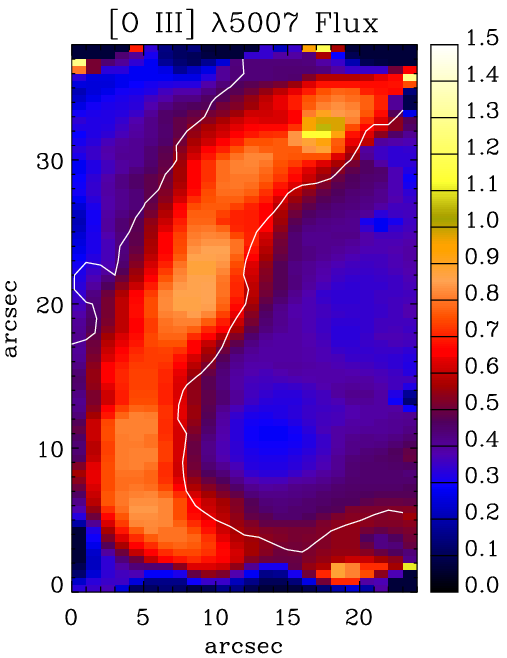}%
\includegraphics[width=1.75in]{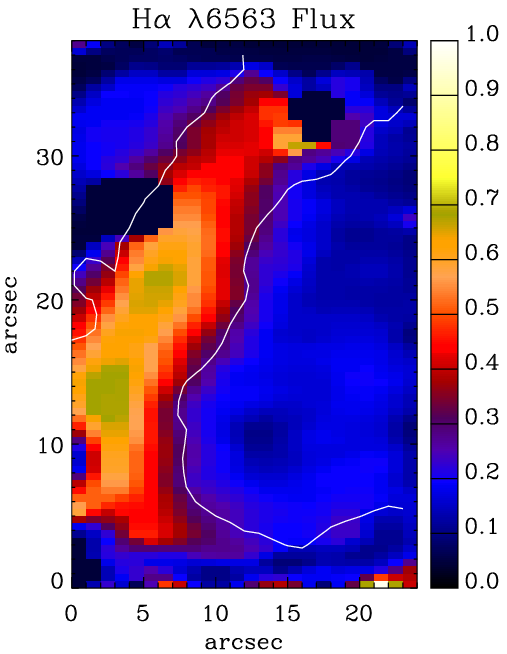}%
\includegraphics[width=1.75in]{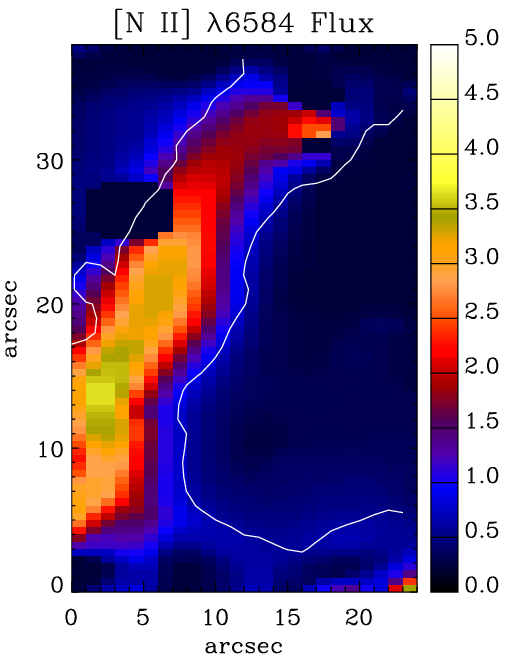}%
\includegraphics[width=1.75in]{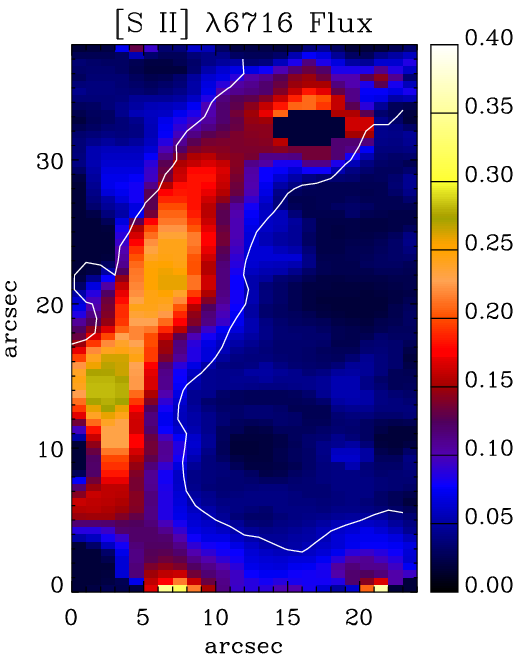}%
\caption{Undereddened flux maps for field 2 (see Fig.~\ref{suwt2:fig1}) of the PN~SuWt~2: $[$O~{\sc iii}$]$ $\lambda$5007, H$\alpha$ $\lambda$6563,
$[$N~{\sc ii}$]$ $\lambda$6584 and $[$S~{\sc ii}$]$ $\lambda$6716. The flux is derived from single Gaussian profile fits to the emission line at each spaxel.  
The white contour lines show the distribution of the narrow-band emission of H$\alpha$ and [N\,\textsc{ii}] in arbitrary unit taken with the ESO 3.6-m telescope. North is up and east is towards the left-hand side. 
Flux unit is in $10^{-15}$~erg\,s${}^{-1}$\,cm${}^{-2}$\,spaxel${}^{-1}$.}%
\label{suwt2:fig5}%
\end{center}
\end{figure*}

For each spatially resolved emission line profile, we extracted flux intensity, central wavelength (or centroid velocity), and FWHM (or velocity dispersion). Each emission line profile for each spaxel is fitted 
to a single Gaussian curve using the \textsc{mpfit} routine \citep{Markwardt2009}, an \textsc{idl} version of the \textsc{minpack-1} \textsc{fortran} code \citep{More1977}, which applies the Levenberg--Marquardt technique to the non-linear least-squares problem.
Flux intensity maps of key emission lines of field 2 are shown in Fig.\,\ref{suwt2:fig5} for $[$O~{\sc iii}$]$ $\lambda$5007, H$\alpha$ $\lambda$6563, $[$N~{\sc ii}$]$ $\lambda$6584 and $[$S~{\sc ii}$]$ $\lambda$6716; the same ring morphology is visible in the $[$N~{\sc ii}$]$ map as seen in Fig.\,\ref{suwt2:fig1}. White contour lines in the figures depict the distribution of the narrow-band
emission of H$\alpha$ and $[$N~{\sc ii}$]$ taken with the ESO 3.6 m telescope, which can be used to distinguish the borders between the ring structure and the inside region. 
We excluded the stellar continuum offset from the final flux maps using \textsc{mpfit}, so spaxels show only the flux intensities of the nebulae.   

\renewcommand{\baselinestretch}{0.9}
\renewcommand{\baselinestretch}{1.5}

The H$\alpha$ and H$\beta$ Balmer emission-line fluxes were used to derive the logarithmic extinction at H$\beta$, $c({\rm H}\beta)=\log[I({\rm H}\beta)/F({\rm H}\beta)]$, for the theoretical line ratio of the case B recombination \citep[$T_{\rm e}=10\,000$\,K and $N_{\rm e}=100$ cm$^{-3}$;][]{Hummer1987}. Each flux at the central wavelength was corrected for reddening using the logarithmic extinction $c({\rm H}\beta)$  according to
\begin{equation}
I(\lambda)=F(\lambda)\,10^{c({\rm H}\beta)[1+f(\lambda)]},
\label{suwt2:eq_reddening}%
\end{equation}
where $F(\lambda)$ and $I(\lambda)$ are the observed and intrinsic line flux, respectively, and $f(\lambda)$ is the standard Galactic
extinction law for a total-to-selective extinction ratio of $R_V \equiv A(V)/E(B-V)=3.1$ \citep{Seaton1979b,Seaton1979a,Howarth1983}.

\renewcommand{\baselinestretch}{0.9}
\renewcommand{\baselinestretch}{1.5}

Accordingly, we obtained an extinction of $c({\rm H}\beta)=0.64$ [$E(B-V) = 0.44$] for the total fluxes (column 9 in Table~\ref{suwt2:tab:obslines}). Our derived nebular extinction is in good agreement with the value found by \citet{Exter2010}, $E(B-V) = 0.40$ for the central star, though they obtained $E(B-V) = 0.56$ for the nebula. It may point to the fact that all reddening is not due to the interstellar medium (ISM), and there is some dust contribution in the nebula.  
Adopting a total observed flux value of log$F$(\ha)\,=\,$-11.69$  erg\,cm${}^{-2}$\,s${}^{-1}$ for the ring and interior structure \citep{Frew2008,Frew2013a,Frew2013b} and using $c({\rm H}\beta)=0.64$, lead to the dereddened H$\alpha$ flux of log$I$(\ha)\,=\,$-11.25$ erg\,cm${}^{-2}$\,s${}^{-1}$.

According to the strength of He~{\sc ii} $\lambda$4686 relative to H$\beta$, the PN SuWt 2 is classified as the intermediate excitation class with ${\rm EC}=6.6$ \citep{Dopita1990} or ${\rm EC}=7.8$ \citep{Reid2010}. The EC is an indicator of the central star effective temperature \citep{Dopita1991,Reid2010}. Using the $T_{\rm eff}$--EC relation of Magellanic Cloud PNe found by \citet{Dopita1991}, we estimate $T_{\rm eff}=143$\,kK for ${\rm EC}=6.6$. However, we get $T_{\rm eff}=177$\,kK for ${\rm EC}=7.8$ according to the transformation given by \citet{Reid2010} for Large Magellanic Cloud PNe.

\section{Kinematics}
\label{suwt2:sec:kinematic}

\renewcommand{\baselinestretch}{1.2}
\begin{table}
\begin{center}
\caption{Kinematic parameters on the SuWt~2's ring and its central star.}
\begin{tabular}{cc}
\hline
\hline
{Parameter\hspace{25 mm}}  & { Value} \\
\hline
$a=r$ (outer radius) \dotfill  & $45 \pm 4$ arcsec\\
$b=r\cos i$ \dotfill  & $17 \pm 2$ arcsec\\
thickness \dotfill  & $13 \pm 2$ arcsec\\
{\rm PA} \dotfill  &  $48^{\circ} \pm 2^{\circ}$\\
{\rm GPA} \dotfill  &  $62^{\circ}16\arcmin \pm 2^{\circ} $\\
inclination ($i$) \dotfill  &  $68^{\circ} \pm 2^{\circ}$\\
$v_{\rm sys}$ (LSR) \dotfill  &  $-29.5\pm5$ km\,s${}^{-1}$\\
$v_{\rm exp}$ \dotfill  &  $28\pm 5$ km\,s${}^{-1}$\\
\hline
\label{suwt2:tab:kinematic:parameters}
\end{tabular}
\end{center}
\end{table}
\renewcommand{\baselinestretch}{1.5}

\renewcommand{\baselinestretch}{0.9}
\begin{figure}
\begin{center}
\includegraphics[width=1.7in]{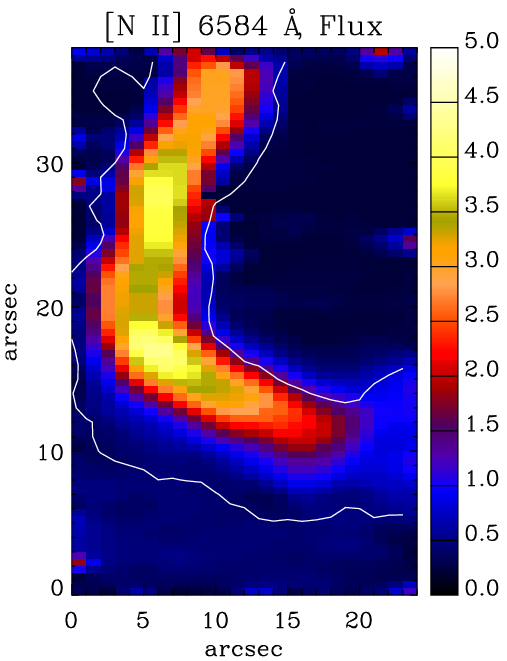}%
\includegraphics[width=1.7in]{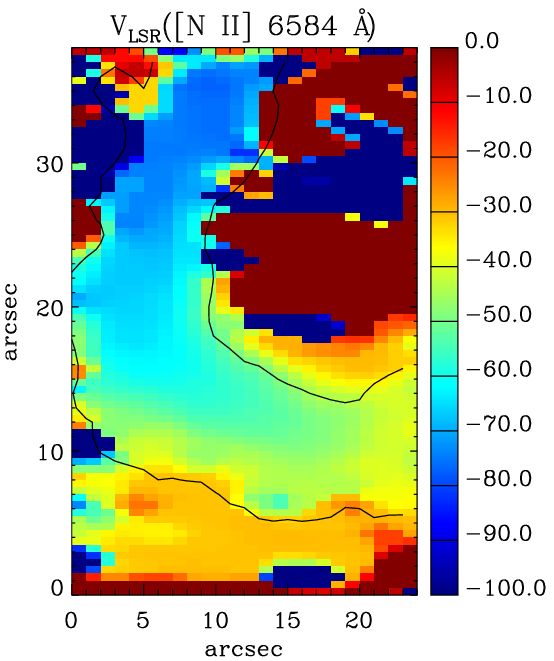}%
\caption{Flux intensity and radial velocity ($V_{\rm LSR}$) map in $[$N~{\sc ii}$]$ $\lambda$\,6584\,{\AA} for \textit{Field 1} (see Table~\ref{suwt2:tab:observations}) of the PN~SuWt~2. 
The white/black contour lines show the distribution of the narrow-band emission of H$\alpha$ and [N\,\textsc{ii}] in arbitrary unit taken with the ESO 3.6-m telescope. North is up and east is toward the left-hand side. Units are in km\,s${}^{-1}$.
}%
\label{suwt2:fig8}%
\end{center}
\end{figure}
\renewcommand{\baselinestretch}{1.5}

Fig.~\ref{suwt2:fig8} presents maps of the flux intensity and the local standard of rest (LSR) radial velocity derived from the Gaussian profile fits for the emission line $[$N~{\sc ii}$]$ $\lambda$\,6584\,{\AA}. We transferred the observed velocity $v_{\rm obs}$ to 
the LSR radial velocity $v_{\rm lsr}$ by determining the radial velocities induced by the motions of the Earth and Sun using the  \textsc{iraf}/\textsc{astutil} task RVCORRECT.
The emission-line profile is also resolved if its velocity dispersion is wider than the instrumental width $\sigma_{\rm ins}$. The instrumental width can be derived from the $[$O~{\sc i}$]\,\lambda$5577{\AA} and $\lambda$6300{\AA} night sky lines; it is typically $\sigma_{\rm ins}\approx42$\,km\,s$^{-1}$ for $R\sim3000$ and $\sigma_{\rm ins}\approx19$\,km\,s$^{-1}$ for $R\sim7000$.  
Fig.~\ref{suwt2:fig8}(right) shows the variation of the LSR radial velocity in the south-east side of the nebula. We see that the radial velocity decreases as moving anti-clockwise on the ellipse. It has a low value of about $-70\pm30$\,km\,s$^{-1}$ on the west co-vertex of the ellipse, and a high value of  $-50\pm25$\,km\,s$^{-1}$ on the south vertex. This variation corresponds to the orientation of this nebula, namely the inclination and projected nebula on the plane of the sky. 
It obviously implies that the east side moves towards us, while the west side escapes from us. 

\begin{figure}
\begin{center}
\includegraphics[width=1.70in]{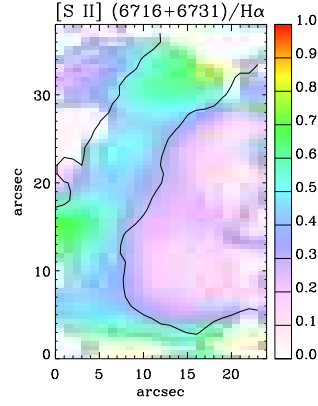}%
\caption{
Flux ratio maps of the $[$S~{\sc ii}$]$ $\lambda$\,6716+6731\,{\AA} to the H$\alpha$ recombination line emission. }%
\label{a48:fig7}%
\end{center}
\end{figure}

Kinematic information of the ring and the central star is summarized in Table~\ref{suwt2:tab:kinematic:parameters}. 
\citet{Jones2010} implemented a morpho-kinematic model using the modelling program \textsc{shape} \citep{Steffen2006} based on the long-slit emission-line spectra at the high resolution of $R\sim40\,000$, which is much higher than the moderate resolution of $R\sim3000$ in our observations. They obtained
the nebular expansion velocity of $v_{\rm exp}=28$ km\,s${}^{-1}$ and the LSR systemic velocity of $v_{\rm sys}=-29.5\pm5$ at the best-fitting inclination of $i=68^{\circ} \pm 2^{\circ}$ between the line of sight and the nebular axisymmetry axis. We notice that the nebular axisymmetric axis has a position angle of ${\rm PA}=48^{\circ}$ projected on to the plane of the sky, and measured from the north towards the east in the equatorial coordinate system (ECS). Transferring the PA in the ECS to the PA in the Galactic coordinate system yields 
the Galactic position angle of ${\rm GPA}=62^{\circ}16\arcmin$, which is the PA of the nebular axisymmetric axis projected on to the plane of the sky, measured from the North Galactic Pole (NGP; ${\rm GPA}=0^{\circ}$) towards the Galactic east (${\rm GPA}=90^{\circ}$). 
We notice an angle of $-27^{\circ}44\arcmin$ between
the nebular axisymmetric axis projected onto the plane of the sky and the Galactic plane. Fig.\,\ref{a48:fig7} shows the flux ratio map for the $[$S~{\sc ii}$]$ doublet to the H$\alpha$ recombination line emission. 
The shock criterion $[$S~{\sc ii}$]$ $\lambda\lambda$6716,6731/H$\alpha\geq 0.5$ indicates the presence of a shock-ionization front in the ring. Therefore, the brightest south-east side of the nebula has a signature of an interaction with ISM. 

\renewcommand{\baselinestretch}{0.9}
\renewcommand{\baselinestretch}{1.5}

The PPMXL catalogue\footnote{Website: \url{http://vo.uni-hd.de/ppmxl}} \citep{Roeser2010} reveals that the A-type stars of SuWt~2 move with the proper motion of $v_l=D\mu_{l}\cos(b)=(-8.09\pm8.46)D$ km\,s$^{-1}$ and $v_b=D\mu_{b}=(11.79\pm8.82) D$ km\,s$^{-1}$, where $D$ is its distance in kpc. 
They correspond to the magnitude of $v_{\mu}=(14.30 \pm 8.83) D$ km\,s$^{-1}$. 
Assuming a distance of $D=2.3$\,kpc \citep[][]{Exter2010} and $v_{\rm sys}=-29.5\pm5$\,km\,s$^{-1}$ \citep[LSR;][]{Jones2010}, this PN moves in the Cartesian Galactocentric frame with peculiar (non-circular) velocity components of ($U_s$,\,$V_s$,\,$W_s)=$\,($35.4 \pm 18.4$,\,$11.0\pm13.7$,\,$33.18\pm26.4$)\,km\,s$^{-1}$, where $U_s$ is towards the Galactic centre, $V_s$ in the local direction of Galactic rotation, and $W_s$ towards the NGP 
\citep[see ][peculiar motion calculations in appendix]{Reid2009}.  We see that SuWt 2 moves towards the NGP with $W_s=33.18$\,km\,s$^{-1}$, and there is an interaction with ISM in the direction of its motion, i.e., the east-side of the nebula. 

We notice a very small peculiar velocity ($V_s=11$ km\,s$^{-1}$) in the local direction of Galactic rotation, so a kinematic distance may also be estimated as
the Galactic latitude is a favorable one for such a determination. 
We used the \textsc{fortran} code for the `revised' kinematic distance prescribed in \citet{Reid2009}, and adopted the IAU standard parameters of the Milky Way, namely the distance to the Galactic centre $R_0=8.5$ kpc and a circular rotation speed $\Theta_0=220$ km\,s${}^{-1}$ for a flat rotation curve (${\rm d}\Theta/{\rm d}R=0$), and the solar motion of ${\rm U}_{\bigodot}=10.30$ km\,s${}^{-1}$, ${\rm V}_{\bigodot}=15.3$ km\,s${}^{-1}$ and ${\rm W}_{\bigodot}=7.7$ km\,s${}^{-1}$. The LSR systemic velocity of $-29.5$ km\,s${}^{-1}$ \citep{Jones2010} gives a kinematic distance of $2.26$~kpc, which is in quite good agreement with the distance of 2.3\,$\pm$\,0.2 kpc found by \citet{Exter2010} based on an analysis of the double-lined eclipsing binary system. This distance implies that SuWt~2 is in the tangent of the Carina-Sagittarius spiral arm of the Galaxy ($l=311\fdg0$, $b=2\fdg4$).    
Our adopted distance of 2.3~kpc means 
the ellipse's major radius of $45$\,arcsec corresponds to a ring radius of $r=0.47\pm0.04$ pc. 
The expansion velocity of the ring then yields a dynamical age of $\tau_{\rm dyn}=r/v_{\rm exp}=17500\pm 1560$ yr, which is defined as the radius divided by the constant expansion velocity. Nonetheless, the true age is more than the dynamical age, since the nebula expansion velocity is not constant through the nebula evolution. \citet{Dopita1996} estimated the true age typically around 1.5 of the dynamical age, so we get $\tau_{\rm true}=26250\pm2330$ yr for SuWt\,2. If we take the asymptotic giant branch (AGB) expansion velocity of $v_{\rm AGB}=v_{\rm exp}/2$ \citep{Gesicki2000}, as the starting velocity of the new evolving PN, we also estimate the true age as $\tau_{\rm true}=2r/(v_{\rm exp}+v_{\rm AGB})= 23360\pm 2080$ yr. 

\section{Plasma diagnostics}
\label{suwt2:sec:tempdens}

\renewcommand{\baselinestretch}{0.9}
\begin{figure*}
\begin{center}
\includegraphics[width=1.75in]{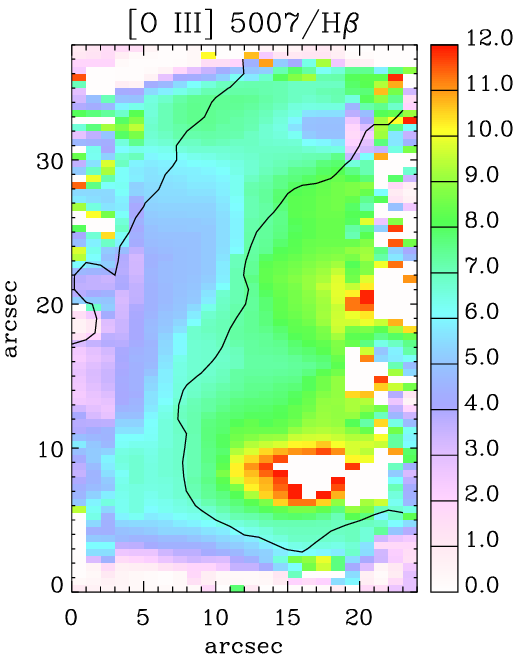}%
\includegraphics[width=1.75in]{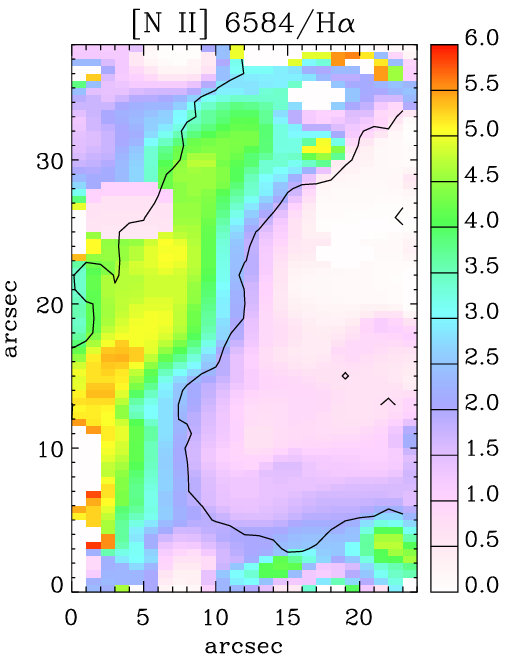}%
\includegraphics[width=1.75in]{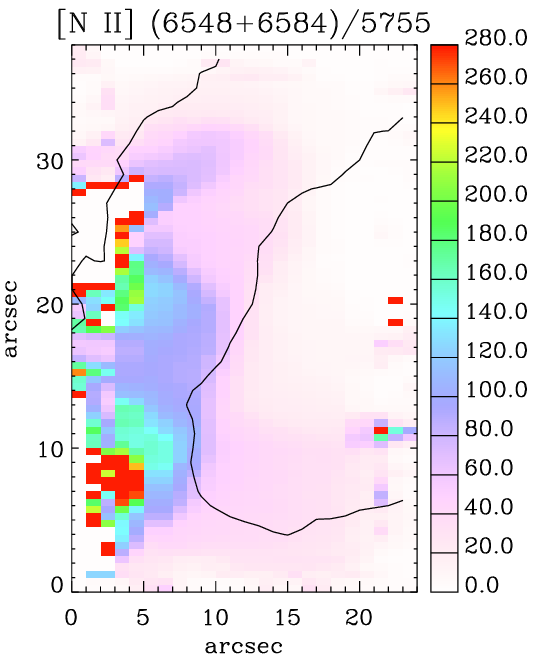}%
\includegraphics[width=1.75in]{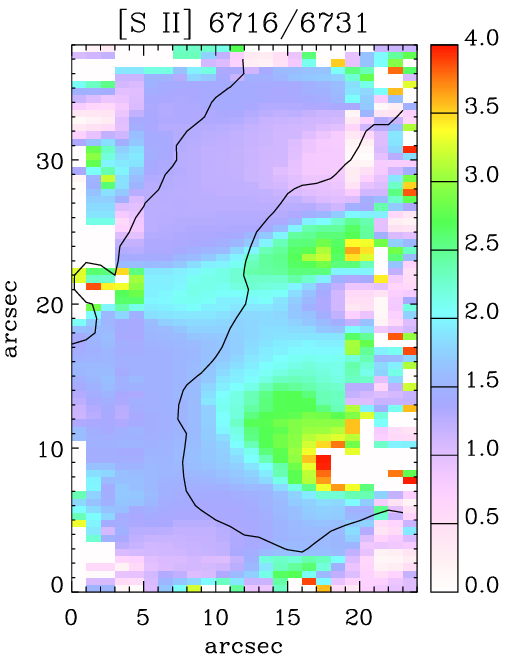}%
\hspace{41 mm}(a)\hspace{41 mm}(b)\hspace{41 mm}(c)\hspace{41 mm}(d)
\caption{Flux ratio maps for field 2 (see Fig.~\ref{suwt2:fig1}) of the PN SuWt 2. From left to right: (a) flux ratio maps of the $[$O~{\sc iii}$]$ $\lambda$\,5007\,{\AA} to the H$\beta$ recombination line emission, (b) flux ratio map of the $[$N~{\sc ii}$]$ $\lambda$\,6584 to the H$\alpha$ recombination line emission, (c) flux ratio map for the temperature-sensitive $[$N~{\sc ii}$]$  $\lambda\lambda$5755, 6548, 6583 lines, and (d) for the density-sensitive $[$S~{\sc ii}$]$ doublet. The black contour lines show the distribution of the narrow-band emission of H$\alpha$ and [N\,\textsc{ii}] in arbitrary units taken with the ESO 3.6-m telescope observations. }%
\label{suwt2:fig11}%
\end{center}
\end{figure*}
\renewcommand{\baselinestretch}{1.5}

We derived the nebular electron temperatures $T_{\rm e}$ and densities $N_{\rm e}$ from the intensities of the collisionally excited lines (CELs) by solving the equilibrium equations for an $n$-level atom ($\geqslant 5$) using \textsc{equib}, a \textsc{fortran} code originally developed by \citet{Howarth1981}. 
Recently, it has been converted to \textsc{fortran}~90, and combined into simpler routines for \textsc{neat} \citep{Wesson2012}. The atomic data sets used for our plasma diagnostics, as well as for the CEL abundance determination in \S\,\ref{suwt2:sec:abundances}, are the same as those used by \citet{Wesson2012}. 

The diagnostics procedure was as follows: we assumed a representative initial electron temperature of 10\,000\,K in order to derive $N_{\rm e}([$S~{\sc ii}$])$; then $T_{\rm e}([$N~{\sc ii}$])$ was derived in conjunction with the mean density derived from $N_{\rm e}([$S~{\sc ii}$])$. The calculations were iterated to give self-consistent results for $N_{\rm e}$ and $T_{\rm e}$. The correct choice of electron density and temperature is essential to determine ionic abundances. 

Fig.~\ref{suwt2:fig11} shows flux ratio maps for the density-sensitive $[$S~{\sc ii}$]$ doublet. It indicates the electron density $N_{\rm e}$ of about $\lesssim 100$~cm${}^{-3}$ in the ring. We see that the interior region has a $[$S~{\sc ii}$]$ $\lambda\lambda$ 6716/6731 flux ratio of more than 1.4, which means the inside of the ring has a very low density ($N_{\rm e}\lesssim 50$ cm${}^{-3}$). Flux ratio maps for the temperature-sensitive $[$N~{\sc ii}$]$  $\lambda\lambda$5755, 6548, 6583 lines indicate that the electron temperature $T_{\rm e}$ varies from  7\,000 to  14\,000~K. As shown in Fig.~\ref{suwt2:fig11}, the brightest part of the ring in $[$N~{\sc ii}$]$ $\lambda$\,6584\,{\AA} has an electron temperature of about 8\,000~K. The inside of the ring has a mean electron temperature of about 11\,800~K. 
We notice that \citet{Smith2007} found $N_{\rm e}=90$~cm${}^{-3}$ and $T_{\rm e}=11\,400$~K using the R-C Spectrograph ($R\sim6000$) on the CTIO 4-m telescope, though they obtained them from a $0.8$\,arcsec slit oriented along the major axis of the ring (${\rm PA}=135^{\circ}$).

Table~\ref{suwt2:tab:tenediagnostics} lists the electron density
($N_{\rm e}$) and the electron temperature ($T_{\rm e}$) of the different regions, 
together with the
ionization potential required to create the emitting ions. We see that
the east part of the ring has a mean electron density of $N_{\rm e}([$S~{\sc
  ii}$])\lesssim 100$ cm${}^{-3}$ and mean temperatures of $T_{\rm e}([$N~{\sc
  ii}$])=8\,140$~K and $T_{\rm e}([$O~{\sc iii}$])=12\,390$~K, while the less dense region inside the ring shows a high mean temperature of 
$T_{\rm e}([$N~{\sc ii}$])=11\,760$~K and $T_{\rm e}([$O~{\sc iii}$])$ less than $20\,000$~K. We point out that the [S~{\sc ii}] $\lambda\lambda$6716/6731 line ratio  of more than 1.40 is associated with the low-density limit of $N_{\rm e}<100$ cm${}^{-3}$, and we cannot accurately determine the electron density less than this limit \citep[see e.g. A\,39;][]{Jacoby2001}. Furthermore,  we cannot resolve the [O~{\sc ii}] $\lambda\lambda$3726,3729 doublet with our moderate spectral resolution ($R\sim3000$). Plasma diagnostics indicates that the interior region is much hotter than the ring region. This implies the presence of a hard ionizing source located at the centre. It is worth to mention that $T_{\rm e}([$N~{\sc ii}$])$ is more appropriate for singly ionized species, while $T_{\rm e}([$O~{\sc iii}$])$ is associated with doubly and more ionized species. \citet{Kingsburgh1994} found that $T_{\rm e}([$O~{\sc iii}$])/T_{\rm e}([$N~{\sc ii}$])=1.25$ for medium-excitation PNe and $T_{\rm e}([$O~{\sc iii}$])/T_{\rm e}([$N~{\sc ii}$])=1.15 + 0.0037 I(4686)$ for high-excitation PNe. Here, we notice that $T_{\rm e}([$O~{\sc iii}$])/T_{\rm e}([$N~{\sc ii}$])=1.52$ for the ring and
$T_{\rm e}([$O~{\sc iii}$])/T_{\rm e}([$N~{\sc ii}$])=1.39$ for the total flux.

\renewcommand{\baselinestretch}{0.9}
\begin{table*}
\begin{center}
\caption{Diagnostic ratios for the electron temperature, $T_{\rm e}$ and the electron density, $N_{\rm e}$. }
\begin{tabular}{lcccccccccc}
\hline
\hline
{ Ion}     & { Diagnostic}     & { I.P.(eV)} &\multicolumn{2}{c}{Interior}  &  \multicolumn{2}{c}{Ring}  &  \multicolumn{2}{c}{Total}        \\
\hline
 &    &   & Ratio & $T_{\rm e}(10^3{\rm K})$ & Ratio & $T_{\rm e}(10^3{\rm K})$ & Ratio & $T_{\rm e}(10^3{\rm K})$ \\
\hline
$[$N~{\sc ii}$]$  & { $\frac{\lambda6548+\lambda6584}{\lambda5755}$}  &  14.53 & 63.71: & 11.76:  & 161.33 &  8.14 & 175.78  &  7.92 \\
\noalign{\smallskip}
$[$O~{\sc iii}$]$ & {$\frac{\lambda4959+\lambda5007}{\lambda4363}$}  &  35.12  & 35.41:: & $\lesssim20.0$::  & 110.93 & 12.39  &   152.49 &  11.07 \\
\hline
 & & & Ratio & $N_{\rm e}({\rm cm}^{-3})$ & Ratio & $N_{\rm e}({\rm cm}^{-3})$ & Ratio & $N_{\rm e}({\rm cm}^{-3})$ \\
\hline
$[$S~{\sc ii}$]$  & { $\frac{\lambda6716}{\lambda6731}$} & 10.36 &  2.02 & $\lesssim50.0$ & 1.46  & $\lesssim100.0$ & 1.51 &  $\lesssim100.0$ \\
\noalign{\smallskip}
\hline \label{suwt2:tab:tenediagnostics}
\end{tabular}
\end{center}
\end{table*}
\renewcommand{\baselinestretch}{1.5}

\renewcommand{\baselinestretch}{1.2}
\begin{table}
\begin{center}
\caption{Ionic and total abundances deduced from empirical analysis of the observed fluxes across different nebula regions of SuWt~2.}
\begin{tabular}{llccc}
\hline
\hline
{ $\lambda$({\AA})}& {Abundance} & {Interior}  & {Ring} & {Total} \\
\hline   
5876 & He${}^{+}$/H${}^{+}$ 	& -- 			& 0.066 		& 0.049 \\
6678 & He${}^{+}$/H${}^{+}$ 	& 0.031  		& 0.057 		& 0.043 \\
Mean & He${}^{+}$/H${}^{+}$ 	& 0.031  		& 0.064 		& 0.048 \\
\noalign{\smallskip}
4686 & He${}^{2+}$/H${}^{+}$ 	& 0.080 		& 0.027 		&  0.036 \\
\noalign{\smallskip}
     & $icf$(He) 				& 1.0 			& 1.0 			& 1.0 \\
     & He/H 					&  0.111 		& 0.091 		& 0.084 \\
\hline   
6548 & N${}^{+}$/H${}^{+}$  	& 7.932($-6$)  	& 1.269($-4$) 	&  1.284($-4$) \\ 
6584 & N${}^{+}$/H${}^{+}$  	& 1.024($-5$) 	& 1.299($-4$)  	&  1.334($-4$) \\ 
Mean & N${}^{+}$/H${}^{+}$  	& 9.086($-6$) 	& 1.284($-4$) 	&  1.309($-4$) \\ 
\noalign{\smallskip}
     & $icf$(N) 				& 16.240 		& 2.014 		& 3.022 \\
     & N/H 						& 1.476($-4$) 	& 2.587($-4$) 	&  3.956($-4$) \\
\hline
3727 & O${}^{+}$/H${}^{+}$  	& 1.109($-5$) 	& 1.576($-4$)  	& 1.597($-4$) \\ 
4959 & O${}^{2+}$/H${}^{+}$  	& 6.201($-5$) 	& 8.881($-5$) 	& 1.615($-4$) \\ 
5007 & O${}^{2+}$/H${}^{+}$  	& 6.998($-5$) 	& 9.907($-5$) 	& 1.808($-4$) \\ 
Mean & O${}^{2+}$/H${}^{+}$  	& 6.599($-5$)  	& 9.394($-5$)  	& 1.711($-4$) \\ 
\noalign{\smallskip}
     & $icf$(O) 				& 2.336 		& 1.262 		& 1.459 \\
     & O/H 						& 1.801($-4$) 	& 3.175($-4$) 	& 4.826($-4$) \\
\hline
3869 & Ne${}^{2+}$/H${}^{+}$  	& 2.635($-5$)  	& 9.608($-5$) 	& 1.504($-4$) \\ 
3968 & Ne${}^{2+}$/H${}^{+}$  	& -- 			& 3.306($-5$) 	& -- \\ 
Mean & Ne${}^{2+}$/H${}^{+}$  	& 2.635($-5$) 	& 6.457($-5$) 	& 1.504($-4$) \\ 
\noalign{\smallskip}
     & $icf$(Ne) 				& 2.728 		& 3.380 		& 2.820 \\
     & Ne/H 					& 7.191($-5$) 	& 2.183($-4$) 	& 4.241($-4$) \\  
\hline  
6716 & S${}^{+}$/H${}^{+}$  	& 3.307($-7$) 	& 2.034($-6$) 	& 2.179($-6$) \\ 
6731 & S${}^{+}$/H${}^{+}$  	& 2.189($-7$) 	& 1.834($-6$) 	& 1.903($-6$)  \\ 
Mean &  S${}^{+}$/H${}^{+}$  	& 2.748($-7$) 	& 1.934($-6$) 	& 2.041($-6$) \\ 
\noalign{\smallskip}
6312  & S${}^{2+}$/H${}^{+}$  	& -- 			& 3.292($-8$) 	&  -- \\ 
9069 & S${}^{2+}$/H${}^{+}$  	& 3.712($-7$) 	& 1.198($-6$) 	& 1.366($-6$) \\ 
Mean & S${}^{2+}$/H${}^{+}$  	& 3.712($-7$) 	& 6.155($-7$) 	& 1.366($-6$) \\ 
\noalign{\smallskip}
     & $icf$(S) 				& 1.793 		& 1.047  		& 1.126 \\
     & S/H 						& 1.158($-6$)  	& 2.668($-6$) 	& 3.836($-6$) \\
\hline  
7136 & Ar${}^{2+}$/H${}^{+}$  	& 3.718($-7$) 	& 8.756($-7$) 	& 1.111($-6$) \\ 
4740 & Ar${}^{3+}$/H${}^{+}$  	& -- 			& -- 			& 4.747($-7$) \\ 
7005  & Ar${}^{4+}$/H${}^{+}$  	& 3.718($-7$) 	& -- 			& --  \\ 
\noalign{\smallskip}
     & $icf$(Ar) 				& 1.066  		& 1.986 		&  1.494 \\
     & Ar/H 					& 5.230($-7$) 	& 1.739($-6$) 	&  2.370($-6$) \\
\hline \label{suwt2:tab:abundances:empirical}
\end{tabular}
\end{center}
\end{table}
\renewcommand{\baselinestretch}{1.5}

\section{Ionic and total abundances}
\label{suwt2:sec:abundances}

We derived ionic abundances for SuWt~2 using the observed CELs and 
the optical recombination lines (ORLs).  
We determined abundances for ionic species of N, O, Ne, S and Ar from CELs.
In our determination, we adopted the mean $T_{\rm e}$([O~{\sc iii}]) and the upper limit of $N_{\rm e}$([S~{\sc ii}]) obtained from our empirical analysis in Table~\ref{suwt2:tab:tenediagnostics}. 
Solving the equilibrium equations, using \textsc{equib}, yields level populations and line sensitivities for given $T_{\rm e}$ and $N_{\rm e}$. Once the level population are solved, the ionic abundances, X${}^{i+}$/H${}^{+}$, can be derived from the observed line intensities of CELs.
We determined ionic abundances for He from the measured intensities of ORLs using the effective recombination coefficients from 
 \citet{Storey1995} and \citet{Smits1996}. 
We derived the total abundances from deduced ionic abundances using the ionization correction factor ($icf$) formulae given by \citet{Kingsburgh1994}: 
\begin{equation}
\footnotesize
\frac{{\rm He}}{{\rm H}}=\left(\frac{{\rm He}^{+}}{{\rm H}^{+}}+\frac{{\rm He}^{2+}}{{\rm H}^{+}}\right)\times
{{icf}}({\rm He}),~~{{icf}}({\rm He})=1,
\label{suwt2:eq_he_orl1}
\end{equation}
\begin{equation}
\footnotesize
\frac{{\rm O}}{{\rm H}}=\left(\frac{{\rm O}^{+}}{{\rm H}^{+}} + \frac{{\rm O}^{2+}}{{\rm H}^{+}} \right)
\times
{{icf}}({\rm O}),~~
{{icf}}({\rm O}) = \left(1+\frac{{\rm He}^{2+} }{{\rm He}^{+}}\right)^{2/3},
\label{suwt2:eq_o_cel2}
\end{equation}
\begin{equation}
\footnotesize
\frac{{\rm N}}{{\rm H}}=\left(\frac{{\rm N}^{+}}{{\rm H}^{+}}\right)
\times {{icf}}({\rm N}),~~{{icf}}({\rm N})=
\left(\frac{{\rm O}}{{\rm O}^{+}}\right),
\label{suwt2:eq_n_cel1}
\end{equation}
\begin{equation}
\footnotesize
\frac{{\rm Ne}}{{\rm H}}= \left(\frac{{\rm Ne}^{2+}}{{\rm H}^{+}} \right)
\times{{icf}}({\rm Ne}),~~{{icf}}({\rm Ne})=
\left(\frac{{\rm O}}{{\rm O}^{2+}}\right),
\label{suwt2:eq_ne_cel1}
\end{equation}
\begin{equation}
\footnotesize
\frac{{\rm S}}{{\rm H}}=\left(\frac{{\rm S}^{+}}{{\rm H}^{+}} + \frac{{\rm S}^{2+}}{{\rm H}^{+}} \right)
\times
{{icf}}({\rm S}),
\label{suwt2:eq_s_cel1}
\end{equation}
\begin{equation}
\footnotesize
{{icf}}({\rm S})=
\left[1-\left(1-\frac{{\rm O}^{+}}{{\rm O}}\right)^{3}\right]^{-1/3},
\label{suwt2:eq_s_cel2}
\end{equation}
\begin{equation}
\footnotesize
\frac{{\rm Ar}}{{\rm H}}= \left(\frac{{\rm Ar}^{2+}}{{\rm H}^{+}} \right)
\times {{icf}}({\rm Ar}),~~{{icf}}({\rm Ar})=
\left(1-\frac{{\rm N}^{+}}{{\rm N}}\right)^{-1}
.\label{suwt2:eq_ar_cel1}
\end{equation}
We derived the ionic and total helium abundances from the observed $\lambda$5876 and $\lambda$6678, and He~{\sc ii} $\lambda$4686 ORLs. We assumed case B recombination for the singlet He~{\sc i} $\lambda$6678 line and case A for other He~{\sc i} $\lambda$5876 line \citep[theoretical recombination models of][]{Smits1996}.
The He${}^{+}$/H${}^{+}$ ionic abundances from the He~{\sc i} lines at $\lambda$5876 and $\lambda$6678 were averaged with weights of 3:1, roughly the intrinsic intensity ratios of the two lines. The He${}^{2+}$/H${}^{+}$ ionic abundances were derived from the He~{\sc ii} $\lambda$4686 line using theoretical case B recombination rates from \citet{Storey1995}. For high- and middle-EC PNe (E.C.\,$> 4$), the total He/H abundance ratio can be obtained by simply taking the sum of singly and doubly ionized helium abundances, and with an $icf$(He) equal or less than 1.0. For PNe with low levels of ionization it is more than 1.0. SuWt~2 is an intermediate-EC PN \citep[${\rm EC}=6.6$;][]{Dopita1990}, so we can use an $icf$(He) of 1.0. 
We determined the O${}^{+}$/H${}^{+}$ abundance ratio from the $[$O~{\sc ii}$]$ $\lambda$3727 doublet, and the O${}^{2+}$/H${}^{+}$ abundance ratio from the $[$O~{\sc iii}$]$ $\lambda$4959 and $\lambda$5007 lines. In optical spectra, only O${}^{+}$ and O${}^{2+}$ CELs are seen, so the singly and doubly ionized helium abundances deduced from ORLs are used to include the higher ionization stages of oxygen abundance.

\begin{table}
\begin{center}
\caption{Parameters of the two best-fitting photoionization models. 
The initial mass, final mass, and Post-AGB age are obtained from the evolutionary tracks calculated for hydrogen- and helium-burning models by \citet{Bloecker1995}
}
\begin{tabular}{llclc}
\hline
\hline
& \multicolumn{2}{c}{Nebula abundances} & \multicolumn{2}{c}{Stellar parameters}  \\
\hline
\multirow{7}{*}{\rotatebox{90}{Model 1}} 
 &  He/H &  0.090  & $T_{\rm eff}$(kK)  & 140 \\ 
 &  C/H  & 4.00($-4$)  & $L_{\star}\,({\rm L}_{\bigodot})$  &  700     \\
 &  N/H  & 2.44($-4$) & $\log\,g$ (cgs)      &  $7.0$ \\
 &  O/H  & 2.60($-4$) & H\,:\,He           &  8\,:\,2    \\
 &  Ne/H & 1.11($-4$) & $M_{\star}\,({\rm M}_{\bigodot})$  &  $\sim 0.605$ \\
 &  S/H  &  1.57($-6$)& $M_{\rm \textsc{zams}}\, ({\rm M}_{\bigodot})$    &  3.0  \\
 &  Ar/H & 1.35($-6$) & $\tau_{\rm post-\textsc{agb}}$ (yr)  &  7\,500 \\
\hline
\multirow{7}{*}{\rotatebox{90}{Model 2}} 
 &  He/H &  0.090   & $T_{\rm eff}$(kK)  & 160 \\ 
 &  C/H  & 4.00($-4$)   & $L_{\star}\,({\rm L}_{\bigodot})$ &  600   \\
 &  N/H  & 2.31($-4$)  & $\log\,g$ (cgs)     &  $7.3$ \\
 &  O/H  & 2.83($-4$)  & He\,:\,C\,:\,N\,:\,O       &  33\,:\,50\,:\,2\,:\,15    \\
 &  Ne/H & 1.11($-4$)  & $M_{\star}\,({\rm M}_{\bigodot})$  &  $\sim 0.64$ \\
 &  S/H  & 1.57($-6$)  & $M_{\rm \textsc{zams}}\, ({\rm M}_{\bigodot})$    &  3.0  \\
 &  Ar/H & 1.35($-6$)  & $\tau_{\rm post-\textsc{agb}}$ (yr)   &  $25\,000$ \\
\hline
& \multicolumn{4}{c}{Nebula physical parameters}\\
\hline
\multicolumn{2}{l}{$M_{i}/{\rm M}_{\bigodot}$}   &  0.21  & $D$\,(pc)   &  2\,300   \\
\multicolumn{2}{l}{$N_{\rm torus}$} & 100\,cm${}^{-3}$ &  $\tau_{\rm true}$\,(yr)  &  $23\,400$--$26\,300$ \\
\multicolumn{2}{l}{$N_{\rm spheroid}$}  & 50\,cm${}^{-3}$ & $[{\rm Ar/H}]$       &  $-0.049$ \\
\hline
\label{suwt2:tab:modelparameters}
\end{tabular}\\
\end{center}
\end{table}

We derived the ionic and total nitrogen abundances from $[$N~{\sc
  ii}$]$ $\lambda$6548 and $\lambda$6584 CELs. For
optical spectra, it is possible to derive only N${}^{+}$, which mostly
comprises only a small fraction ($\sim10$-30\%) of the total nitrogen
abundance. Therefore, the oxygen abundances were used to correct the
nitrogen abundances for unseen ionization stages of N${}^{2+}$ and
N${}^{3+}$. Similarly, the total Ne/H abundance was corrected for undetermined Ne${}^{3+}$ by using the oxygen abundances. The $\lambda\lambda$6716,6731 lines usually detectable in PN are preferred to be used for the determination of S${}^{+}$/H${}^{+}$, since the $\lambda\lambda$4069,4076 lines are usually enhanced by recombination contribution,
and also blended with O~{\sc ii} lines. We notice that the $\lambda\lambda$6716,6731
doublet is affected by shock excitation of the ISM interaction, so the S${}^{+}$/H${}^{+}$ ionic abundance must be lower.  
When the observed S${}^{+}$ is not appropriately determined, it is possible to use the expression given by \citet{Kingsburgh1994} in the calculation, i.e. $({\rm S}^{2+}/{\rm S}^{+})=4.677+({\rm O}^{2+}/{\rm O}^{+})^{0.433}$. 

\begin{table}
\begin{center}
\caption{Model line fluxes for SuWt~2.}
\begin{tabular}{lcccc}
\hline
\hline
Line  &  \rotatebox{45}{Observ.} & \rotatebox{45}{Model 1} & \rotatebox{45}{Model 2}  \\
\hline
3726 $[$O~{\sc ii}$]$ 	&  $702$::  & 309.42  		& 335.53  \\
3729 $[$O~{\sc ii}$]$ 	&  *  			& 408.89 		& 443.82 \\
3869 $[$Ne~{\sc iii}$]$ & 204.57::  		& 208.88  		& 199.96 \\
4069 $[$S~{\sc ii}$]$ 	&  1.71:: 		&  1.15  		&  1.25 \\
4076 $[$S~{\sc ii}$]$ 	&  -- 		    &  0.40  		&  0.43 \\
4102 H$\delta$ 			&  22.15:  		& 26.11   		&  26.10 \\
4267 C~{\sc ii} 		& --   			&  0.27  		&  0.26 \\
4340 H$\gamma$ 			&  38.18:  		& 47.12 		&  47.10 \\
4363 $[$O~{\sc iii}$]$  &  6.15  		& 10.13  		& 9.55 \\
4686 He~{\sc ii}  		&  43.76  		&  42.50  		&  41.38  \\
4740 $[$Ar~{\sc iv}$]$ 	&  1.94  		& 2.27 			&  2.10 \\
4861 H$\beta$ 			&  100.00  		& 100.00  		& 100.0 \\
4959 $[$O~{\sc iii}$]$ 	&  216.72   	& 243.20 		&  238.65 \\
5007 $[$O~{\sc iii}$]$ 	&  724.02   	& 725.70 		& 712.13 \\
5412 He~{\sc ii}  		&  5.68   		& 3.22  		&  3.14 \\
5755 $[$N~{\sc ii}$]$ 	&  7.64  		& 21.99  		&  21.17 \\
5876 He~{\sc i} 		&  6.54 		& 8.01  		&  8.30 \\
6548 $[$N~{\sc ii}$]$ 	& 321.94    	& 335.22		&  334.67 \\
6563 H$\alpha$ 			&  286.00  		& 281.83   		& 282.20 \\
6584 $[$N~{\sc ii}$]$ 	&  1021.68   	& 1023.78 		&  1022.09 \\
6678 He~{\sc i} 		&  1.63 		& 2.25  		&  2.33 \\
6716 $[$S~{\sc ii}$]^{\rm \ast}$ 	&  70.36  		&  9.17			&  10.21 \\
6731 $[$S~{\sc ii}$]^{\rm \ast}$ 	&  46.47  		&  6.94 		&  7.72 \\
7065 He~{\sc i} 		&  1.12  		&  1.59  		& 1.63 \\
7136 $[$Ar~{\sc iii}$]$ & 15.51  		& 15.90  		& 15.94 \\
7320 $[$O~{\sc ii}$]$ 	&  5.93  		& 10.60			& 11.17 \\
7330 $[$O~{\sc ii}$]$ 	&  3.37  		&  8.64 		& 9.11 \\
7751 $[$Ar~{\sc iii}$]$ &  9.60  		&  3.81 		& 3.82 \\
9069 $[$S~{\sc iii}$]$ 	&  5.65  		&   5.79  		& 5.58 \\
\hline
{$I($H$\beta)$/10$^{-12}$\,$\frac{\rm erg}{{\rm cm}^2{\rm s}}$} &  1.95  & 2.13 &  2.12  \\
\hline \label{suwt2:tab:modelresults}
\end{tabular}\\
\end{center}
\begin{flushleft}
{\footnotesize \textbf{Note.} $^{\rm \ast}$~{The shock-excitation largely enhances the observed $[$S~{\sc ii}$]$ doublet.}}
\end{flushleft}
\end{table}

\renewcommand{\baselinestretch}{0.9}
\begin{figure*}
\centering
\includegraphics[width=1.75in]{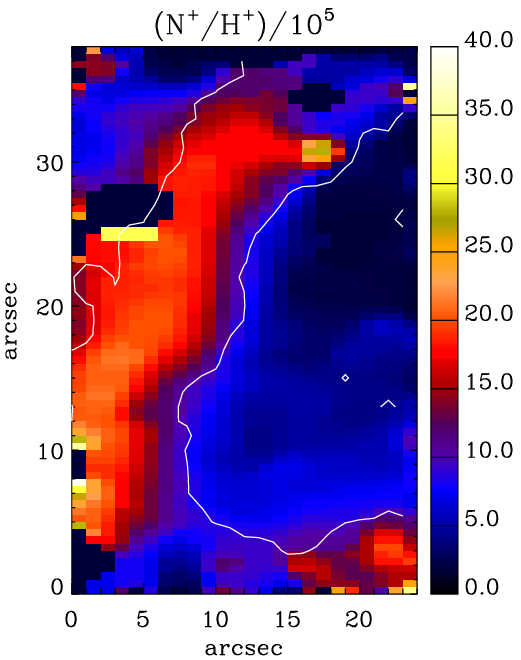}%
\includegraphics[width=1.75in]{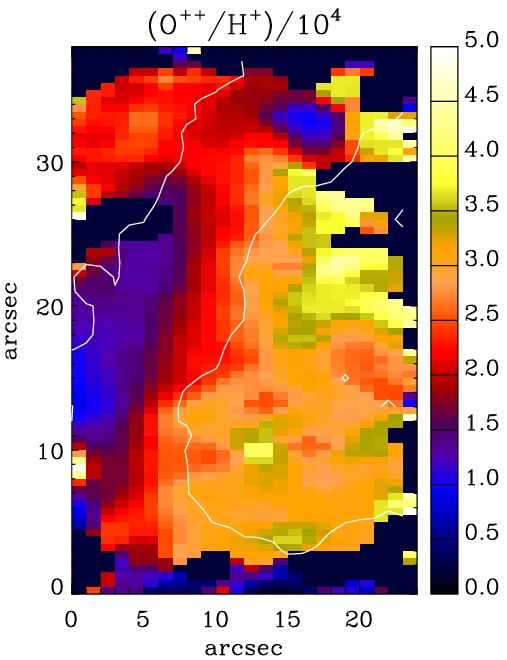}%
\includegraphics[width=1.75in]{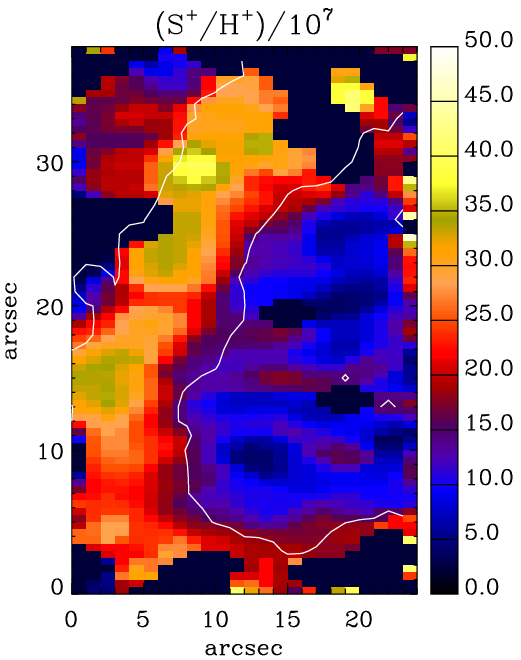}
\caption{Spatial distribution maps of 
ionic nitrogen abundance ratio N${}^{+}$/H${}^{+}$ ($\times 10^{-5}$) from $[$N~{\sc ii}$]$ CELs (6548, 6584); ionic oxygen abundance ratio O${}^{++}$/H${}^{+}$ ($\times 10^{-4}$) from $[$O~{\sc iii}$]$ CELs (4959, 5007); and ionic Sulfur abundance ratio S${}^{+}$/H${}^{+}$ ($\times 10^{-7}$) from $[$S~{\sc ii}$]$ CELs (6716, 6731) for $T_{\rm e}=10\,000$\,K and $N_{\rm e}=100$\,cm$^{-3}$.
White contour lines show the distribution of the narrow-band emission of H$\alpha$ and [N\,\textsc{ii}] in arbitrary unit taken with the ESO 3.6-m telescope.}%
\label{suwt2:fig6}%
\end{figure*}
\renewcommand{\baselinestretch}{1.5}

The total abundances of He, N, O, Ne, S, and Ar derived from our empirical analysis for selected regions of the nebula are given in Table~\ref{suwt2:tab:abundances:empirical}. 
From Table~\ref{suwt2:tab:abundances:empirical}
we see that SuWt~2 is a nitrogen-rich PN, which may be evolved from a massive progenitor ($M \geq 5$). However, the nebula's age (23\,400--26\,300 yr) cannot be associated with faster evolutionary time-scale of a massive progenitor, since the evolutionary time-scale of $7{\rm M}_{\bigodot}$ calculated by \citet{Bloecker1995} implies a short time-scale (less than 8000\,yr) for the effective temperatures  and the stellar luminosity (see Table\,\ref{suwt2:tab:obslines}) 
that are required to ionize the surrounding nebula.  
So, another mixing mechanism occurred during AGB nucleosynthesis, which further increased the Nitrogen abundances in SuWt 2. 
Mass transfer to the two A-type companions may explain this typical abundance pattern.
 
Fig.\,\ref{suwt2:fig6} shows the spatial distribution of ionic abundance ratio N${}^{+}$/H${}^{+}$, O${}^{++}$/H${}^{+}$ and S${}^{+}$/H${}^{+}$ derived for given $T_{\rm e}=10\,000$\,K and $N_{\rm e}=100$\,cm$^{-3}$. We notice that O${}^{++}$/H${}^{+}$ ionic abundance is very high in the inside shell; through the assumption of homogeneous electron temperature and density is not correct. The values in Table\,\ref{suwt2:tab:abundances:empirical} are obtained using the mean $T_{\rm e}$([O~{\sc iii}]) and $N_{\rm e}$([S~{\sc ii}]) listed in Table~\ref{suwt2:tab:tenediagnostics}. We notice 
that O$^{2+}$/O$^{+}=5.9$ for the interior and O$^{2+}$/O$^{+}=0.6$ for the ring. 
Similarly, He$^{2+}$/He$^{+}=2.6$ for the interior and He$^{2+}$/He$^{+}=0.4$ for the ring. This means that there are many more ionizing photons in the inner region than in the outer region, which hints at the presence of a hot ionizing source in the centre of the nebula.

\section{Photoionization model}
\label{suwt2:sec:photoionization}

\renewcommand{\baselinestretch}{0.9}
\begin{figure}
\begin{center}
\includegraphics[width=2.2in]{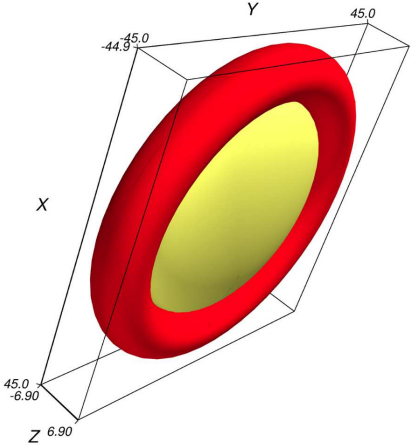}%
\caption{3-D isodensity plot of the dense torus adopted for photoionization modelling of SuWt~2. The torus has a homogeneous density of 100 cm${}^{-3}$, a radius of $38.1$\,arcsec from its centre to the tube centre, and a tube radius of $6.9$\,arcsec. The less dense oblate spheroid has a homogeneous density of 50 cm${}^{-3}$, a semi-major axis of $31.2$\,arcsec and a semi-minor axis of $6.9$\,arcsec. Axis units are arcsec, where 1 arcsec is equal to $1.12\times 10^{-2}$ pc based on the distance determined by our photoionization models.
}%
\label{suwt2:fig14}%
\end{center}
\end{figure}
\renewcommand{\baselinestretch}{1.5}

We used the 3 D photoionization code \textsc{mocassin} (version 2.02.67) to study the ring
of the PN SuWt~2. The code,
described in detail by \citet{Ercolano2003a,Ercolano2005,Ercolano2008}, applies a Monte Carlo method to solve
self-consistently the 3 D radiative transfer of the stellar and diffuse field in a gaseous and/or dusty nebula having asymmetric/symmetric density distribution and inhomogeneous/homogeneous chemical abundances, so it can deal with any structure and morphology. It also allows us to include multiple ionizing sources located in any arbitrary position in the nebula. It produces several outputs that can be compared with observations, namely a nebular emission-line spectrum, projected emission-line maps, temperature structure and fractional ionic abundances. This code has already been used for a number of axisymmetric PNe, such as NGC~3918 \citep{Ercolano2003b}, NGC~7009 \citep{Gonccalves2006}, NGC~6781 \citep{Schwarz2006}, NGC~6302 \citep{Wright2011} and NGC~40 \citep{Monteiro2011}. 
To save computational time, we began with the gaseous model of a $22\times22\times3$ Cartesian grid, with the ionizing source being placed in a corner in order to take advantage of the axisymmetric morphology used.
This initial low-resolution grid helped us explore the parameter space of the
photoionization models, namely ionizing source, nebula abundances and distance. Once we found the best fitting model, the final simulation was done using a density distribution model constructed in
$45\times45\times7$ cubic grids with the same size, corresponding to
14\,175 cubic cells of length 1 arcsec each. 
Due to computational restrictions on time, we did not run any model with higher number of cubic cells.
The atomic data set used for the photoionization modelling,
includes the CHIANTI database \citep[version 5.2;][]{Landi2006}, the improved coefficients of the H~{\sc i}, He~{\sc i} and He~{\sc ii} free--bound continuous emission \citep{Ercolano2006} and the photoionization cross-sections and ionic ionization energies \citep{Verner1993,Verner1995}.

\renewcommand{\baselinestretch}{0.9}
\begin{figure*}
\begin{center}
\includegraphics[width=6.5in]{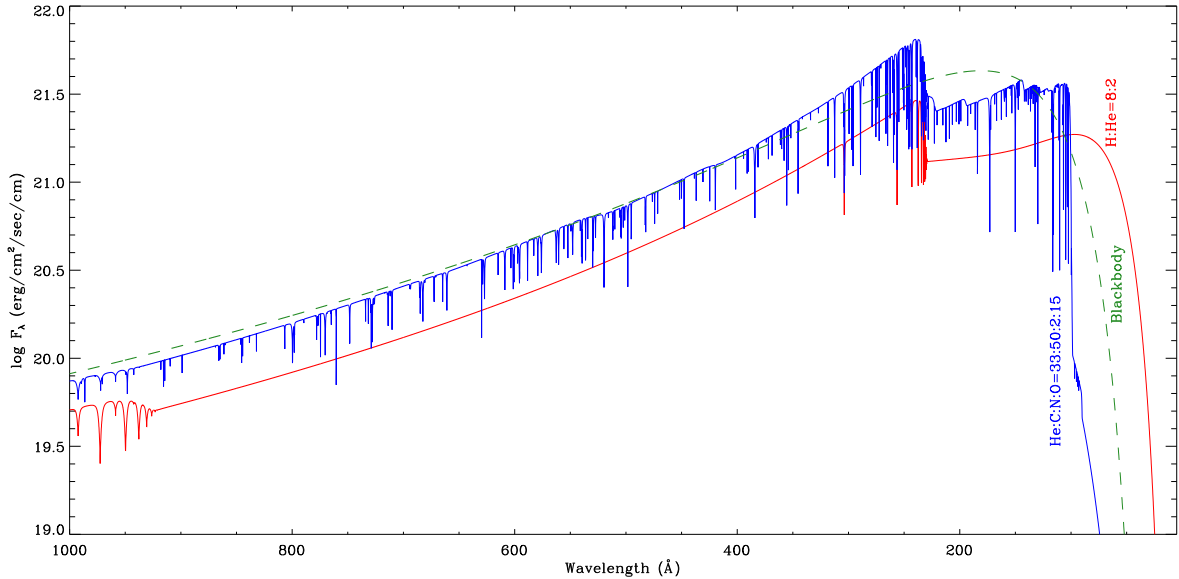}%
\caption{Comparison of two NLTE model atmosphere fluxes \citep{Rauch2003} used as ionizing inputs in our two models. Red line: H-rich model with an abundance ratio of H\,:\,He\,=\,8\,:\,2 by mass, $\log g =7$ (cgs) and $T_{\rm eff}=140\,000$~K. Blue line: PG~1159 model with He\,:\,C\,:\,N\,:\,O\,=\,33\,:\,50\,:\,2\,:\,15, $\log g =7$ and $T_{\rm eff}=160\,000$~K. Dashed green line: the flux of a blackbody with $T_{\rm eff}=160\,000$~K.
}%
\label{suwt2:fig17}%
\end{center}
\end{figure*}
\renewcommand{\baselinestretch}{1.5}

The modelling procedure consists of an iterative process during which the calculated H$\beta$ luminosity $L_{{\rm H}\beta}$(erg\,s${}^{-1}$), the ionic abundance ratios (i.e. He${}^{2+}$/He${}^{+}$, N${}^{+}$/H${}^{+}$,  O${}^{2+}$/H${}^{+}$) and
the flux intensities of some important lines, relative to H$\beta$ (such as He~{\sc ii} $\lambda$4686, $[$N~{\sc ii}$]$
$\lambda$6584 and $[$O~{\sc iii}$]$ $\lambda$5007) are compared with
the observations. We performed a maximum of 20 iterations per
simulation and the minimum convergence level achieved was 95\%.
The free parameters included distance and stellar characteristics, such as
luminosity and effective temperature. Although we adopted the density
and abundances derived in Sections~\ref{suwt2:sec:tempdens} and \ref{suwt2:sec:abundances}, we gradually scaled up/down the
abundances  in Table~\ref{suwt2:tab:abundances:empirical} until the
observed emission-line fluxes were reproduced by the model. Due to the
lack of infrared data we did not model the dust component of this
object. We notice however some variations among the values of $c({\rm H}\beta)$ between the ring and the inner region in
 Table~\ref{suwt2:tab:obslines}. It means that all of the observed reddening may not be due to the ISM.  
We did not include the outer bipolar lobes in our model, 
since the geometrical dilution reduces radiation beyond the ring.  
The faint bipolar lobes projected on the sky are far from the UV radiation field, and are dominated by the photodissociation region (PDR). There is a
transition region between the photoionized region and PDR. Since \textsc{mocassin} cannot currently treat a PDR fully, we are unable to model the region beyond the ionization front, i.e. the ring.
This low-density PN is extremely faint, and not highly optically thick (i.e. some UV radiations escape from the ring), so it is difficult to estimate a stellar luminosity from the total nebula H$\beta$ intrinsic line flux. 
The best-fitting model depends upon the effective temperature ($T_{\rm eff}$) and the stellar luminosity ($L_{\star}$), though both are related to the evolutionary stage of the central star. Therefore, it is necessary to restrict our stellar parameters to the evolutionary tracks of the post-AGB stellar models, e.g., `late thermal pulse', `very late thermal pulse' (VLTP), or `asymptotic giant branch final thermal pulse' \citep[see e.g.][]{Iben1983,Schoenberner1983,Vassiliadis1994,Bloecker1995,Herwig2001,MillerBertolami2006}. To constrain $T_{\rm eff}$ and $L_{\star}$, we employed a set of evolutionary tracks for initial masses between $1$ and $7{\rm M}_{\bigodot}$ calculated by \citet[Tables 3-5]{Bloecker1995}. Assuming a density model shown in Fig.~\ref{suwt2:fig14}, we first estimated the effective temperature and luminosity of the central star by matching the H$\beta$ luminosity $L({\rm H}\beta)$ and the ionic helium abundance ratio He${}^{2+}$/He${}^{+}$ with the values derived from observation and empirical analysis. Then, we scaled up/down abundances to get the best values for ionic abundance ratios and the flux intensities.

\subsection{Model input parameters}

\subsubsection{Density distribution}

The dense torus used for the ring was developed from the higher spectral resolution kinematic model of \citet{Jones2010} and our plasma diagnostics (Section~\ref{suwt2:sec:tempdens}).
Although the density cannot be more than the low-density limit of $N_{\rm e}<100$\,cm$^{-3}$ due to the [S~{\sc ii}] $\lambda\lambda$6716/6731 line ratio of $\gtrsim1.40$, it was slightly adjusted to produce the total H$\beta$ Balmer intrinsic line flux $I({\rm H}\beta)$ derived for the ring and interior structure or the H$\beta$ luminosity $L({\rm H}\beta)=4\pi\,D^2\,I({\rm H}\beta)$ at the specified distance $D$. The three-dimensional density distribution used for the torus and interior structure is shown in Fig.~\ref{suwt2:fig14}. The central star is located in the centre of the torus. The torus has a radius of $38.1$\,arcsec from its centre to the centre of the tube (1 arcsec is equal to $1.12\times 10^{-2}$ pc based on the best-fitting photoionization models). The radius of the tube of the ring is $6.9$\,arcsec. The hydrogen number density of the torus is taken to be homogeneous and equal to $N_{\rm H}=100$~cm${}^{-3}$. 
\citet{Smith2007} studied similar objects, including SuWt~2, and found that the ring itself can be a swept-up thin disc, and the interior of the
ring is filled with a uniform equatorial disc. Therefore, inside the ring, there is a less dense oblate spheroid with a homogeneous density of 50~cm${}^{-3}$, a semimajor axis of $31.2$\,arcsec and a semiminor axis of $6.9$\,arcsec. The H number density of the oblate spheroid is chosen to match the total $L({\rm H}\beta)$ and be a reasonable fit for H$^{2+}$/H$^{+}$ compared to the empirical results. The dimensions of the model were estimated from the kinematic model of \citet{Jones2010} with an adopted inclination of 68${}^{\circ}$. 
The distance was estimated over a range 2.1--2.7 kpc, which
corresponds to a reliable range based on the H$\alpha$ surface brightness--radius relation of \citet{Frew2006} and \citet{Frew2008}. The distance was allowed to vary to find
the best-fitting model. The value of 2.3 kpc adopted in this work
yielded the best match to the observed H$\beta$ luminosity and it is also in very good agreement with \citet{Exter2010}.

\subsubsection{Nebular abundances}

All major contributors to the thermal balance of the gas were included
in our model. We used a homogeneous elemental abundance distribution consisting of eight elements. 
The initial abundances of He, N, O, Ne, S and Ar were taken from the
observed empirically derived total abundances listed in
Table~\ref{suwt2:tab:abundances:empirical}. The abundance of C was a
free parameter, typically varying between $5\times10^{-5}$ and
$8\times10^{-3}$ in PNe.  We initially used the typical value of ${\rm C}/{\rm H}=5.5\times10^{-4}$ 
\citep{Kingsburgh1994}, and adjusted it to preserve the thermal balance of the nebula. 
We kept the
initial abundances fixed while the stellar parameters and distance were
being scaled to produce the best fit for the H$\beta$ luminosity and
He${}^{2+}$/He${}^{+}$ ratio, and then we gradually varied them to obtain
the finest match between the predicted and observed emission-line
fluxes, as well as ionic abundance ratios from the empirical analysis. 

The flux intensity of He~{\sc ii} $\lambda$4686 {\AA} and
the He${}^{2+}$/He${}^{+}$ ratio highly depend on the temperature and
luminosity of the central star. Increasing either $T_{\rm eff}$ or
$L_{\star}$ or both increases the He${}^{2+}$/He${}^{+}$
ratio. Our method was to match the He${}^{2+}$/He${}^{+}$ ratio, and
then scale the He/H abundance ratio to produce the observed intensity of He~{\sc ii} $\lambda$4686 {\AA}.

The abundance ratio of oxygen was adjusted to match the intensities of $[$O~{\sc iii}$]$ $\lambda\lambda$4959,5007 and to a lesser degree $[$O~{\sc ii}$]$ $\lambda\lambda$3726,~3729. In particular, the intensity of the $[$O~{\sc ii}$]$ doublet is unreliable due to 
the contribution of recombination and the uncertainty of about 30\% at the
extreme blue of the WiFeS. 
So we gradually modified the abundance ratio O/H until the best match
for $[$O~{\sc iii}$]$ $\lambda\lambda$4959,5007 and
O${}^{2+}$/H${}^{+}$ was produced. The abundance ratio of nitrogen
was adjusted to match the intensities of $[$N~{\sc ii}$]$
$\lambda\lambda$6548,6584 and N${}^{+}$/H${}^{+}$. Unfortunately, the
weak $[$N~{\sc ii}$]$ $\lambda$5755 emission line does not have a high S/N ratio in our data. 

The abundance ratio of sulphur was adjusted to match the intensities of $[$S~{\sc iii}$]$ $\lambda$9069. The intensities of $[$S~{\sc ii}$]$ $\lambda\lambda$6716,6731 and S${}^{+}$/H calculated by our models are about seven and ten times lower than those values derived from observations and empirical analysis, respectively. The intensity of $[$S~{\sc ii}$]$ $\lambda\lambda$6716,6731 is largely increased due to shock-excitation effects.  

\renewcommand{\baselinestretch}{0.9}
\begin{figure*}
\begin{center}
\includegraphics[width=3.5in]{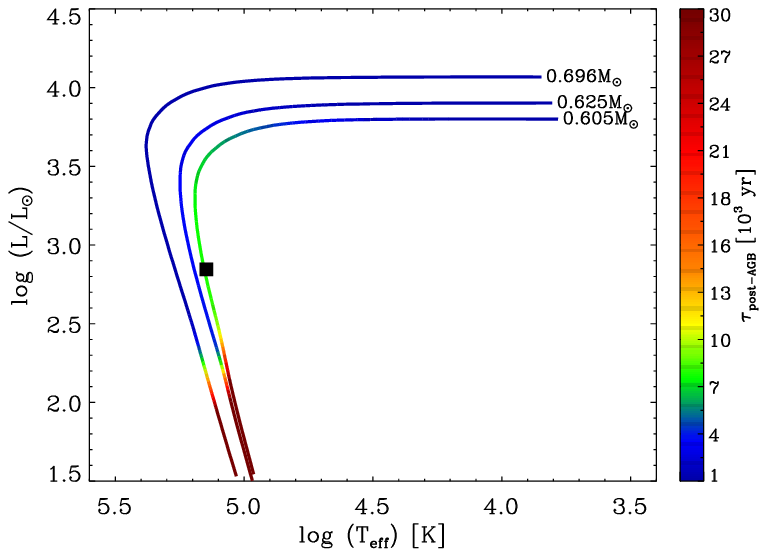}%
\includegraphics[width=3.5in]{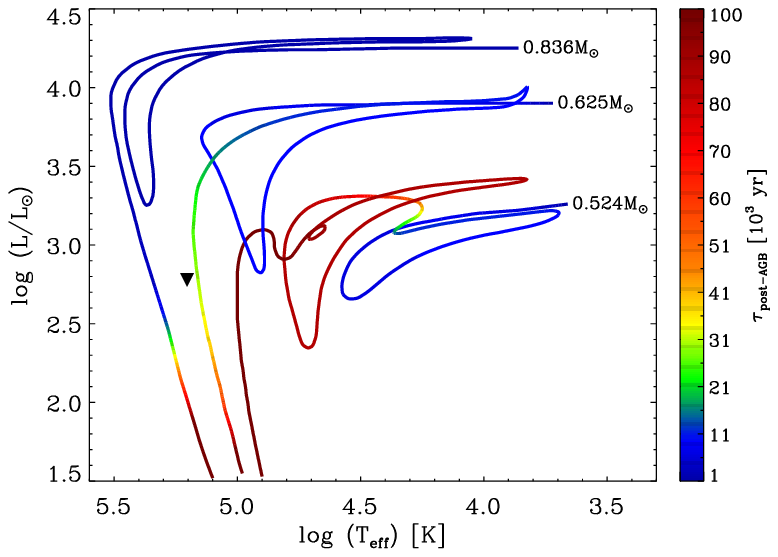}%
\caption{Hertzsprung--Russell diagrams for hydrogen-burning models (left-hand panel) with $(M_{\rm ZAMS},M_{\star})=$ $(3{\rm M}_{\bigodot}, 0.605{\rm M}_{\bigodot})$, $(3{\rm M}_{\bigodot}, 0.625{\rm M}_{\bigodot})$ and $(4{\rm M}_{\bigodot}, 0.696{\rm M}_{\bigodot})$, and helium-burning models (right-hand panel) with $(M_{\rm ZAMS},M_{\star})=$ $(1{\rm M}_{\bigodot}, 0.524{\rm M}_{\bigodot})$, $(3{\rm M}_{\bigodot}, 0.625{\rm M}_{\bigodot})$ and $(5{\rm M}_{\bigodot}, 0.836{\rm M}_{\bigodot})$ 
from \citet{Bloecker1995} compared to the position of the central star of SuWt~2 derived from two different photoionization models, namely Model 1 (denoted by $\blacksquare$) and Model 2 ($\blacktriangledown$). 
On the right, the evolutionary tracks contain the first evolutionary phase, the VLTP (born-again scenario), and the second evolutionary phase. The colour scales indicate the post-AGB ages ($\tau_{\rm post-\textsc{agb}}$) in units of $10^3$ yr.
}%
\label{suwt2:fig15}%
\end{center}
\end{figure*}
\renewcommand{\baselinestretch}{1.5}

Finally, the differences between the total abundances from our
photoionization model and those derived from our empirical analysis
can be explained by the $icf$ errors resulting from a non-spherical
morphology and properties of the exciting
source. \citet{Gonccalves2012} found that additional corrections are
necessary compared to those introduced by \citet{Kingsburgh1994} due
to geometrical effects.  Comparison with results from photoionization
models shows that the empirical analysis overestimated the neon
abundances. The neon abundance must be lower than the value found by
the empirical analysis to reproduce the observed intensities of $[$Ne~{\sc
  iii}$]$ $\lambda\lambda$3869,3967. It means that the $icf$(Ne) of
\citet{Kingsburgh1994} overestimates the unseen ionization
stages. \citet{Bohigas2008} suggested to use an alternative empirical method for correcting unseen ionization stages of neon. 
It is clear that with the typical Ne${}^{2+}$/Ne\,=\,O${}^{2+}$/O assumption of the $icf$ method, the neon total abundance is overestimated by the empirical analysis. 

\begin{figure*}
\begin{center}
\includegraphics[width=2.3in]{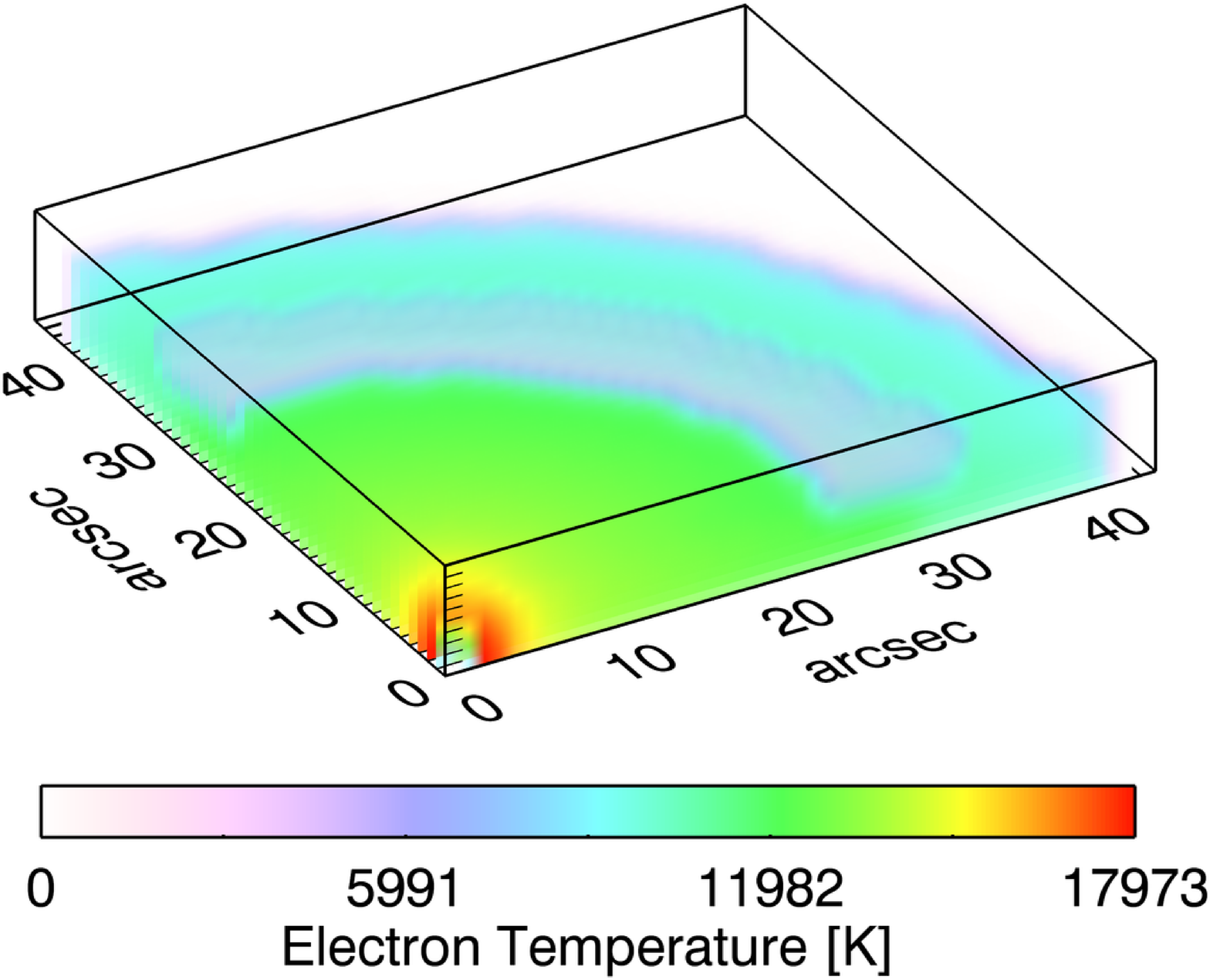}%
\includegraphics[width=2.3in]{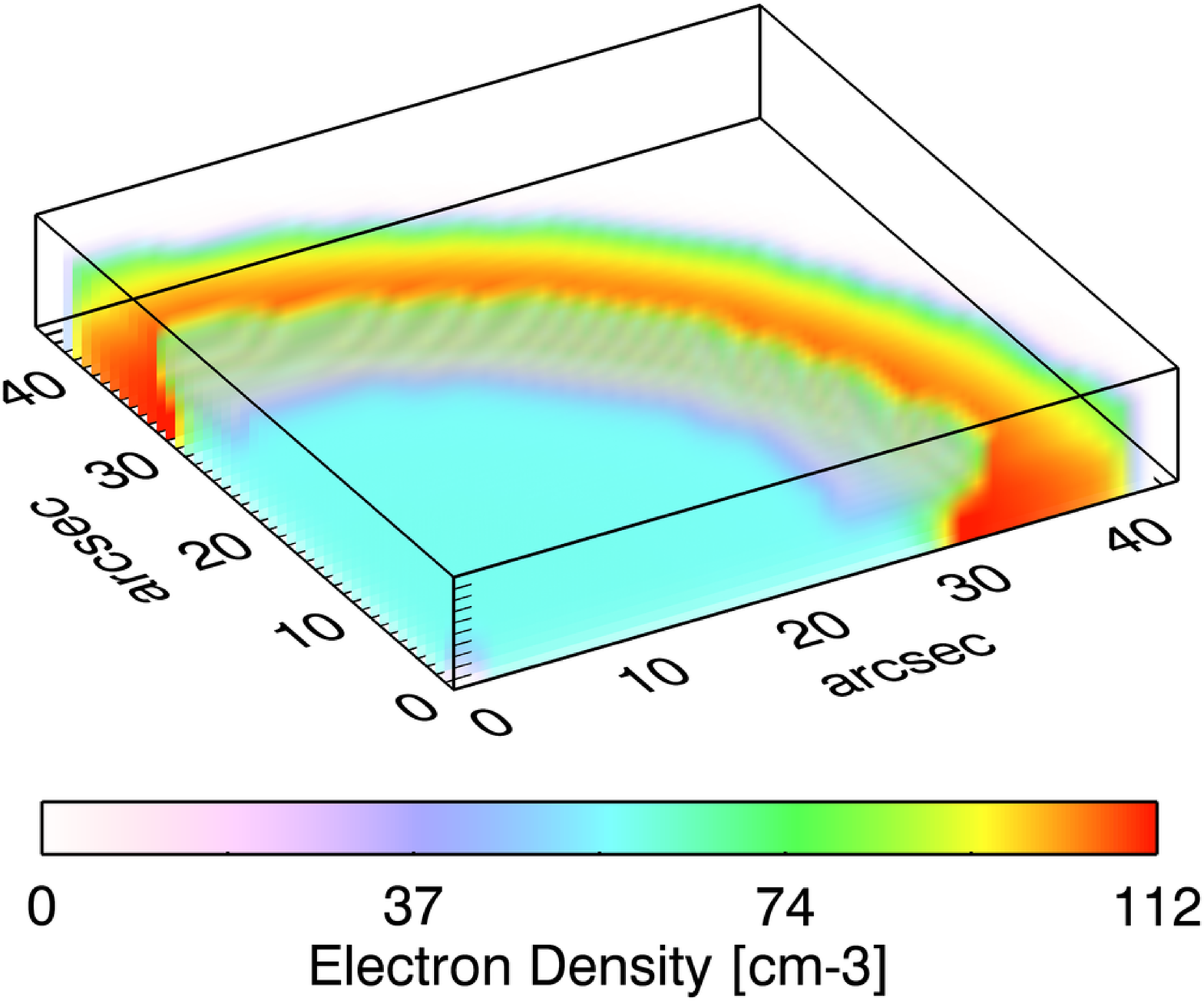}%
\includegraphics[width=2.3in]{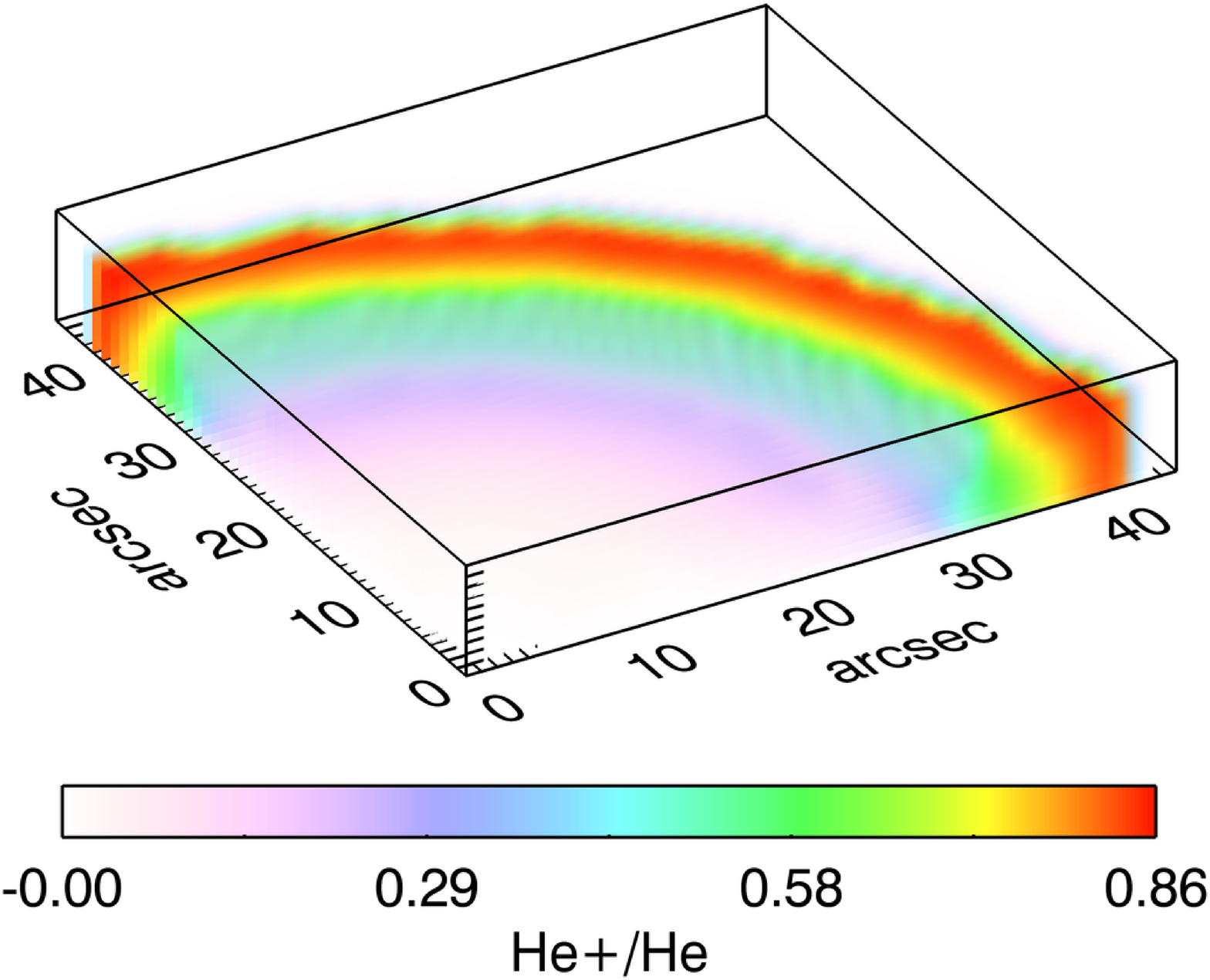}\\
\includegraphics[width=2.3in]{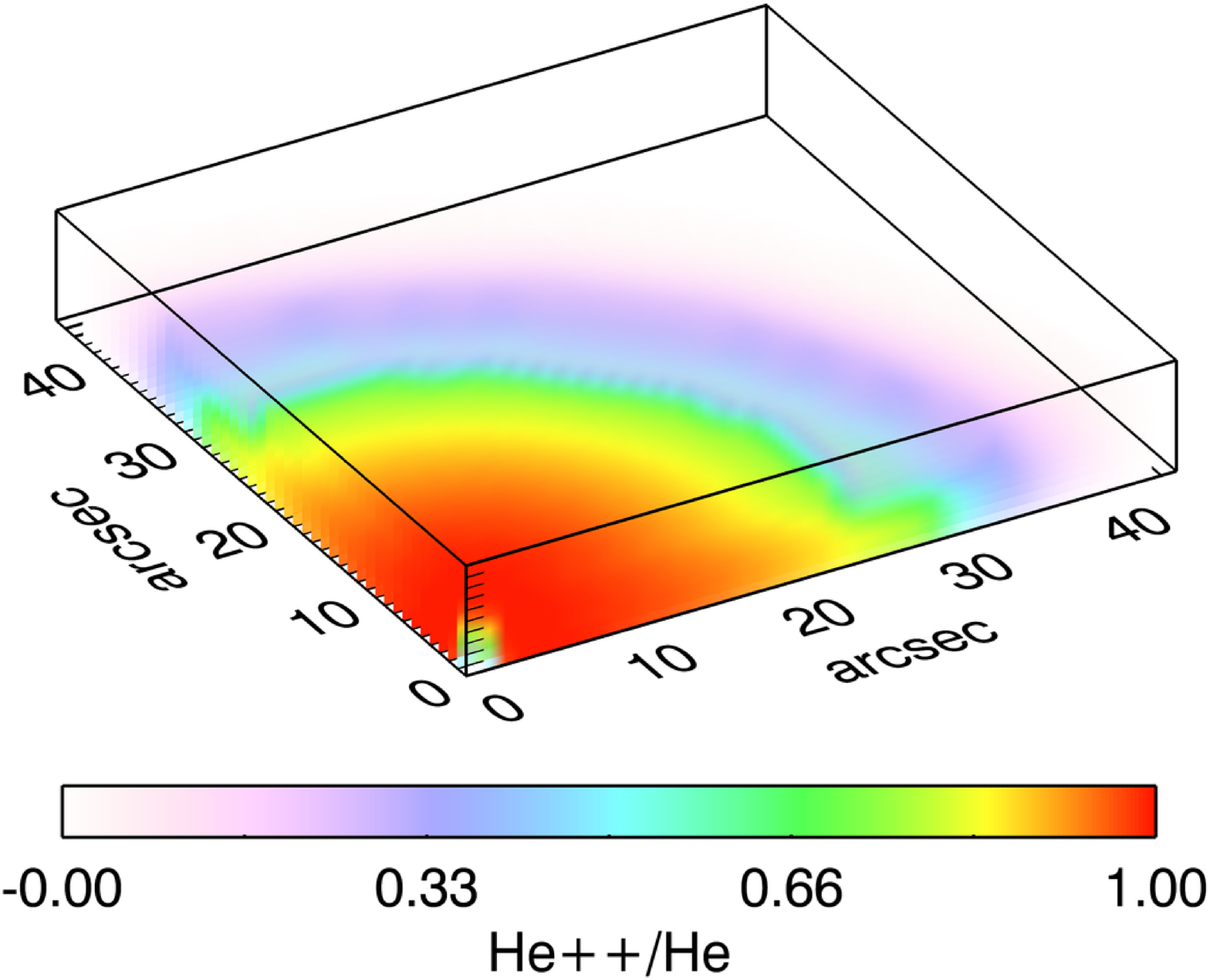}%
\includegraphics[width=2.3in]{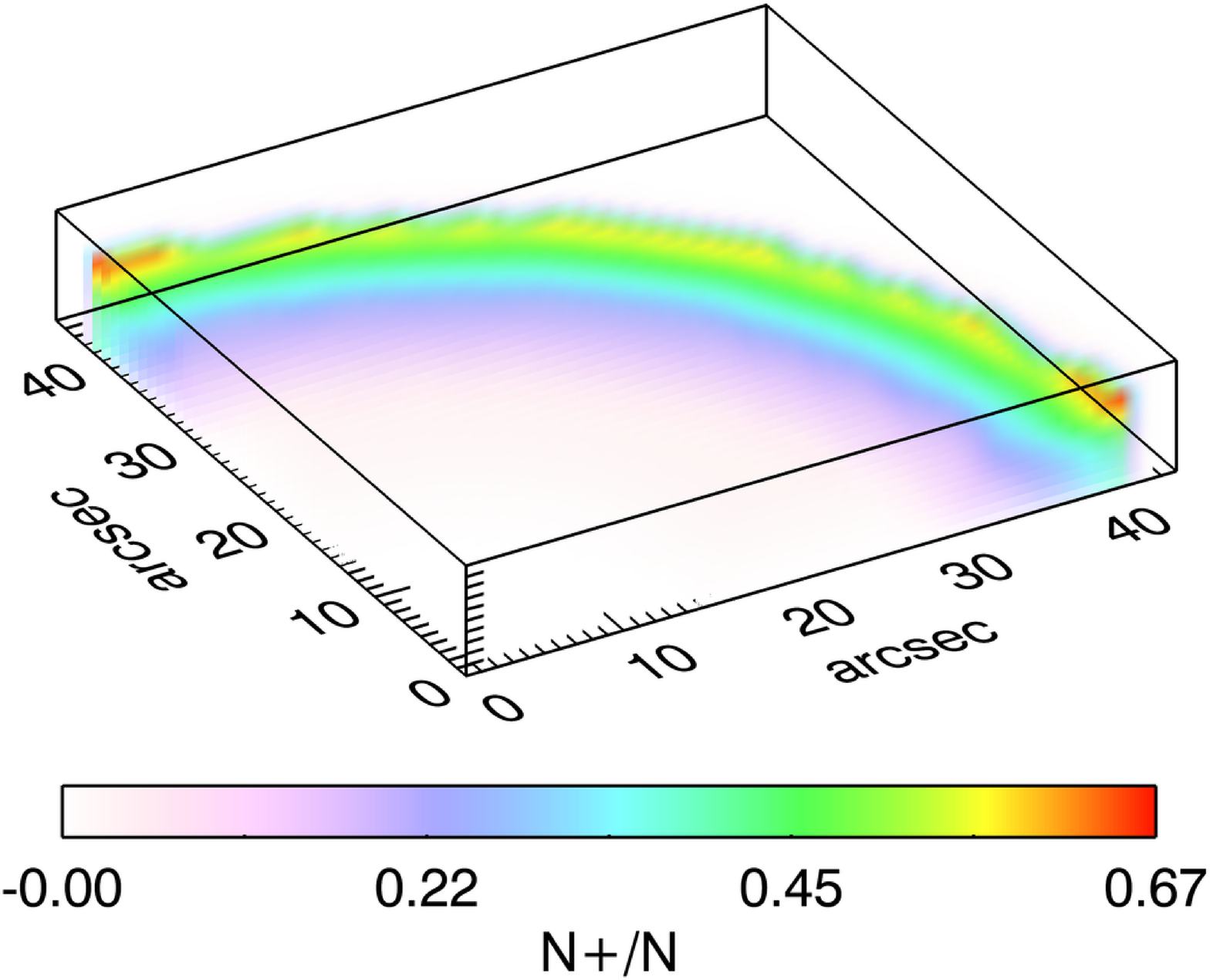}%
\includegraphics[width=2.3in]{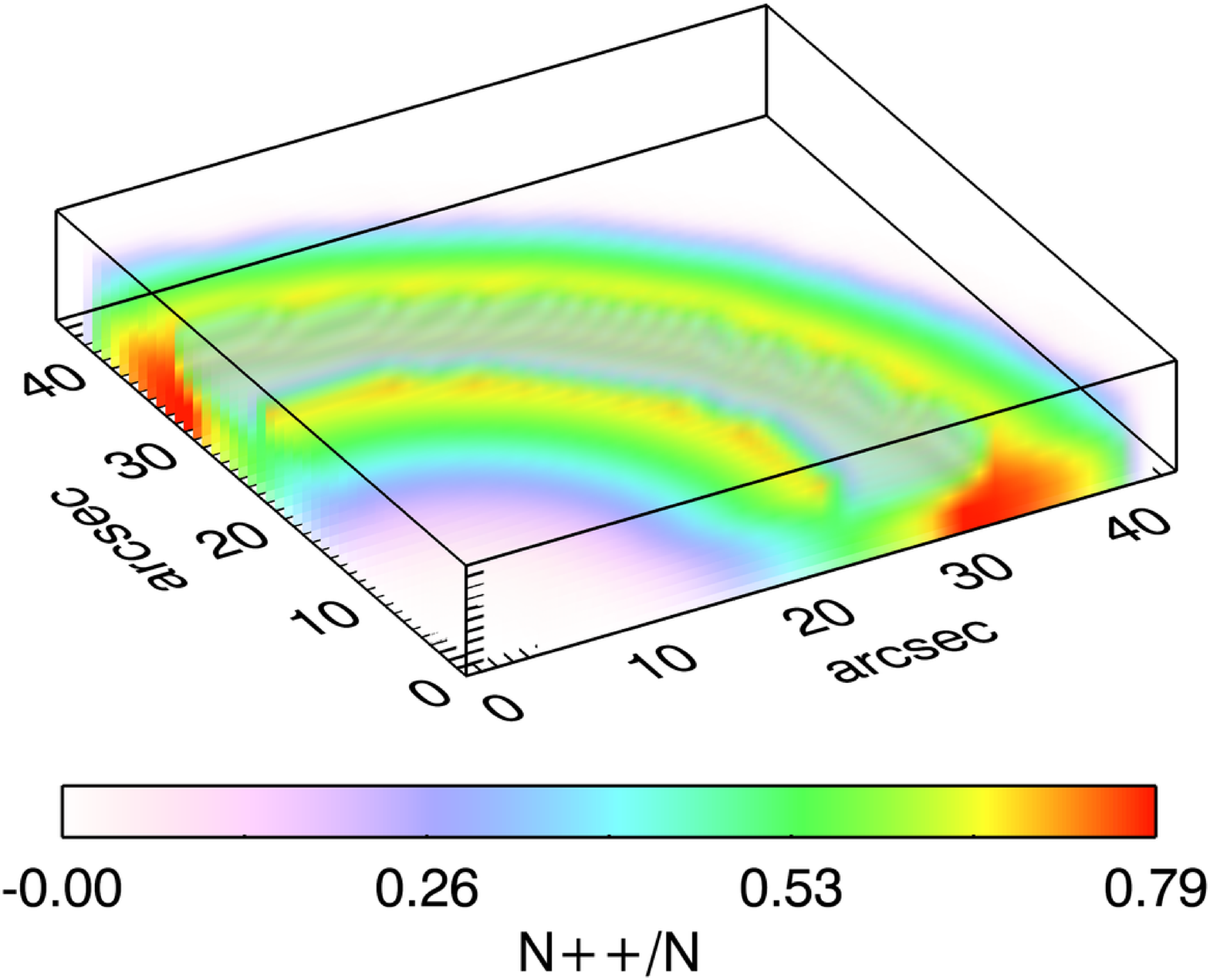}\\
\includegraphics[width=2.3in]{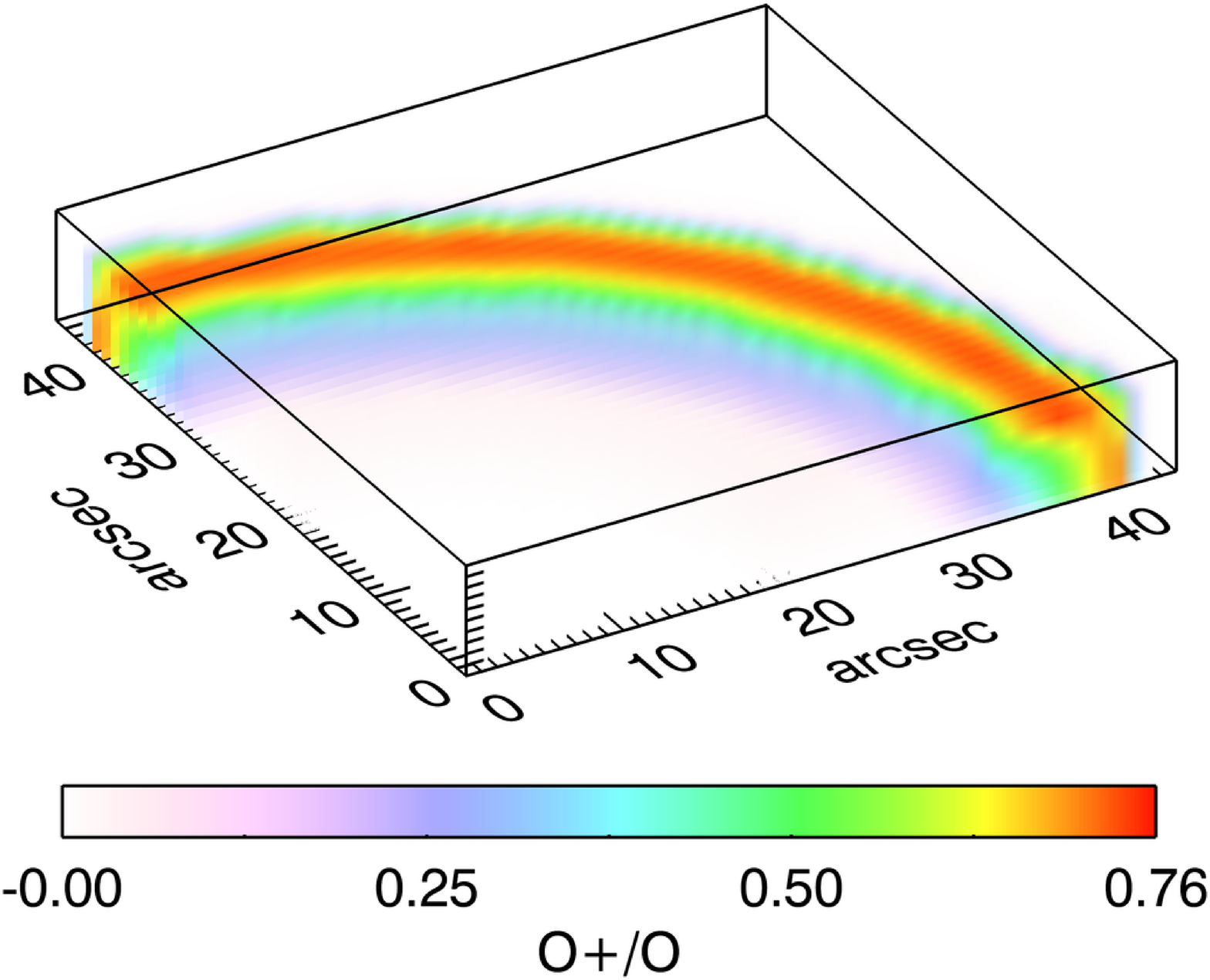}%
\includegraphics[width=2.3in]{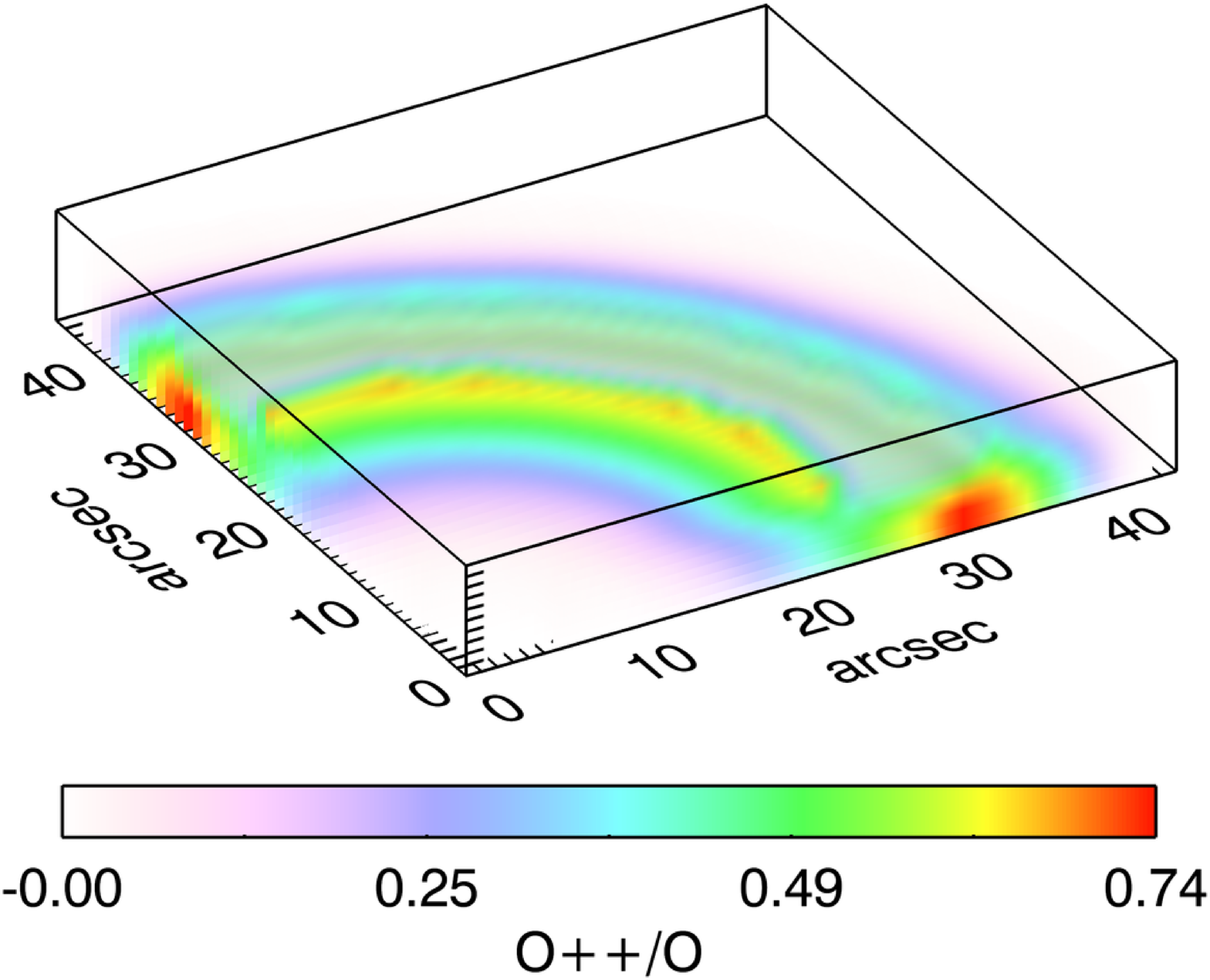}%
\includegraphics[width=2.3in]{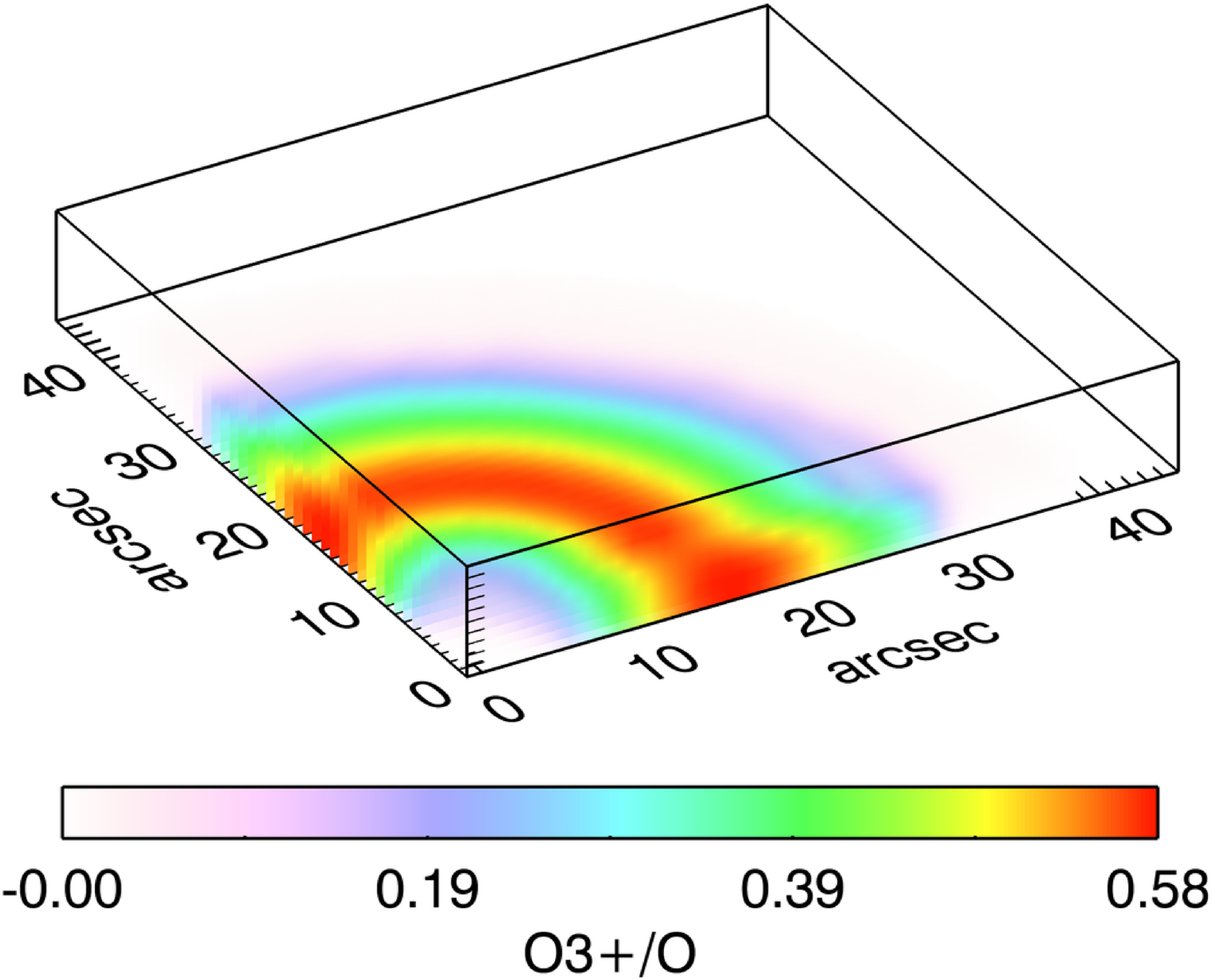}\\
\includegraphics[width=2.3in]{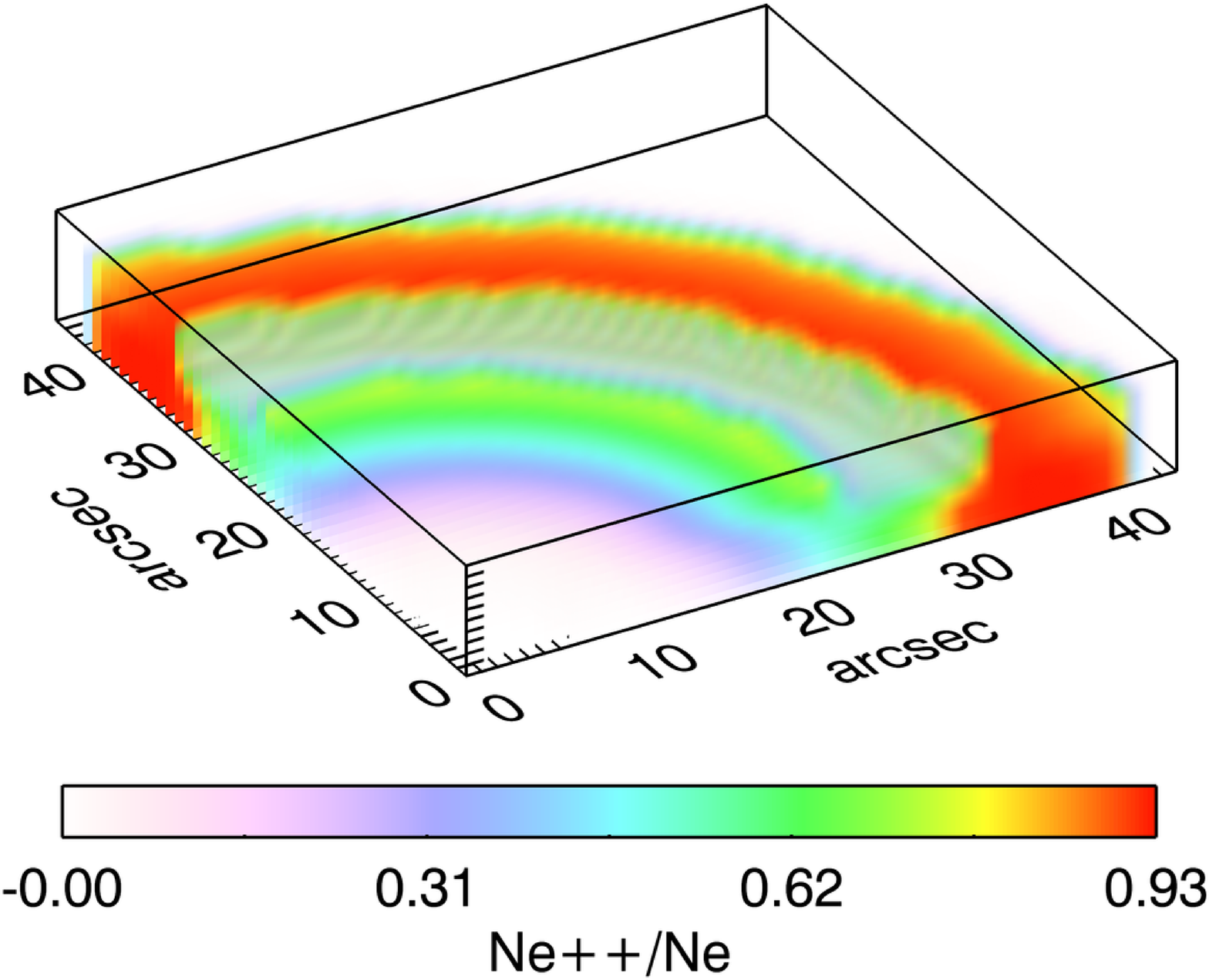}%
\includegraphics[width=2.3in]{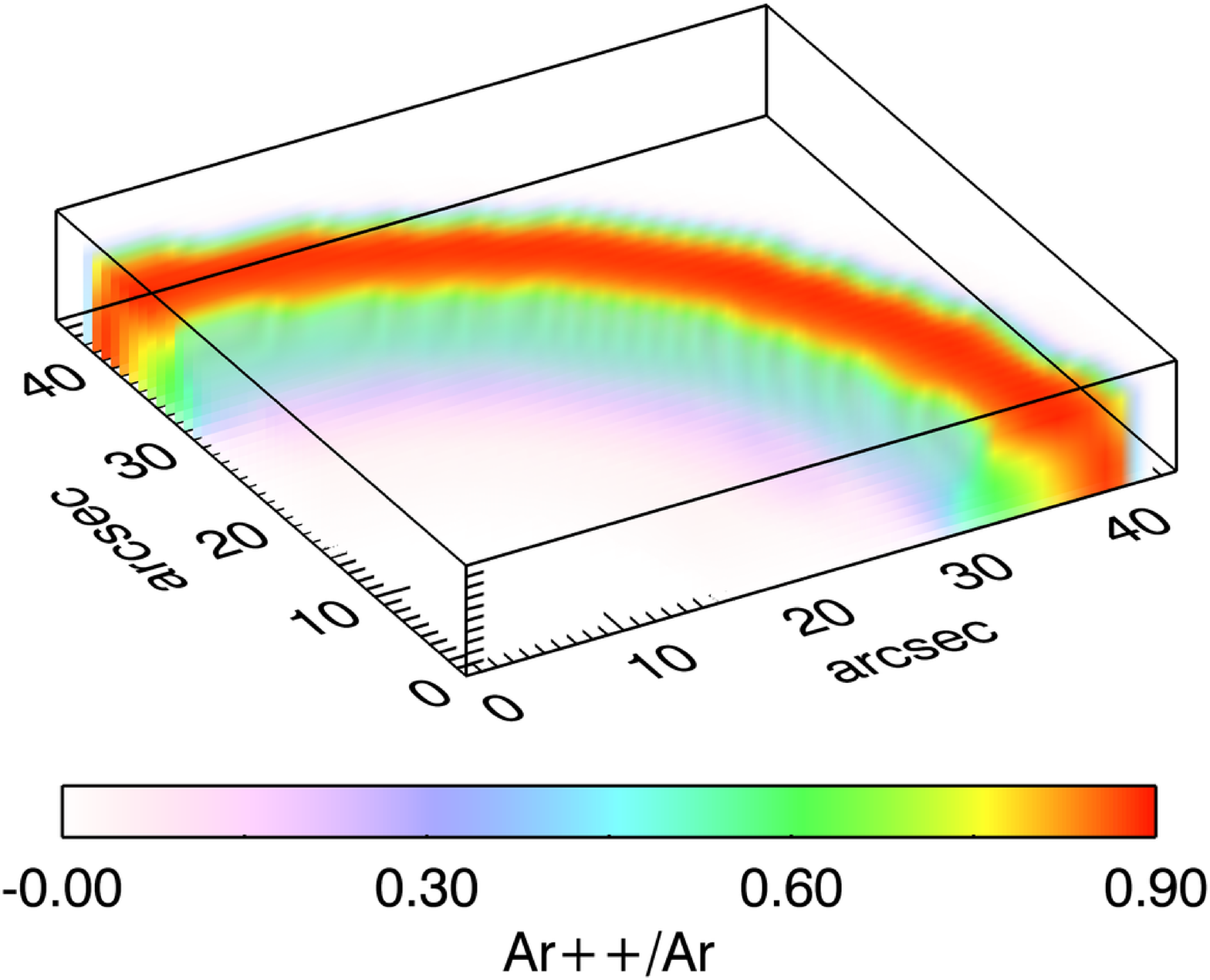}%
\includegraphics[width=2.3in]{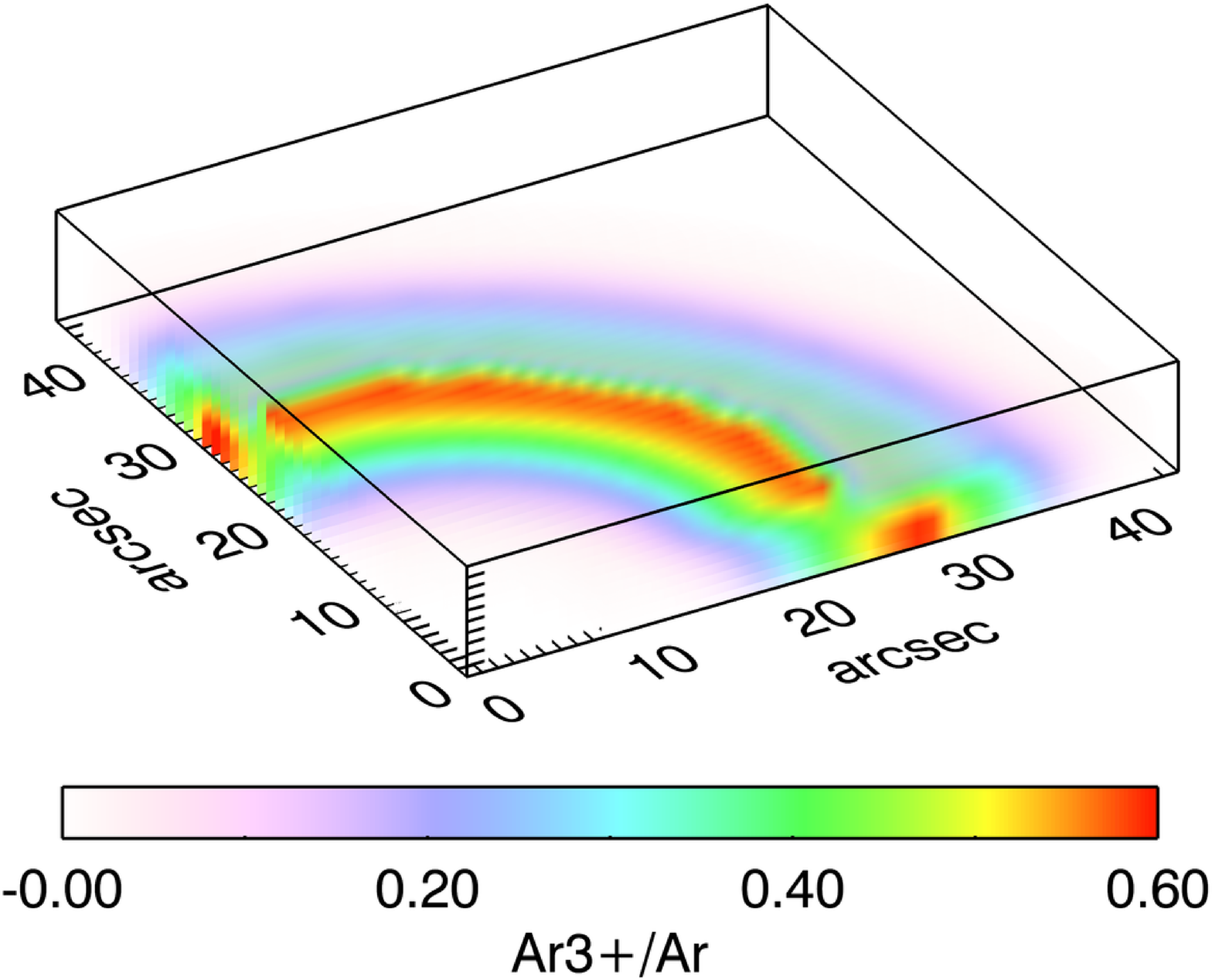}%
\caption{The 3 D distributions of electron temperature, electron density and ionic fractions from the adopted Model 2 constructed in $45 \times 45 \times 7$ cubic grids, and the ionizing source being placed in the corner (0,0,0). Each cubic cell has a length of $1.12 \times 10^{-2}$ pc, which corresponds to the actual PN  ring size.
}%
\label{suwt2:fig16}%
\end{center}
\end{figure*}
\renewcommand{\baselinestretch}{1.5}

\subsubsection{Ionizing source}

The central ionizing source of SuWt~2 was modelled using different
non-local thermodynamic equilibrium \citep[NLTE;][]{Rauch2003} model
atmospheres listed in Table~\ref{suwt2:tab:modelparameters}, as they
resulted in the best fit of the nebular emission-line
fluxes. Initially, we tested a set of blackbody fluxes with the effective temperature ($T_{\rm
  eff}$) ranging from $80\,000$ to $190\,000$ K, the stellar luminosity compared to that of the Sun ($L_{\star}/{\rm L}_{\bigodot}$) ranging from
50-800 and different evolutionary tracks \citep{Bloecker1995}. A
blackbody spectrum provides a rough estimate of the ionizing source
required to photoionize the PN SuWt~2. The assumption of
a blackbody spectral energy distribution (SED) is not quite correct as
indicated by \citet{Rauch2003}. The strong differences between a blackbody
SED and a stellar atmosphere are mostly noticeable at energies higher
than 54 eV (He~{\sc ii} ground state). We thus successively used the
NLTE T\"{u}bingen Model-Atmosphere 
Fluxes Package\footnote{Website: http://astro.uni-tuebingen.de/~rauch/TMAF/TMAF.html} \citep[TMAF;][]{Rauch2003} for hot compact stars. 
We initially chose the stellar temperature and luminosity (gravity) of the best-fitting blackbody model, and changed them to get the best observed ionization properties of the nebula. 

Fig.~\ref{suwt2:fig17} shows the NTLE model atmosphere fluxes used
to reproduce the observed nebular emission-line spectrum by our
photoionization models. We first used a hydrogen-rich model atmosphere with an
abundance ratio of H\,:\,He\,=\,8\,:\,2 by mass, $\log g =7$ (cgs), and $T_{\rm
  eff}=140\,000$~K (Model 1), corresponding to the final stellar mass of $M_{\star}=0.605\,{\rm M}_{\bigodot}$ and the zero-age main sequence (ZAMS) mass of $M_{\rm \textsc{zams}}=3\,{\rm M}_{\bigodot}$, where ${\rm M}_{\bigodot}$ is the solar mass. However, its post-AGB age ($\tau_{\rm post-\textsc{agb}}$) of 7\,500\,yr,
as shown in Fig.~\ref{suwt2:fig15} (left-hand panel), is too short to
explain the nebula's age. We therefore moved to a hydrogen-deficient model, which includes Wolf-Rayet central stars ([WC]) and the hotter PG~1159 stars. 
[WC]-type central stars are mostly associated with carbon-rich nebula \citep{Zijlstra1994}. The evolutionary tracks of the VLTP for H-deficient models, as shown in Fig.\,\ref{suwt2:fig15} (right-hand panel), imply a surface gravity of $\log g= 7.2$ for given $T_{\rm eff}$ and $L_{\star}$. 
From the high temperature and high surface gravity, we decided to use `typical' PG~1159 model atmosphere fluxes (He\,:\,C\,:\,N\,:\,O\,=\,33\,:\,50\,:\,2\,:\,15) with $T_{\rm eff}=160\,000$~K and $L_{\star}/{\rm L}_{\bigodot}=600$ (Model 2), corresponding to the post-AGB age of about $\tau_{\rm post-\textsc{agb}}=2$5\,000\,yr, $M_{\star}=0.64\,{\rm M}_{\bigodot}$ and $M_{\rm \textsc{zams}}=3\,{\rm M}_{\bigodot}$. 
The stellar mass found here is in agreement with the $0.7\,{\rm M}_{\bigodot}$ estimate of \citet{Exter2010}. 
Fig.\,\ref{suwt2:fig17} compares the two model atmosphere fluxes with a blackbody with $T_{\rm eff}=160\,000$~K. 

Table~\ref{suwt2:tab:modelparameters} lists the parameters used for our final simulations in two different NTLE model atmosphere fluxes. The ionization structure of this nebula was best reproduced using these best two models. Each model has different effective temperature, stellar luminosity and abundances (N/H, O/H and Ne/H). The results of our two models are compared in Tables~\ref{suwt2:tab:modelresults}--\ref{suwt2:tab:ionratio} to those derived from the observation and empirical analysis.

\renewcommand{\baselinestretch}{1.2}
\begin{table*}
\begin{center}
\caption{Mean electron temperatures (K) weighted by ionic species for the whole nebula obtained from the photoionization model. For each element the upper row is for Model 1 and the lower row is for Model 2. The bottom lines present the mean electron temperatures and electron densities for the torus (ring) and the oblate spheroid (inside).
}
\begin{tabular}{lcccccccc}
\multicolumn{2}{c}{} \\
\hline
\hline
     & \multicolumn{7}{c}{Ion}\\
\cline{2-9}
  Element & {\sc i}   &{\sc ii}   &{\sc iii}&{\sc iv}&{\sc v} &{\sc vi}&{\sc vii}\\
\hline
H & 11696 & 12904  &   &   &   &   &   \\ 
  & 11470 & 12623 &   &   &   &   &   \\ 
He & 11628 & 12187 & 13863 &   &   &   &   \\ 
   & 11405 & 11944  & 13567  &   &   &   &   \\ 
C & 11494  & 11922  & 12644  & 15061  & 17155 & 17236 & 12840 \\ 
 & 11289  & 11696 & 12405  & 14753 & 16354 & 16381 & 12550  \\ 
N & 11365 & 11864 & 12911 & 14822  & 16192 & 18315  & 18610  \\ 
  & 11170 & 11661  & 12697  & 14580  & 15836  & 17368 & 17475  \\ 
O & 11463 & 11941 & 12951 & 14949 & 15932  & 17384 & 20497 \\ 
  & 11283  & 11739 & 12744  & 14736 & 15797  & 17559  & 19806 \\ 
Ne & 11413  & 11863 & 12445 & 14774 & 16126 & 18059  & 22388  \\ 
   & 11196  & 11631  & 12215 & 14651 & 16166 & 18439 & 20488 \\
S & 11436 & 11772  & 12362  & 14174 & 15501 & 16257  & 18313  \\ 
  & 11239 & 11557 & 12133  & 13958  & 15204  & 15884  & 17281  \\ 
Ar & 11132 & 11593 & 12114  & 13222 & 14908 & 15554 & 16849  \\ 
  & 10928  & 11373  & 11894 & 13065  & 14713 & 15333  & 16392 \\ 
\hline
& \multicolumn{4}{c}{Torus}& \multicolumn{4}{c}{Spheroid}\\
\cline{2-9}
& \multicolumn{2}{c}{$T_{\rm e}[$O\,{\sc iii}$]$}&\multicolumn{2}{c}{$N_{\rm e}[$S\,{\sc ii}$]$}&\multicolumn{2}{c}{$T_{\rm e}[$O\,{\sc iii}$]$}&\multicolumn{2}{c}{$N_{\rm e}[$S\,{\sc ii}$]$}\\
\hline
{M.1} & \multicolumn{2}{c}{12187\,K}& \multicolumn{2}{c}{105\,cm${}^{-3}$\,}&\multicolumn{2}{c}{15569\,K}&\multicolumn{2}{c}{58\,cm${}^{-3}$}\\
{M.2} & \multicolumn{2}{c}{11916\,K}& \multicolumn{2}{c}{103\,cm${}^{-3}$}&\multicolumn{2}{c}{15070\,K}&\multicolumn{2}{c}{58\,cm${}^{-3}$}\\
\hline 
\label{suwt2:tab:temperatures}
\end{tabular}
\end{center}
\end{table*}
\renewcommand{\baselinestretch}{1.5}

\renewcommand{\baselinestretch}{1.2}
\begin{table*}
\begin{center}
\caption{Fractional ionic abundances for SuWt~2 obtained from the photoionization models.
For each element the upper row is for the torus (ring) and the lower row is for the oblate spheroid (inside).}
\begin{tabular}{llccccccc}
\multicolumn{8}{c}{}\\
\hline  
\hline 
   &   & \multicolumn{7}{c}{Ion}\\
\cline{3-9}
   &Element & {\sc i}   &{\sc ii}   &{\sc iii}&{\sc iv}&{\sc v} &{\sc vi}&{\sc vii}\\
\hline 
\multirow{16}{*}{\rotatebox{90}{Model 1}} 
   &H  &   6.53($-2$) &    9.35($-1$) &   &   &   &   &   \\
   &   &   3.65($-3$) &    9.96($-1$) &   &   &   &   &   \\
   &He &   1.92($-2$) &    7.08($-1$) &    2.73($-1$)  &   &   &   &   \\
   &   &   3.05($-4$) &    1.27($-1$) &    8.73($-1$)  &   &   &   &   \\
   &C  &   5.92($-3$) &    2.94($-1$) &    6.77($-1$) &    2.33($-2$) &    1.86($-4$) &    7.64($-16$) &    1.00($-20$) \\
   &   &   3.49($-5$) &    1.97($-2$) &    3.97($-1$) &    4.50($-1$) &    1.33($-1$) &    1.09($-12$) &    1.00($-20$) \\
   &N  &   7.32($-3$) &    4.95($-1$) &    4.71($-1$) &    2.62($-2$) &    4.18($-4$) &    6.47($-6$) &    2.76($-17$) \\
   &   &   1.02($-5$) &    1.30($-2$) &    3.65($-1$) &    3.97($-1$) &    1.59($-1$) &    6.69($-2$) &    6.89($-13$) \\
   &O  &   6.15($-2$) &    4.98($-1$) &    4.21($-1$) &    1.82($-2$) &    7.09($-4$) &    1.34($-5$) &    7.28($-8$) \\
   &   &   6.96($-5$) &    1.26($-2$) &    3.31($-1$) &    4.03($-1$) &    1.69($-1$) &    6.00($-2$) &    2.42($-2$) \\
   &Ne &   3.46($-4$) &    6.70($-2$) &    9.10($-1$) &    2.26($-2$) &    3.56($-4$) &    4.25($-6$) &    2.11($-9$) \\
   &   &   1.39($-6$) &    3.32($-3$) &    3.71($-1$) &    3.51($-1$) &    2.05($-1$) &    6.55($-2$) &    4.49($-3$) \\
   &S  &   1.13($-3$) &    1.67($-1$) &    7.75($-1$) &    5.52($-2$) &    1.15($-3$) &    6.20($-5$) &    8.53($-7$) \\
   &   &   3.18($-6$) &    3.89($-3$) &    1.73($-1$) &    3.53($-1$) &    2.43($-1$) &    1.57($-1$) &    6.91($-2$) \\
   &Ar &   4.19($-4$) &    3.15($-2$) &    7.51($-1$) &    2.10($-1$) &    5.97($-3$) &    1.13($-3$) &    5.81($-5$) \\
   &   &   1.12($-7$) &    2.33($-4$) &    5.81($-2$) &    2.83($-1$) &    1.85($-1$) &    2.73($-1$) &    2.01($-1$) \\
\hline
\multirow{16}{*}{\rotatebox{90}{Model 2}} 
   &H  &   7.94($-2$) &    9.21($-1$) &   &   &   &   &   \\
   &   &   4.02($-3$) &    9.96($-1$) &   &   &   &   &   \\
   &He &   2.34($-2$) &    7.25($-1$) &    2.51($-1$)  &   &   &   &   \\
   &   &   3.51($-4$) &    1.33($-1$) &    8.67($-1$)  &   &   &   &   \\
   &C  &   7.97($-3$) &    3.23($-1$) &    6.49($-1$) &    1.93($-2$) &    1.29($-4$) &    5.29($-16$) &    1.00($-20$) \\
   &   &   4.45($-5$) &    2.23($-2$) &    4.13($-1$) &    4.41($-1$) &    1.23($-1$) &    1.00($-12$) &    1.00($-20$) \\
   &N  &   1.00($-2$) &    5.44($-1$) &    4.24($-1$) &    2.15($-2$) &    2.62($-4$) &    2.20($-6$) &    9.23($-18$) \\
   &   &   1.31($-5$) &    1.52($-2$) &    3.84($-1$) &    4.07($-1$) &    1.50($-1$) &    4.40($-2$) &    4.34($-13$) \\
   &O  &   7.91($-2$) &    5.29($-1$) &    3.78($-1$) &    1.40($-2$) &    4.27($-4$) &    2.05($-6$) &    6.62($-11$) \\
   &   &   9.34($-5$) &    1.50($-2$) &    3.60($-1$) &    4.20($-1$) &    1.75($-1$) &    2.97($-2$) &    1.85($-4$) \\
   &Ne &   4.54($-4$) &    7.35($-2$) &    9.09($-1$) &    1.73($-2$) &    1.41($-4$) &    1.94($-8$) &    2.25($-14$) \\
   &   &   1.75($-6$) &    3.85($-3$) &    4.19($-1$) &    3.86($-1$) &    1.89($-1$) &    1.73($-3$) &    6.89($-7$) \\
   &S  &   1.64($-3$) &    1.95($-1$) &    7.58($-1$) &    4.47($-2$) &    7.84($-4$) &    3.39($-5$) &    3.05($-7$) \\
   &   &   4.23($-6$) &    4.86($-3$) &    1.96($-1$) &    3.61($-1$) &    2.39($-1$) &    1.47($-1$) &    5.16($-2$) \\
   &Ar &   7.22($-4$) &    3.99($-2$) &    7.74($-1$) &    1.81($-1$) &    3.95($-3$) &    5.62($-4$) &    1.60($-5$) \\
   &   &   1.72($-7$) &    3.22($-4$) &    7.30($-2$) &    3.30($-1$) &    1.96($-1$) &    2.62($-1$) &    1.39($-1$) \\
\hline \label{suwt2:tab:ionfraction}
\end{tabular}
\end{center}
\end{table*}
\renewcommand{\baselinestretch}{1.5}

\renewcommand{\baselinestretch}{1.2}
\begin{table*}
\begin{center}
\caption{Integrated ionic abundance ratios for the entire nebula obtained from the photoionization models. }
\begin{tabular}{lccccc}
\hline
\hline
            &             &   \multicolumn{2}{c}{Model 1 }  & \multicolumn{2}{c}{Model 2 }  \\
\cline{3-6}
Ionic ratio  		& Empirical   &   Abundance  &  Ionic Fraction&   Abundance  &  Ionic Fraction\\
\hline
He${}^{+}$/H${}^{+}$   & 4.80($-2$) & 5.308($-2$)  &  58.97\%  	& 5.419($-2$)   	&   60.21\%  	 \\
He${}^{2+}$/H${}^{+}$  & 3.60($-2$) & 3.553($-2$)  &  39.48\% 	& 3.415($-2$)    	&   37.95\%  	\\
C${}^{+}$/H${}^{+}$    & --     	& 9.597($-5$)  &  23.99\% 	& 1.046($-4$)   	&   26.16\%  	\\
C${}^{2+}$/H${}^{+}$   & --     	& 2.486($-4$)  &  62.14\% 	& 2.415($-4$)  		&   60.38\%   	\\
N${}^{+}$/H${}^{+}$    & 1.309($-4$) & 9.781($-5$) &  40.09\% 	& 1.007($-4$)    	&   43.58\%  	\\
N${}^{2+}$/H${}^{+}$   & --     	& 1.095($-4$)  &  44.88\% 	& 9.670($-5$)  		&   41.86\%   	\\
N${}^{3+}$/H${}^{+}$   & --     	& 2.489($-5$)  &  10.20\% 	& 2.340($-5$)  		&   10.13\%   	\\
O${}^{+}$/H${}^{+}$    & 1.597($-4$) & 1.048($-4$) &  40.30\% 	& 1.201($-4$)    	&   42.44\%   	\\
O${}^{2+}$/H${}^{+}$   & 1.711($-4$) & 1.045($-4$) &  40.20\%  	& 1.065($-4$)  		&   37.64\%  	\\
O${}^{3+}$/H${}^{+}$   & --     	& 2.526($-5$)  &  9.72\%  	& 2.776($-5$)  		&   9.81\%    	\\
Ne${}^{+}$/H${}^{+}$   & --     	& 6.069($-6$)  &  5.47\%  	& 6.571($-6$)   	&   5.92\%  	\\
Ne${}^{2+}$/H${}^{+}$  & 1.504($-4$) & 8.910($-5$) &  80.27\%  	& 9.002($-5$)  		&   81.10\%  	\\
Ne${}^{3+}$/H${}^{+}$  & --     	& 1.001($-5$)  &  9.02\%  	& 1.040($-5$)  		&   9.37\%     	\\
S${}^{+}$/H${}^{+}$\,$^{a}$    & 2.041($-6$) & 2.120($-7$) &  13.50\%  	& 2.430($-7$)  		&   15.48\%   	\\
S${}^{2+}$/H${}^{+}$   & 1.366($-6$) & 1.027($-6$) &  65.44\%  	& 1.013($-6$)  		&   64.55\%   	\\
S${}^{3+}$/H${}^{+}$   & --     	& 1.841($-5$)  &  11.73\%  	& 1.755($-7$)  		&   11.18\%  	\\
Ar${}^{+}$/H${}^{+}$   & --     	& 3.429($-8$)  &  2.54\%  	& 4.244($-8$) 		&   3.14\%  	\\
Ar${}^{2+}$/H${}^{+}$  & 1.111($-6$) & 8.271($-7$) &  61.26\%  	& 8.522($-7$)   	&   63.13\%  	\\
Ar${}^{3+}$/H${}^{+}$  & 4.747($-7$) & 3.041($-7$) &  22.52\%  	& 2.885($-7$) 		&   21.37\%   	\\
Ar${}^{4+}$/H${}^{+}$  & --     	& 5.791($-8$)  &  4.29\%   	& 5.946($-8$)  		&   4.40\%  	\\
Ar${}^{5+}$/H${}^{+}$  & --     	& 7.570($-8$)  &  5.61\%  	& 7.221($-8$)   	&   5.35\%  	\\
\hline \label{suwt2:tab:ionratio}
\end{tabular}\\
\end{center}
\begin{flushleft}
{\footnotesize $^{a}$~{Shock excitation largely enhances the S${}^{+}$/H${}^{+}$ ionic abundance ratio.}
}
\end{flushleft}
\end{table*}
\renewcommand{\baselinestretch}{1.5}

\subsection{Model results}

\subsubsection{Emission-line fluxes}

Table~\ref{suwt2:tab:modelresults} compares the flux intensities calculated by our models with those from the observations. The fluxes are given relative to H$\beta$, on a scale where H$\beta=100$. Most predicted line fluxes from each model are in fairly good agreement with the observed values and the two models produce very similar fluxes for most observed species. There are still some discrepancies in the few lines, e.g. $[$O~{\sc ii}$]$ $\lambda\lambda$3726,3729 and $[$S~{\sc ii}$]$ $\lambda\lambda$6716,6731.
The discrepancies in $[$O~{\sc ii}$]$ $\lambda\lambda$3726,3729 can be
explained by either recombination contributions or intermediate phase
caused by a complex density distribution \citep[see e.g. discussion
in][]{Ercolano2003c}. $[$S~{\sc ii}$]$
$\lambda\lambda$6716,6731 was affected by shock-ionization and its
true flux intensity is much lower without the shock fronts. Meanwhile, $[$Ar~{\sc iii}$]$
7751 was enhanced by the telluric line. The recombination
line H$\delta$ $\lambda$4102 and He~{\sc ii} $\lambda$5412 were also blended
with the O~{\sc ii} recombination lines. There are also some recombination contributions in the $[$O~{\sc ii}$]$ $\lambda\lambda$7320,7330 doublet. 
Furthermore, the discrepancies in the faint auroral line [N~{\sc ii}] $\lambda$5755 and [O~{\sc iii}] $\lambda$4363 can be explained by the recombination excitation contribution \citep[see section 3.3 in][]{Liu2000}.

\subsubsection{Temperature structure}

Table~\ref{suwt2:tab:temperatures} represents mean electron
temperatures weighted by ionic abundances for Models~1 and 2, as well as the ring region and the
inside region of the PN. We also see each ionic temperature
corresponding to the temperature-sensitive line ratio of a specified
ion. The definition for the mean temperatures was given in
\citet{Ercolano2003b}; and in detail by \citet{Harrington1982}. 
 Our model results for $T_{\rm e}[$O~{\sc iii}$]$ compare well with the value
 obtained from the empirical analysis in
 \S\,\ref{suwt2:sec:tempdens}. Fig.~\ref{suwt2:fig16} (top left) shows
 $T_{\rm e}$ obtained for Model 2 (adopted best-fitting model)
 constructed in $45 \times 45 \times 7$ cubic grids, and with the ionizing
 source being placed in the corner.  It replicates the situation where
 the inner region has much higher $T_{\rm e}$ in comparison to the ring
 $T_{\rm e}$ as previously found by plasma diagnostics in
 \S\,\ref{suwt2:sec:tempdens}. In particular the mean values of
 $T_{\rm e}[$O~{\sc iii}$]$  for the ring (torus of the actual nebula) and
 the inside (spheroid) regions are around $12\,000$ and
 $15\,000$\,K in all two models, respectively. They can be compared to
 the values of Table~\ref{suwt2:tab:tenediagnostics} that is
 $T_{\rm e}[$O~{\sc iii}$]=12\,300$\,K (ring) and $\lesssim20\,000$\,K
 (interior). Although the average temperature of $T_{\rm e}[$N~{\sc ii}$]\simeq11\,700$\,K
 over the entire nebula is higher than that given in
 Table~\ref{suwt2:tab:tenediagnostics}, the average temperature of
 $T_{\rm e}[$O~{\sc iii}$]\simeq13,000$\,K is in decent agreement with that found
 by our plasma diagnostics. 

It can be seen in Table~\ref{suwt2:tab:tenediagnostics} that the
temperatures for the two main regions of the nebula are very
different, although we assumed a homogeneous elemental abundance
distribution for the entire nebula relative to hydrogen. The
temperature  variations in the model can also be seen in
Fig.~\ref{suwt2:fig16}. The gas density structure and the location
of the ionizing source play a major role in heating the central
regions, while the outer regions remain cooler as expected. Overall,
the average electron temperature of the entire nebula increases by
increasing the helium abundance and decreasing the oxygen, carbon and
nitrogen abundances, which are efficient coolants.
We did not include any dust grains in our simulation, although we note
that a large dust-to-gas ratio may play a role in the heating of the
nebula via photoelectric emissions from the surface of grains.

\subsubsection{Ionization structure}

Results for the fractional ionic abundances in the ring (torus) and
inner (oblate spheroid) regions of our two models are shown in
Table~\ref{suwt2:tab:ionfraction} and Fig.~\ref{suwt2:fig16}. It is
clear from the figure and table that the ionization structures from
the models vary through the nebula due to the complex density and
radiation field distribution in the gas.
As shown in Table~\ref{suwt2:tab:ionfraction} , He${}^{2+}$/He is much
higher in the inner regions, while He${}^{+}$/He is larger in the
outer regions, as expected. Similarly, we find that the higher
ionization stages of each element are larger in the inner
regions. From Table~\ref{suwt2:tab:ionfraction} we see that hydrogen
and helium are both fully ionized and neutrals are less than 8\% by number in these best-fitting models. Therefore, our assumption of $icf({\rm He})=1$ is correct in our empirical method. 

Table~\ref{suwt2:tab:ionratio} lists the nebular average ionic
abundance ratios calculated from the photoionization models. The
values that our models predict for the helium ionic ratio are fairly comparable with those from
the empirical methods given in \S\,\ref{suwt2:sec:abundances}, 
though there are a number of significant differences in other ions.
The  O$^{+}$/H$^{+}$ ionic abundance ratio is about 33 per cent lower, 
while O$^{2+}$/H$^{+}$ is about 60\% lower in Model 2 than the empirical
observational value.
The empirical value of S${}^{+}$ differs by a factor of 8 compared to our result in Model 2, 
explained by the shock-excitation effects on the $[$S~{\sc ii}$]$ $\lambda\lambda$6716,6731 doublet. 
Additionally, the Ne${}^{2+}$/H$^{+}$ ionic abundance ratio was underestimated by roughly 67\% in
Model 2 compared to observed results, explained by the properties of the ionizing source.
The Ar$^{3+}$/H$^{+}$ ionic abundance ratio in Model 2 is 56\% lower than the empirical results.
Other ionic fractions do not show major discrepancies; differences remain below 35\%.
We note that the N${}^{+}$/N ratio is roughly equal to the O${}^{+}$/O
ratio, similar to what is generally assumed in the $icf$(N)
method. However, the Ne${}^{2+}$/Ne ratio is nearly a factor of 2
larger than the O${}^{2+}$/O ratio, in contrast to the general
assumption for $icf$(Ne) (see equation~\ref{suwt2:eq_ne_cel1}).  It has
already been noted by \citet{Bohigas2008} that an alternative
ionization correction method is necessary for correcting the unseen
ionization stages for the neon abundance.  

\subsubsection{Evolutionary tracks}

In Fig.~\ref{suwt2:fig15} we compared the values of the effective
temperature $T_{\rm eff}$ and luminosity $L_{\star}$ obtained from our
two models listed in Table~\ref{suwt2:tab:modelparameters} to
evolutionary tracks of hydrogen-burning and helium-burning
models calculated by \citet{Bloecker1995}.  We compared the
post-AGB age of these different models with the dynamical age of the
ring found in \S\,\ref{suwt2:sec:kinematic}. The kinematic analysis
indicates that the nebula was ejected about 23\,400--26\,300 yr ago. 
The post-AGB age of the hydrogen-burning model (left-hand panel in
Fig.~\ref{suwt2:fig15}) is considerably shorter than the nebula's age, suggesting that the helium-burning model (VLTP; right-hand panel in Fig.~\ref{suwt2:fig15}) may be favoured to explain the age.

The physical parameters of the two A-type stars also yield a further constraint. The stellar evolutionary tracks
of the rotating models for solar metallicity calculated by \citet{Ekstrom2012} imply that the A-type stars, both
with masses close to $2.7{\rm M}_{\bigodot}$ and $T_{\rm eff}\simeq9200$\,K, have ages of $\sim500$~Myr. We see that they are in the evolutionary phase of the ``blue hook''; a very short-lived phase just before the Hertzsprung gap. Interestingly, the initial mass of $3{\rm M}_{\bigodot}$ found for the ionizing source has the same age. As previously suggested by \citet{Exter2010}, the PN progenitor with an initial mass slightly greater than $2.7{\rm M}_{\bigodot}$ can be coeval with the A-type stars, and it recently left the AGB phase. But, they adopted the system age of about 520~Myr according to the Y$^2$ evolutionary tracks \citep{Yi2003,Demarque2004}.
 
The effective temperature and stellar luminosity obtained for both models correspond to the progenitor mass of $3{\rm M}_{\bigodot}$.
However, the strong nitrogen enrichment seen in the nebula is inconsistent with this initial mass, so 
another mixing process rather than the hot-bottom burning (HBB) occurs at substantially lower initial
masses than the stellar evolutionary theory 
suggests for AGB-phase
\citep{Herwig2005,Karakas2007,Karakas2009}. The stellar models developed by \citet{Karakas2007} 
indicate that HBB occurs in intermediate-mass AGB stars with the initial mass of $\geqslant5{\rm M}_{\bigodot}$ for the metallicity of $Z=0.02$; and
$\geqslant4{\rm M}_{\bigodot}$ for $Z=0.004$--$0.008$. However, they found that 
a low-metallicity AGB star ($Z=0.0001$) with the progenitor mass of $3{\rm M}_{\bigodot}$ can also
experience HBB. Our determination of the argon abundance in SuWt 2 (see Table\,\ref{suwt2:tab:modelparameters}) indicates that it does not belong to the low-metallicity stellar population; thus, another non-canonical mixing process made the abundance pattern of this PN.

The stellar evolution also depends on the chemical composition of the progenitor, namely the helium content ($Y$) and the metallicity ($Z$), as well as the efficiency of convection \citep[see e.g.][]{Salaris2005}. More helium increases the H-burning efficiency, and more metallicity makes the stellar structure fainter and cooler. Any change in the outer layer convection affects the effective temperature.  There are other non-canonical physical processes such as rotation, magnetic field and mass-loss during Roche lobe overflow (RLOF) in a binary system, which significantly affect stellar evolution. \citet{Ekstrom2012} calculated a grid of stellar evolutionary tracks with rotation, and found that N/H at the surface in rotating models is higher than non-rotating models in the stellar evolutionary tracks 
until the end of the central hydrogen- and helium-burning phases prior to the AGB stage. The Modules for Experiments in Stellar Astrophysics (\textsc{mesa}) code developed by \citet{Paxton2011,Paxton2013} indicates that an increase in the rotation rate (or angular momentum) enhances the mass-loss rate. The rotationally induced and magnetically induced mixing processes certainly influence the evolution of intermediate-mass stars, which need further studies by \textsc{mesa}. The mass-loss in a binary or even triple system is much more complicated than a single rotating star, and many non-canonical physical parameters are involved \citep[see e.g. \textsc{binstar} code by][]{Siess2006,Siess2013}. \citet{Chen2002} used the Cambridge stellar evolution (\textsc{stars}) code developed by \citet{Eggleton1971,Eggleton1972,Eggleton1973} to study numerically evolution of Population I binaries, and produced a helium-rich outer layer.  Similarly, \citet{Benvenuto2003,Benvenuto2005} developed a helium white dwarf from a low mass progenitor in a close binary system. A helium enrichment in the our layer can considerably influence other elements through the helium-burning mixing process.

\section{Conclusion}
\label{suwt2:sec:conclusions}

In this paper we have analysed  new optical integral-field
spectroscopy of the PN SuWt~2 to study detailed
ionized gas properties, and to infer the properties of the unobserved
hot ionizing source located in the centre of the nebula. 
The spatially resolved emission-line maps in the light of $[$N~{\sc
  ii}$]$ $\lambda$6584 have described the kinematic structure of the
ring. The previous kinematic model \citep{Jones2010} allowed us to estimate the nebula's age and large-scale kinematics in the Galaxy. An empirical
analysis of the emission line spectrum led to our initial
determination of the ionization structure of the nebula. The plasma
diagnostics revealed as expected that the inner region is hotter and more excited than the outer regions of the nebula, and is less dense. 
The ionic abundances of He, N, O, Ne, S and Ar were derived based on the empirical methods and adopted mean electron temperatures estimated from the observed $[$O~{\sc iii}$]$ emission lines and electron densities from the observed $[$S~{\sc ii}$]$ emission lines. 

We constructed photoionization models for the ring and interior of SuWt~2. This model consisted of a higher density torus (the ring) surrounding a low-density oblate spheroid (the interior disc). We assumed a homogeneous abundance distribution consisting of eight abundant elements.  The initial aim was to find a model that could reproduce the flux intensities, thermal balance structure and ionization structure as derived from by the observations. 
We incorporated NLTE model atmospheres to model the ionizing flux of the central star. Using a hydrogen-rich model atmosphere, we first fitted all the observed line fluxes, but the time-scale of the evolutionary track was not consistent with the nebula's age. Subsequently, we decided to use hydrogen-deficient stellar atmospheres implying a VLTP (born-again scenario), and  longer time-scales were likely to be in better agreement with the dynamical age of the nebula. 
Although the results obtained by the two models of SuWt~2 are in broad agreement with the observations, each model has slightly different chemical abundances and very different stellar parameters. We found a fairly good fit to a hydrogen-deficient central star with a mass of $\sim0.64{\rm M}_{\bigodot}$ with an initial (model) mass of $\sim3{\rm M}_{\bigodot}$.  The evolutionary track of \citet{Bloecker1995} implies that this central star has a post-AGB age of about 25\,000 yr.   
Interestingly, our kinematic analysis \citep[based on $v_{\rm exp}$ from][]{Jones2010} implies a nebular true age of about 23\,400--26\,300 yr. 

Table~\ref{suwt2:tab:modelparameters} lists two best-fitting photoionization models obtained for SuWt~2. The hydrogen-rich model atmosphere (Model 1) has a normal 
evolutionary path and yields a progenitor mass of $3{\rm M}_{\bigodot}$, a dynamical age of 7,500 yr and nebular ${\rm N}/{\rm O}= 0.939$ (by number). 
The PG 1159 model atmosphere (Model 2) is the most probable solution, which can be explained by a VLTP phase or born-again scenario: VLTP $\rightarrow$ [WCL] $\rightarrow$ [WCE] $\rightarrow$ [WC]-PG\,1159 $\rightarrow$PG\,1159 \citep{Blocker2001,Herwig2001,MillerBertolami2006a,Werner2006}. The PG\,1159 model yields ${\rm N}/{\rm O}=0.816$ and a stellar temperature of $T_{\rm eff}=160$ kK corresponding to the progenitor mass of $3{\rm M}_{\bigodot}$ and much longer evolutionary time-scale.  
The VLTP can be characterized as the helium-burning model, but this cannot purely explain the fast stellar winds ($V_{\infty}=2000$\,km\,s$^{-1}$) of typical [WCE] stars.  It is possible that an external mechanism such as the tidal force of a companion and mass transfer to an accretion disc, or the strong stellar magnetic field of a companion can trigger (late) thermal pulses during post-AGB evolution.

The abundance pattern of SuWt~2 is representative of a nitrogen-rich PN, which is normally considered to be the product of a relatively massive progenitor star \citep{Becker1980,Kingsburgh1994}.    
Recent work suggests that HBB, which enhances the helium and nitrogen, and decreases oxygen and carbon, occurs only for initial masses of $\geq$5\,${\rm M}_{\bigodot}$  \citep[$Z=0.02$;][]{Karakas2007,Karakas2009}; hence, the nitrogen enrichment seen in the nebula appears to result from an additional mixing process active in stars down to a mass of 3${\rm M}_{\bigodot}$. Additional physical processes such as rotation increase the mass-loss rate \citep{Paxton2013} and nitrogen abundance at the stellar surface \citep[end of the core H- and He-burning phases;][]{Ekstrom2012}. The mass-loss via RLOF in a binary (or triple) system can produce a helium-rich outer layer \citep[][]{Chen2002,Benvenuto2005}, which significantly affects other elements at the surface. 

\section*{Acknowledgments}

AD warmly acknowledges the award of an international Macquarie University
Research Excellence Scholarship (iMQRES). 
QAP acknowledges support from Macquarie University and the Australian Astronomical Observatory.
We wish to thank Nick Wright and Michael Barlow for interesting
discussions. 
We are grateful to David J. Frew for the initial help and discussion. 
We thank Anna V. Kovacevic for carrying out the May 2009 2.3 m observing run, and Lizette Guzman-Ramirez for assisting her with it.
We also thank Travis Stenborg for assisting AD with the 2012 August 2.3 m observing run. 
AD thanks Milorad Stupar for his assistance in the reduction process. 
We would like to thank the staff at the ANU Siding Spring Observatory for their support, especially Donna Burton.
This work was supported by the NCI National Facility at the ANU.
We would also like to thank an anonymous referee for helpful
suggestions that greatly improved the paper.

\label{lastpage}


\begin{thebibliography}{85}
\expandafter\ifx\csname natexlab\endcsname\relax\def\natexlab#1{#1}\fi

\bibitem[{{Aller} \& {Liller}(1968)}]{Aller1968}
{Aller} L.~H., {Liller} W., 1968, {Planetary Nebulae}, {Middlehurst} B.~M.,
  {Aller} L.~H., eds., the University of Chicago Press, p. 483

\bibitem[{{Becker} \& {Iben}(1980)}]{Becker1980}
{Becker} S.~A., {Iben}, Jr. I., 1980, \apj, 237, 111

\bibitem[{{Benvenuto} \& {De Vito}(2003)}]{Benvenuto2003}
{Benvenuto} O.~G., {De Vito} M.~A., 2003, \mnras, 342, 50

\bibitem[{{Benvenuto} \& {De Vito}(2005)}]{Benvenuto2005}
{Benvenuto} O.~G., {De Vito} M.~A., 2005, \mnras, 362, 891

\bibitem[{{Bl\"{o}cker}(1995)}]{Bloecker1995}
{Bl\"{o}cker} T., 1995, \aap, 299, 755

\bibitem[{{Bl{\"o}cker}(2001)}]{Blocker2001}
{Bl{\"o}cker} T., 2001, \apss, 275, 1

\bibitem[{{Bohigas}(2008)}]{Bohigas2008}
{Bohigas} J., 2008, \apj, 674, 954

\bibitem[{{Bond}(2000)}]{Bond2000}
{Bond} H.~E., 2000, in Astronomical Society of the Pacific Conference Series,
  Vol. 199, Asymmetrical Planetary Nebulae II: From Origins to Microstructures,
  {J.~H.~Kastner, N.~Soker, \& S.~Rappaport}, ed., p. 115

\bibitem[{{Bond} {et~al}\mbox{.}(2002){Bond}, {O'Brien}, {Sion}, {Mullan},
  {Exter}, {Pollacco}, \& {Webbink}}]{Bond2002}
{Bond} H.~E., {O'Brien} M.~S., {Sion} E.~M., {Mullan} D.~J., {Exter} K.,
  {Pollacco} D.~L., {Webbink} R.~F., 2002, in Astronomical Society of the
  Pacific Conference Series, Vol. 279, Exotic Stars as Challenges to Evolution,
  {C.~A.~Tout \& W.~van Hamme}, ed., p. 239

\bibitem[{{Bond} {et~al}\mbox{.}(2003){Bond}, {Pollacco}, \&
  {Webbink}}]{Bond2003}
{Bond} H.~E., {Pollacco} D.~L., {Webbink} R.~F., 2003, \aj, 125, 260

\bibitem[{{Chen} \& {Han}(2002)}]{Chen2002}
{Chen} X., {Han} Z., 2002, \mnras, 335, 948

\bibitem[{{Demarque} {et~al}\mbox{.}(2004){Demarque}, {Woo}, {Kim}, \&
  {Yi}}]{Demarque2004}
{Demarque} P., {Woo} J.-H., {Kim} Y.-C., {Yi} S.~K., 2004, \apjs, 155, 667

\bibitem[{{Dopita} {et~al}\mbox{.}(2007){Dopita}, {Hart}, {McGregor}, {Oates},
  {Bloxham}, \& {Jones}}]{Dopita2007}
{Dopita} M., {Hart} J., {McGregor} P., {Oates} P., {Bloxham} G., {Jones} D.,
  2007, \apss, 310, 255

\bibitem[{{Dopita} {et~al}\mbox{.}(2010){Dopita}, {Rhee}, {Farage}, {McGregor},
  {Bloxham}, {Green}, {Roberts}, {Neilson}, {Wilson}, {Young}, {Firth},
  {Busarello}, \& {Merluzzi}}]{Dopita2010}
{Dopita} M. {et~al.}, 2010, \apss, 327, 245

\bibitem[{{Dopita} \& {Meatheringham}(1990)}]{Dopita1990}
{Dopita} M.~A., {Meatheringham} S.~J., 1990, \apj, 357, 140

\bibitem[{{Dopita} \& {Meatheringham}(1991)}]{Dopita1991}
{Dopita} M.~A., {Meatheringham} S.~J., 1991, \apj, 377, 480

\bibitem[{{Dopita} {et~al}\mbox{.}(1996){Dopita}, {Vassiliadis},
  {Meatheringham}, {Bohlin}, {Ford}, {Harrington}, {Wood}, {Stecher}, \&
  {Maran}}]{Dopita1996}
{Dopita} M.~A. {et~al.}, 1996, \apj, 460, 320

\bibitem[{{Eggleton}(1971)}]{Eggleton1971}
{Eggleton} P.~P., 1971, \mnras, 151, 351

\bibitem[{{Eggleton}(1972)}]{Eggleton1972}
{Eggleton} P.~P., 1972, \mnras, 156, 361

\bibitem[{{Eggleton}(1973)}]{Eggleton1973}
{Eggleton} P.~P., 1973, \mnras, 163, 279

\bibitem[{{Ekstr{\"o}m} {et~al}\mbox{.}(2012){Ekstr{\"o}m}, {Georgy},
  {Eggenberger}, {Meynet}, {Mowlavi}, {Wyttenbach}, {Granada}, {Decressin},
  {Hirschi}, {Frischknecht}, {Charbonnel}, \& {Maeder}}]{Ekstrom2012}
{Ekstr{\"o}m} S. {et~al.}, 2012, \aap, 537, A146

\bibitem[{{Ercolano} {et~al}\mbox{.}(2005){Ercolano}, {Barlow}, \&
  {Storey}}]{Ercolano2005}
{Ercolano} B., {Barlow} M.~J., {Storey} P.~J., 2005, \mnras, 362, 1038

\bibitem[{{Ercolano} {et~al}\mbox{.}(2003a){Ercolano}, {Barlow}, {Storey}, \&
  {Liu}}]{Ercolano2003a}
{Ercolano} B., {Barlow} M.~J., {Storey} P.~J., {Liu} X.-W., 2003a, \mnras, 340,
  1136

\bibitem[{{Ercolano} {et~al}\mbox{.}(2003c){Ercolano}, {Barlow}, {Storey},
  {Liu}, {Rauch}, \& {Werner}}]{Ercolano2003c}
{Ercolano} B., {Barlow} M.~J., {Storey} P.~J., {Liu} X.-W., {Rauch} T.,
  {Werner} K., 2003c, \mnras, 344, 1145

\bibitem[{{Ercolano} {et~al}\mbox{.}(2003b){Ercolano}, {Morisset}, {Barlow},
  {Storey}, \& {Liu}}]{Ercolano2003b}
{Ercolano} B., {Morisset} C., {Barlow} M.~J., {Storey} P.~J., {Liu} X.-W.,
  2003b, \mnras, 340, 1153

\bibitem[{{Ercolano} \& {Storey}(2006)}]{Ercolano2006}
{Ercolano} B., {Storey} P.~J., 2006, \mnras, 372, 1875

\bibitem[{{Ercolano} {et~al}\mbox{.}(2008){Ercolano}, {Young}, {Drake}, \&
  {Raymond}}]{Ercolano2008}
{Ercolano} B., {Young} P.~R., {Drake} J.~J., {Raymond} J.~C., 2008, \apjs, 175,
  534

\bibitem[{{Exter} {et~al}\mbox{.}(2003){Exter}, {Bond}, {Pollacco}, \&
  {Dufton}}]{Exter2003}
{Exter} K., {Bond} H., {Pollacco} D., {Dufton} P., 2003, in IAU Symposium, Vol.
  209, Planetary Nebulae: Their Evolution and Role in the Universe, {S.~Kwok,
  M.~Dopita, \& R.~Sutherland}, ed., p. 234

\bibitem[{{Exter} {et~al}\mbox{.}(2010){Exter}, {Bond}, {Stassun}, {Smalley},
  {Maxted}, \& {Pollacco}}]{Exter2010}
{Exter} K., {Bond} H.~E., {Stassun} K.~G., {Smalley} B., {Maxted} P.~F.~L.,
  {Pollacco} D.~L., 2010, \aj, 140, 1414

\bibitem[{{Frew}(2008)}]{Frew2008}
{Frew} D.~J., 2008, PhD thesis, Department of Physics, Macquarie University,
  NSW 2109, Australia

\bibitem[{{Frew} {et~al}\mbox{.}(2013a){Frew}, {Boji{\v c}i{\'c}}, \&
  {Parker}}]{Frew2013a}
{Frew} D.~J., {Boji{\v c}i{\'c}} I.~S., {Parker} Q.~A., 2013a, \mnras, 431, 2

\bibitem[{{Frew} {et~al}\mbox{.}(2013b){Frew}, {Boji\v{c}i\'c}, {Parker},
  {Pierce}, {Gunawardhana}, \& {Reid}}]{Frew2013b}
{Frew} D.~J., {Boji\v{c}i\'c} I.~S., {Parker} Q.~A., {Pierce} M.~J.,
  {Gunawardhana} M.~L.~P., {Reid} W.~A., 2013b, MNRAS, submitted, e-print:
  arXiv:1303.4555

\bibitem[{{Frew} \& {Parker}(2006)}]{Frew2006}
{Frew} D.~J., {Parker} Q.~A., 2006, in IAU Symposium, Vol. 234, Planetary
  Nebulae in our Galaxy and Beyond, {Barlow} M.~J., {M{\'e}ndez} R.~H., eds.,
  pp. 49--54

\bibitem[{{Gesicki} \& {Zijlstra}(2000)}]{Gesicki2000}
{Gesicki} K., {Zijlstra} A.~A., 2000, \aap, 358, 1058

\bibitem[{{Gon{\c c}alves} {et~al}\mbox{.}(2006){Gon{\c c}alves}, {Ercolano},
  {Carnero}, {Mampaso}, \& {Corradi}}]{Gonccalves2006}
{Gon{\c c}alves} D.~R., {Ercolano} B., {Carnero} A., {Mampaso} A., {Corradi}
  R.~L.~M., 2006, \mnras, 365, 1039

\bibitem[{{Gon{\c c}alves} {et~al}\mbox{.}(2012){Gon{\c c}alves}, {Wesson},
  {Morisset}, {Barlow}, \& {Ercolano}}]{Gonccalves2012}
{Gon{\c c}alves} D.~R., {Wesson} R., {Morisset} C., {Barlow} M., {Ercolano} B.,
  2012, in IAU Symposium, Vol. 283, IAU Symposium, pp. 144--147

\bibitem[{{Harrington} {et~al}\mbox{.}(1982){Harrington}, {Seaton}, {Adams}, \&
  {Lutz}}]{Harrington1982}
{Harrington} J.~P., {Seaton} M.~J., {Adams} S., {Lutz} J.~H., 1982, \mnras,
  199, 517

\bibitem[{{Herwig}(2001)}]{Herwig2001}
{Herwig} F., 2001, \apss, 275, 15

\bibitem[{{Herwig}(2005)}]{Herwig2005}
{Herwig} F., 2005, \araa, 43, 435

\bibitem[{{Howarth}(1983)}]{Howarth1983}
{Howarth} I.~D., 1983, \mnras, 203, 301

\bibitem[{{Howarth} \& {Adams}(1981)}]{Howarth1981}
{Howarth} I.~D., {Adams} S., 1981, Program {EQUIB}. University College London,
  ({Wesson} R., 2009, Converted to FORTRAN 90)

\bibitem[{{Hummer} \& {Storey}(1987)}]{Hummer1987}
{Hummer} D.~G., {Storey} P.~J., 1987, \mnras, 224, 801

\bibitem[{{Iben} \& {Renzini}(1983)}]{Iben1983}
{Iben}, Jr. I., {Renzini} A., 1983, \araa, 21, 271

\bibitem[{{Jacoby} {et~al}\mbox{.}(2001){Jacoby}, {Ferland}, \&
  {Korista}}]{Jacoby2001}
{Jacoby} G.~H., {Ferland} G.~J., {Korista} K.~T., 2001, \apj, 560, 272

\bibitem[{{Jones} {et~al}\mbox{.}(2010){Jones}, {Lloyd}, {Mitchell},
  {Pollacco}, {O'Brien}, \& {Vaytet}}]{Jones2010}
{Jones} D., {Lloyd} M., {Mitchell} D.~L., {Pollacco} D.~L., {O'Brien} T.~J.,
  {Vaytet} N.~M.~H., 2010, \mnras, 401, 405

\bibitem[{{Karakas} \& {Lattanzio}(2007)}]{Karakas2007}
{Karakas} A., {Lattanzio} J.~C., 2007, \pasa, 24, 103

\bibitem[{{Karakas} {et~al}\mbox{.}(2009){Karakas}, {van Raai}, {Lugaro},
  {Sterling}, \& {Dinerstein}}]{Karakas2009}
{Karakas} A.~I., {van Raai} M.~A., {Lugaro} M., {Sterling} N.~C., {Dinerstein}
  H.~L., 2009, \apj, 690, 1130

\bibitem[{{Kingsburgh} \& {Barlow}(1994)}]{Kingsburgh1994}
{Kingsburgh} R.~L., {Barlow} M.~J., 1994, \mnras, 271, 257

\bibitem[{{Landi} {et~al}\mbox{.}(2006){Landi}, {Del Zanna}, {Young}, {Dere},
  {Mason}, \& {Landini}}]{Landi2006}
{Landi} E., {Del Zanna} G., {Young} P.~R., {Dere} K.~P., {Mason} H.~E.,
  {Landini} M., 2006, \apjs, 162, 261

\bibitem[{{Liu} {et~al}\mbox{.}(2000){Liu}, {Storey}, {Barlow}, {Danziger},
  {Cohen}, \& {Bryce}}]{Liu2000}
{Liu} X.-W., {Storey} P.~J., {Barlow} M.~J., {Danziger} I.~J., {Cohen} M.,
  {Bryce} M., 2000, \mnras, 312, 585

\bibitem[{{Markwardt}(2009)}]{Markwardt2009}
{Markwardt} C.~B., 2009, in Astronomical Society of the Pacific Conference
  Series, Vol. 411, Astronomical Data Analysis Software and Systems XVIII,
  {Bohlender} D.~A., {Durand} D., {Dowler} P., eds., p. 251

\bibitem[{{Miller Bertolami} \& {Althaus}(2006)}]{MillerBertolami2006a}
{Miller Bertolami} M.~M., {Althaus} L.~G., 2006, \aap, 454, 845

\bibitem[{{Miller Bertolami} {et~al}\mbox{.}(2006){Miller Bertolami},
  {Althaus}, {Serenelli}, \& {Panei}}]{MillerBertolami2006}
{Miller Bertolami} M.~M., {Althaus} L.~G., {Serenelli} A.~M., {Panei} J.~A.,
  2006, \aap, 449, 313

\bibitem[{{Monteiro} \& {Falceta-Gon{\c c}alves}(2011)}]{Monteiro2011}
{Monteiro} H., {Falceta-Gon{\c c}alves} D., 2011, \apj, 738, 174

\bibitem[{{Mor\'{e}}(1977)}]{More1977}
{Mor\'{e}} J., 1977, in Numerical Analysis, Watson G.~A., ed., Vol. vol. 630,
  Springer-Verlag: Berlin, p. 105

\bibitem[{{Parker} {et~al}\mbox{.}(2005){Parker}, {Phillipps}, {Pierce}, \&
  et~al.}]{Parker2005}
{Parker} Q.~A., {Phillipps} S., {Pierce} M., et~al., 2005, \mnras, 362, 689

\bibitem[{{Paxton} {et~al}\mbox{.}(2011){Paxton}, {Bildsten}, {Dotter},
  {Herwig}, {Lesaffre}, \& {Timmes}}]{Paxton2011}
{Paxton} B., {Bildsten} L., {Dotter} A., {Herwig} F., {Lesaffre} P., {Timmes}
  F., 2011, \apjs, 192, 3

\bibitem[{{Paxton} {et~al}\mbox{.}(2013){Paxton}, {Cantiello}, {Arras},
  {Bildsten}, {Brown}, {Dotter}, {Mankovich}, {Montgomery}, {Stello}, {Timmes},
  \& {Townsend}}]{Paxton2013}
{Paxton} B. {et~al.}, 2013, ArXiv e-print: 1301.0319

\bibitem[{{Rauch}(2003)}]{Rauch2003}
{Rauch} T., 2003, \aap, 403, 709

\bibitem[{{Reid} {et~al}\mbox{.}(2009){Reid}, {Menten}, {Zheng}, {Brunthaler},
  {Moscadelli}, {Xu}, {Zhang}, {Sato}, {Honma}, {Hirota}, {Hachisuka}, {Choi},
  {Moellenbrock}, \& {Bartkiewicz}}]{Reid2009}
{Reid} M.~J. {et~al.}, 2009, \apj, 700, 137

\bibitem[{{Reid} \& {Parker}(2010)}]{Reid2010}
{Reid} W.~A., {Parker} Q.~A., 2010, \pasa, 27, 187

\bibitem[{{Roeser} {et~al}\mbox{.}(2010){Roeser}, {Demleitner}, \&
  {Schilbach}}]{Roeser2010}
{Roeser} S., {Demleitner} M., {Schilbach} E., 2010, \aj, 139, 2440

\bibitem[{{Salaris} \& {Cassisi}(2005)}]{Salaris2005}
{Salaris} M., {Cassisi} S., 2005, {Evolution of Stars and Stellar Populations}

\bibitem[{{Sch\"{o}nberner}(1983)}]{Schoenberner1983}
{Sch\"{o}nberner} D., 1983, \apj, 272, 708

\bibitem[{{Schwarz} {et~al}\mbox{.}(1992){Schwarz}, {Corradi}, \&
  {Melnick}}]{Schwarz1992}
{Schwarz} H.~E., {Corradi} R.~L.~M., {Melnick} J., 1992, \aaps, 96, 23

\bibitem[{{Schwarz} \& {Monteiro}(2006)}]{Schwarz2006}
{Schwarz} H.~E., {Monteiro} H., 2006, \apj, 648, 430

\bibitem[{{Seaton}(1979{\natexlab{a}})}]{Seaton1979a}
{Seaton} M.~J., 1979{\natexlab{a}}, \mnras, 187, 785

\bibitem[{{Seaton}(1979{\natexlab{b}})}]{Seaton1979b}
{Seaton} M.~J., 1979{\natexlab{b}}, \mnras, 187, 73P

\bibitem[{{Siess}(2006)}]{Siess2006}
{Siess} L., 2006, \aap, 448, 717

\bibitem[{{Siess} {et~al}\mbox{.}(2013){Siess}, {Izzard}, {Davis}, \&
  {Deschamps}}]{Siess2013}
{Siess} L., {Izzard} R.~G., {Davis} P.~J., {Deschamps} R., 2013, \aap, 550,
  A100

\bibitem[{{Smith} {et~al}\mbox{.}(2007){Smith}, {Bally}, \&
  {Walawender}}]{Smith2007}
{Smith} N., {Bally} J., {Walawender} J., 2007, \aj, 134, 846

\bibitem[{{Smits}(1996)}]{Smits1996}
{Smits} D.~P., 1996, \mnras, 278, 683

\bibitem[{{Steffen} \& {L{\'o}pez}(2006)}]{Steffen2006}
{Steffen} W., {L{\'o}pez} J.~A., 2006, \rmxaa, 42, 99

\bibitem[{{Storey} \& {Hummer}(1995)}]{Storey1995}
{Storey} P.~J., {Hummer} D.~G., 1995, \mnras, 272, 41

\bibitem[{{Tylenda} {et~al}\mbox{.}(2003){Tylenda}, {Si{\'o}dmiak},
  {G{\'o}rny}, {Corradi}, \& {Schwarz}}]{Tylenda2003}
{Tylenda} R., {Si{\'o}dmiak} N., {G{\'o}rny} S.~K., {Corradi} R.~L.~M.,
  {Schwarz} H.~E., 2003, \aap, 405, 627

\bibitem[{{van Dokkum}(2001)}]{Dokkum2001}
{van Dokkum} P.~G., 2001, \pasp, 113, 1420

\bibitem[{{Vassiliadis} \& {Wood}(1994)}]{Vassiliadis1994}
{Vassiliadis} E., {Wood} P.~R., 1994, \apjs, 92, 125

\bibitem[{{Verner} \& {Yakovlev}(1995)}]{Verner1995}
{Verner} D.~A., {Yakovlev} D.~G., 1995, \aaps, 109, 125

\bibitem[{{Verner} {et~al}\mbox{.}(1993){Verner}, {Yakovlev}, {Band}, \&
  {Trzhaskovskaya}}]{Verner1993}
{Verner} D.~A., {Yakovlev} D.~G., {Band} I.~M., {Trzhaskovskaya} M.~B., 1993,
  Atomic Data and Nuclear Data Tables, 55, 233

\bibitem[{{Werner} \& {Herwig}(2006)}]{Werner2006}
{Werner} K., {Herwig} F., 2006, \pasp, 118, 183

\bibitem[{{Wesson} {et~al}\mbox{.}(2012){Wesson}, {Stock}, \&
  {Scicluna}}]{Wesson2012}
{Wesson} R., {Stock} D.~J., {Scicluna} P., 2012, \mnras, 422, 3516

\bibitem[{{West}(1976)}]{West1976}
{West} R.~M., 1976, \pasp, 88, 896

\bibitem[{{Wright} {et~al}\mbox{.}(2011){Wright}, {Barlow}, {Ercolano}, \&
  {Rauch}}]{Wright2011}
{Wright} N.~J., {Barlow} M.~J., {Ercolano} B., {Rauch} T., 2011, \mnras, 418,
  370

\bibitem[{{Yi} {et~al}\mbox{.}(2003){Yi}, {Kim}, \& {Demarque}}]{Yi2003}
{Yi} S.~K., {Kim} Y.-C., {Demarque} P., 2003, \apjs, 144, 259

\bibitem[{{Zijlstra} {et~al}\mbox{.}(1994){Zijlstra}, {van Hoof}, {Chapman}, \&
  {Loup}}]{Zijlstra1994}
{Zijlstra} A.~A., {van Hoof} P.~A.~M., {Chapman} J.~M., {Loup} C., 1994, \aap,
  290, 228

\end{thebibliography}
\end{document}